\documentclass[twocolumn,twocolappendix]{aastex631}

\newcommand{\SFR}{\textrm{SFR}}

\newcommand{\Mgas}{M_\mathrm{gas}}

\newcommand{\yOcc}{y_\mathrm{O}^\mathrm{CC}}
\newcommand{\MO}{M_\mathrm{O}}
\newcommand{\ZO}{Z_\mathrm{O}}
\newcommand{\ZOinf}{Z_\mathrm{O}^\mathrm{inf}}
\newcommand{\ZOeq}{Z_\mathrm{O}^\mathrm{eq}}
\newcommand{\yFecc}{y_\mathrm{Fe}^\mathrm{CC}}
\newcommand{\yFeIa}{y_\mathrm{Fe}^\mathrm{Ia}}
\newcommand{\MFe}{M_\mathrm{Fe}}
\newcommand{\ZFe}{Z_\mathrm{Fe}}
\newcommand{\ZFeinf}{Z_\mathrm{Fe}^\mathrm{inf}}
\newcommand{\ZFeeq}{Z_\mathrm{Fe}^\mathrm{eq}}
\newcommand{\KFeIa}{K_\mathrm{Fe}^\mathrm{Ia}}

\newcommand{\fIa}{f_\mathrm{Ia}}

\usepackage{ulem}
\usepackage{comment}

\received{}
\revised{}
\accepted{}
\submitjournal{ApJ}

\shorttitle{zCOSMOS-deep: stellar metallicity and O/Fe}
\shortauthors{Kashino et al.}

\begin{document}

\title{The stellar mass versus stellar metallicity relation of star-forming galaxies at $1.6\le z\le3.0$ and implications for the evolution of the $\alpha$-enhancement}

\correspondingauthor{Daichi Kashino}
\email{kashino.daichi@b.mbox.nagoya-u.ac.jp}\tabletypesize{\footnotesize}

\author[0000-0001-9044-1747]{Daichi Kashino}
\affiliation{Institute for Advanced Research, Nagoya University, Nagoya 464-8601, Japan}
\affiliation{Division of Particle and Astrophysical Science, Graduate School of Science, Nagoya University, Nagoya 464-8602, Japan}

\author[0000-0002-6423-3597]{Simon J.~Lilly}
\affiliation{Department of Physics, ETH Z{\"u}rich, Wolfgang-Pauli-Strasse 27, CH-8093 Z{\"u}rich, Switzerland}

\author[0000-0002-7093-7355]{Alvio Renzini}
\affiliation{INAF--Osservatorio Astronomico di Padova, Vicolo dell'Osservatorio 5, I-35122, Padova, Italy}

\author[0000-0002-3331-9590]{Emanuele Daddi}
\affiliation{
CEA, Irfu, DAp, AIM, Universit\'e Paris-Saclay, Universit\'e de Paris, CNRS, F-91191 Gif-sur-Yvette, France}

\author[0000-0002-2318-301X]{Giovanni Zamorani}
\affiliation{
INAF -- Osservatorio di Astrofisica e Scienza dello Spazio di Bologna, via Gobetti 93/3, I-40129, Bologna, Italy
}

\author[0000-0002-0000-6977]{John D.~Silverman}
\affiliation{Kavli Institute for the Physics and Mathematics of the Universe, The University of Tokyo, Kashiwa, Japan 277-8583 (Kavli IPMU, WPI)}
\affiliation{Department of Astronomy, School of Science, The University of Tokyo, 7-3-1 Hongo, Bunkyo, Tokyo 113-0033, Japan}

\author[0000-0002-7303-4397]{Olivier Ilbert}
\affiliation{
Aix Marseille Universit{\'e}, CNRS, LAM, Laboratoire d'Astrophysique de Marseille, Marseille, France
}

\author{Yingjie Peng}
\affiliation{
Kavli Institute for Astronomy and Astrophysics, Peking University, 5 Yiheyuan Road, Beijing 100871, China}
\affiliation{
Department of Astronomy, School of Physics, Peking University, 5 Yiheyuan Road, Beijing 100871, China}

\author[0000-0002-1047-9583]{Vincenzo Mainieri}
\affiliation{
European Southern Observatory, Karl-Schwarzschild-Strasse 2, Garching bei M{\"u}nchen, Germany
}

\author[0000-0002-8900-0298]{Sandro Bardelli}
\affiliation{
INAF -- Osservatorio di Astrofisica e Scienza dello Spazio di Bologna, via Gobetti 93/3, I-40129, Bologna, Italy
}

\author[0000-0002-5845-8132]{Elena Zucca}
\affiliation{
INAF -- Osservatorio di Astrofisica e Scienza dello Spazio di Bologna, via Gobetti 93/3, I-40129, Bologna, Italy
}

\author[0000-0001-9187-3605]{Jeyhan S.~Kartaltepe}
\affiliation{
School of Physics and Astronomy, Rochester Institute of Technology, 84 Lomb Memorial Drive, Rochester NY 14623, USA
}

\author[0000-0002-1233-9998]{David B.~Sanders}
\affiliation{
Institute for Astronomy, University of Hawaii, 2680 Woodlawn Drive, Honolulu, HI 96822, USA
}

\begin{abstract}

We measure the relationship between stellar mass and stellar metallicity, the \textit{stellar} mass--metallicity relation (MZR), for 1336 star-forming galaxies at $1.6\le z\le3.0~(\left<z\right>=2.2)$ using rest-frame far-ultraviolet spectra from the zCOSMOS-deep survey.  High signal-to-noise composite spectra containing stellar absorption features are fit with population synthesis model spectra of a range of metallicity.   We find stellar metallicities, which mostly reflect iron abundances, scaling as $(Z_{\mathrm{Fe},\ast}/Z_{\mathrm{Fe},\odot})=-(0.81\pm0.01)+(0.32+0.03)\log(M_\ast/10^{10}M_\odot)$ across the mass range of $10^9\lesssim M_\ast/M_\odot\lesssim10^{11}$, being $\approx6\times$ lower than seen locally at the same masses. The instantaneous oxygen-to-iron ratio ($\alpha$-enhancement) inferred using the gas-phase oxygen MZRs, is on average found to be $\left[\mathrm{O/Fe}\right]\approx0.47$, being higher than the local $\left[\mathrm{O/Fe}\right]\approx0$.  The observed changes in [O/Fe] and [Fe/H] are reproduced in simple flow-through gas-regulator models with steady star-formation histories (SFHs) that follow the evolving main sequence.   Our models show that the [O/Fe] is determined almost entirely by the instantaneous specific star formation rate alone while being independent of the SFHs, mass, and the gas-regulation characteristics of the systems. We find that the locations of $\sim10^{10}M_\odot$ galaxies at $z\sim2$ in the [O/Fe]--metallicity planes are in remarkable agreement with the sequence of low-metallicity thick-disk stars in our Galaxy.  This manifests a beautiful concordance between the results of Galactic archaeology and observations of high-redshift Milky Way progenitors. However, there remains a question of how and when the old metal-rich, low-$\alpha$/Fe stars seen in the bulge had formed by $z\sim2$ because such a stellar population is not seen in our data and difficult to explain in the context of our models.

\end{abstract}



\section{Introduction}
\label{sec:intro}

The metallicity of galaxies can be measured for either stars or the interstellar medium (ISM) through analysis of galaxy spectra, either the stellar absorption lines in the integrated light of the stellar population(s) or the nebular emission lines from gaseous H\,{\sc ii} regions (see \citealt{2019A&ARv..27....3M} for a recent review).  The gas-phase and stellar metallicities reflect different aspects of the evolutionary history of the galaxies.

The gas-phase metallicity refers usually to the abundance of oxygen relative to hydrogen, which is often measured using emission lines in the rest-frame optical waveband that are produced by the ionized gas in star-forming regions.  The so-called ``direct method'' is based on the detection of faint auroral lines (e.g., [O\,{\sc iii}]$\lambda$4363) and determines the electron temperature and metallicity with high accuracy \citep[e.g.,][]{2013ApJ...765..140A,2014ApJ...780..122L,2019MNRAS.486.1053K,2020MNRAS.491.1427S,2020ApJ...898..142K}.  
The so-called ``strong-line'' methods, which use empirical relations between the metallicity and the ratios of strong optical emission lines (e.g., ([O\,{\sc ii}]+[O\,{\sc iii}])/H$\beta$), have been widely applied to estimate metallicities from spectra with low or moderate signal-to-noise ratio (S/N) \citep[e.g.,][]{2004MNRAS.348L..59P,2006A&A...459...85N,2008A&A...488..463M,2017MNRAS.465.1384C}.
The gas-phase metallicities reflect the ``instantaneous'' oxygen abundance in star-forming regions at the time of observation.

The stellar metallicity can be measured through absorption lines caused by metal ions, such as iron and magnesium, in the photospheres of stars.  Measurement is generally carried out by comparing observed spectra with synthetic spectra from stellar population synthesis models.  Some standardized indices, which represent the absorption depths for particular, relatively strong absorption features, or combination of absorption features, have been conventionally used for estimating the average stellar metallicity (e.g., the 1978~{\AA} index; \citealt{2004ApJ...615...98R}; see also \citealt{2008A&A...479..417H,2015ApJ...808..161O}).  
More recently, full spectral fitting that uses all the information contained in the spectra, has been employed \citep{2016ApJ...826..159S,2015ApJ...808..161O,2017ApJ...847...18Z,2018MNRAS.479...25M,2018ApJ...856...15L,2019ApJ...880L..31K,2020ApJ...902..117H,2020MNRAS.495.4430T,2020MNRAS.499.1652T}.  In any case, these measurements require a high S/N detection of the continuum emission, and are thus generally more expensive than the gas-phase metallicity measurements using the strong-line methods.  

An important point is that, in contrast to the gas-phase metallicity, the stellar metallicity is measured as the luminosity-weighted average value across all the different stellar populations that contribute to the integrated light of the galaxy at a particular wavelength.  The inferred metallicity may therefore also depend on which portion of the spectrum is used.  For example, the metallicity derived from the rest-frame optical light reflects the light from older (i.e., possibly lower metallicity) populations, whereas that from the far-ultraviolet (FUV) spectrum is more weighted towards younger (i.e., possibly higher metallicity) populations, and should thus be closer to the abundance in the gas phase.

Another key aspect is that oxygen and iron, usually traced by gas-phase and stellar metallicities respectively, form through different channels: oxygen, or the $\alpha$-elements, are supplied mainly through core-collapse supernovae (CCSNe) while the Type Ia supernovae (SNe~Ia) is the main supplier of the iron-peak elements.  Therefore the past SFH of the galaxies is imprinted in the abundance pattern between these species, often called $\alpha$-enhancement, due to the time delay of SNe~Ia (from 40~Myr to several Gyr) since the formation of their progenitor stars.  

The overall relationship between galaxy stellar mass ($M_\ast$) and metallicity, often called the mass--metallicity relation (MZR), has long been thought to be a fundamental measurement to constrain models of galaxy evolution \citep[e.g.,][]{1979A&A....80..155L}.
In the local universe, a tight correlation between these two quantities has been robustly established both for the gas-phase metallicity \citep{2004ApJ...613..898T,2013ApJ...765..140A,2020MNRAS.491..944C} and for the metallicity of the stellar component \citep{2005MNRAS.362...41G,2017ApJ...847...18Z} using the Sloan Digital Sky Survey \citep{2000AJ....120.1579Y}.  
At high redshifts, the gas-phase MZR has been measured back to $z\sim4$ by many authors, mostly using strong-line methods \citep[e.g.,][]{2006ApJ...644..813E,2011ApJ...730..137Z,2012PASJ...64...60Y,2014ApJ...791..130Z,2014ApJ...792...75Z,2015ApJ...799..138S,2017ApJ...835...88K} with only a few cases where the direct method has been used \citep{2016ApJ...828...67L,2020MNRAS.491.1427S}.  The evolution of the gas MZR is established with the metallicity monotonically decreasing at fixed (observed) $M_\ast$ with redshift.  

In contrast, the measurement of the stellar mass--stellar metallicity ($Z_\ast$) relation (hereafter stellar MZR) beyond the local universe is to date very limited \citep{2019MNRAS.487.2038C,2021A&A...646A..39C}.  A notable work was recently carried out by \citet{2019MNRAS.487.2038C}, who presented a $M_\ast$--$Z_\ast$ correlation over $10^{8.5} \lesssim M_\ast/M_\odot \lesssim 10^{10.5}$ using a large statistical sample of star-forming galaxies at $z=2.5\textrm{--}5.0$.
We are, however, still a long way from being able to constrain the evolution of the stellar MZR through cosmic time. Given the limited number of the existing measurements, independent measurements based on a different data set are highly desired.  

In this work, we measure the stellar MZR for a large sample of star-forming galaxies at $1.6\le z \le 3.0$ by utilizing the rest-frame FUV spectra obtained with the VIsible Multi-Object Spectrograph (VIMOS) mounted on the Very Large Telescope (VLT) UT3 in the zCOSMOS-deep survey (\citealt{2007ApJS..172...70L}; S.~J.~Lilly et al., in preparation).  We then explore the evolution of the oxygen-to-iron abundance pattern which is inferred from the comparison with the gas-phase metallicity measurements.

The paper is organized as follows.
Section \ref{sec:galaxy_sample} presents an overview of the observations and describes the sample selection.
Section \ref{sec:measure_metallicities} describes our spectral analysis for estimating the stellar metallicities.
The results are presented in Section \ref{sec:results}.  
Section \ref{sec:modeling} presents further attempts for interpreting the observations by using gas-regulated chemical evolution models to track the iron and oxygen chemical enrichment. 
We then compare our results and models with data of the Galactic stars to explore the link with the Galactic archaeology in Section \ref{sec:Galactis_stars}.
Section \ref{sec:summary} provides a summary of the paper.

We adopt the solar metallicity values of $12+\log(\mathrm{O/H})_\odot=8.69$ and $Z_\odot=0.0142$ \citep{2009ARA&A..47..481A}.  Here $Z$ denotes the overall metal mass fraction.  We use $\ZFe$ and $\ZO$ when specifying the element, either iron or oxygen.
The solar oxygen and iron mass fractions are $Z_\mathrm{O,\odot}=0.00561$ and $Z_\mathrm{Fe,\odot}=0.00126$, respectively.  Magnitudes are quoted on the AB system.  The \citet{2003PASP..115..763C} initial mass function (IMF) is used throughout. This paper uses a standard flat cosmology $(h=0.7, \Omega_\mathrm{M}=0.3, \Omega_\Lambda=0.7)$.

\section{Data and Galaxy sample}
\label{sec:galaxy_sample}

\subsection{Observations}
\label{sec:observations}

The zCOSMOS-deep redshift survey has observed around $10^4$ galaxies in the central $\sim~0.8~\mathrm{deg^2}$ of the COSMOS field \citep{2007ApJS..172...38S}.  Here we provide a brief description and refer the reader to \citet{2007ApJS..172...70L} and \citet{2021ApJ...909..213K} for more details. 

The observations were carried out using VLT/VIMOS \citep{2003SPIE.4841.1670L} with the low-resolution blue grism with 1$\arcsec\!\!$.0~arcsec slits, yielding a spectral resolution of $R \sim 200$ and a spectral coverage of $\approx3600\textrm{--}6700~\textrm{\AA}$.  The selection of the targets was performed based on a then-current version of the COSMOS photometric catalog.  All of the objects were color-selected through a $BzK$ \citep{2004ApJ...617..746D} or $ugr$ \citep{2004ApJ...604..534S} method with a blue magnitude cut $B_\mathrm{AB}<25.25$.  These selection criteria isolate star-forming galaxies in a range $1.4\lesssim z\lesssim3.0$ \citep{2007ApJS..172...70L}.  
Redshifts were visually inspected in 2D and 1D reduced spectra by identifying multiple prominent spectral features in the rest-frame FUV window or Ly$\alpha$ emission line and break.

\subsection{Sample selection}
\label{sec:selection}

We constructed the sample used in this paper from the full catalog of the zCOSMOS-deep survey (S.~J.~Lilly et al., in preparation).  The sample is limited to those having a clear photometric counterpart in the COSMOS2015 photometric catalog \citep{2016ApJS..224...24L}.  Galaxies detected in X-rays are also excluded to remove possible active galactic nuclei from the sample.   

The redshift range for the current analysis is limited to $1.6\le z \le 3.0$ so that the VIMOS spectrum covers the range of $\lambda_\mathrm{rest}\approx1400\textrm{--}1700~\textrm{\AA}$ for all the galaxies.  We adopt all objects with a very secure zCOSMOS-deep redshift (Confidence Class $=$ 3 or 4\footnote{The definition of the quality flags follows \citet[][see Section 4.3]{2007ApJS..172...70L}.  The evaluation of the reliability will be detailed in S.~J.~Lilly et al., in preparation.}) within the redshift range.  For those with Class~$=2$, we use only those that are consistent to within $|z_\mathrm{phot}-z_\mathrm{spec}|/(1+z_\mathrm{spec})\le0.1$ of the photometric redshift in the COSMOS2015 catalog.  The redshifts in these two categories are both estimated to be $\ge 99\%$ reliable (Lilly et al. in preparation).  We do not use any of the objects with less secure redshifts, nor any of those with broad emission lines (i.e., Class $+10$).  Finally, we excluded 34 sources (2.5\% of the remaining sample) for which the FUV continuum is barely detected or that suffered from severe spectral contamination.  

\begin{figure}[tbp]
    \begin{center}
    \includegraphics[width=3.5in]{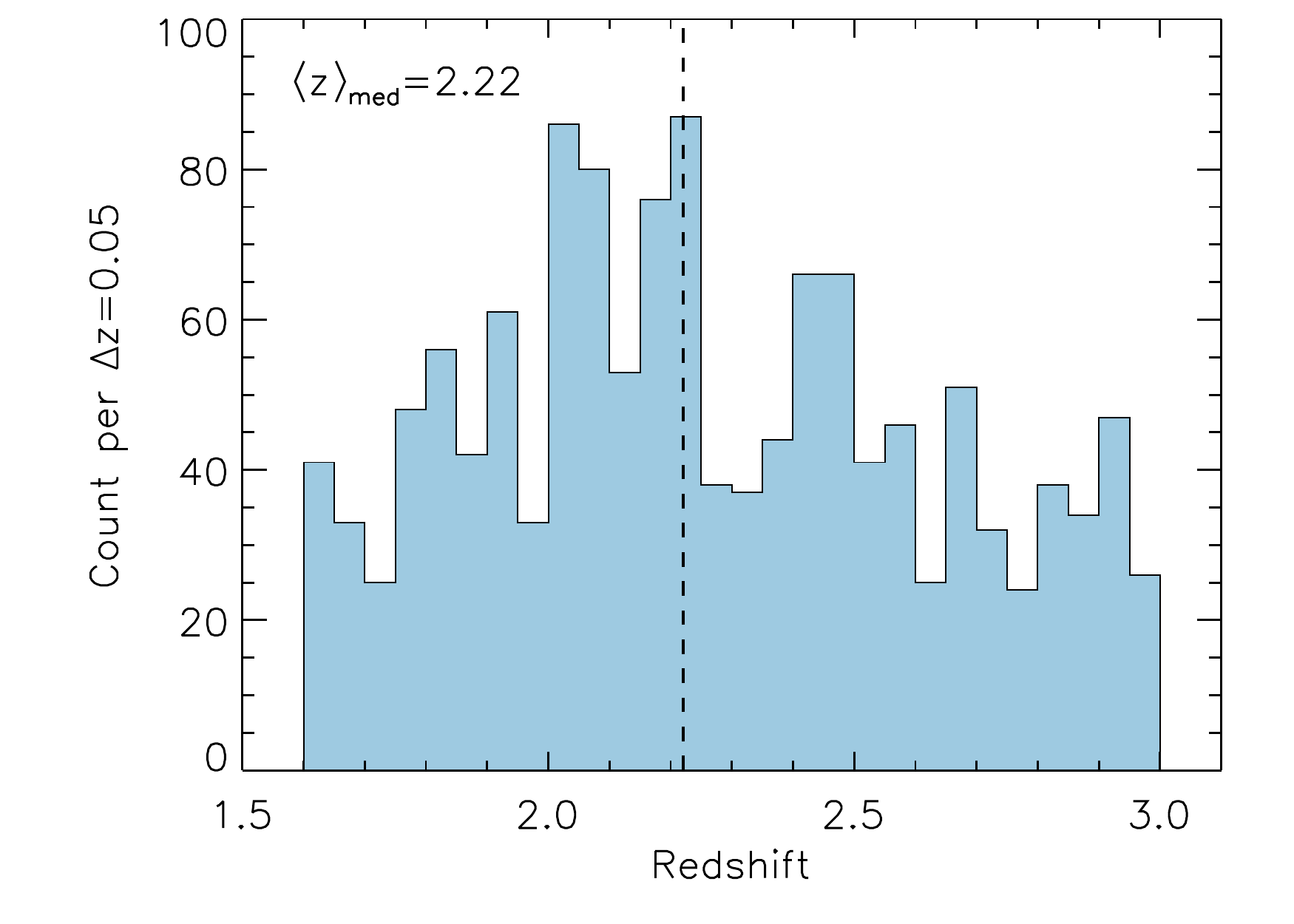}
    \caption{Spectroscopic redshift distribution for the entire sample of 1336 galaxies.  Counts per $\Delta z=0.05$ are shown. The vertical dashed line indicates the median redshift of $\left< z \right>=2.22$.
    \label{fig:zhist}}
    \end{center}
\end{figure}

The final sample consists of 1336 galaxies. The sample has a roughly flat distribution of redshift across the range of $1.6 \le z \le 3.0$ with the median redshift $\left<z\right>_\mathrm{med}=2.22$, as indicated in Figure \ref{fig:zhist}.  
We note that the redshifts were all determined with strong absorption lines due to mostly carbon and/or silicon ions in the ISM, but not with any kind of stellar iron lines.  Here, we ignore any possibility that our sample with secure redshifts may be biased toward those with strong iron features, although the strengths of the ISM absorption lines could be correlated at some level with the overall gas-phase metallicity, and thus the ISM absorption strengths \citep{2016ApJ...822...29F}.  

\subsection{Stellar mass estimation}
\label{sec:mass_estimation}

\begin{deluxetable*}{l|l}
\tablecaption{Input parameters of the SED fitting with \texttt{CIGALE} \label{tb:cigale}}
\tablehead{
    \colhead{Parameter}&
    \colhead{Values}}
\startdata
\multicolumn{2}{c}{delayed$+$burst SFH } \\
\hline
age (main population) & 1000--4500 in steps of 250~[Myr] \\ 
$e$-folding time of the delayed SFH, $\tau_0$~[Myr] & 1000, 2000, 3000~[Myr] \\
age (starburst population) & 50, 100, 150, 200~[Myr] \\ 
$e$-folding time of the late starburst, $\tau_1$~[Myr] & 10000~[Myr] \\
mass fraction of the late burst population  & 0.001, 0.003, 0.010, 0.020, 0.040, 0.100, 0.200, 0.400 \\
\hline
\multicolumn{2}{c}{Stellar population: \citet{2003MNRAS.344.1000B}} \\
\hline
initial mass function & \citet{2003PASP..115..763C} \\ 
metallicity, $Z$ & 0.008 \\
separation age between young and old populations & 10~[Myr] \\
\hline
\multicolumn{2}{c}{dust attenuation: \citet{2000ApJ...539..718C}} \\
\hline
$A_V$ in the diffuse ISM & 0.0, 0.05, 0.1, 0.2, 0.3, 0.4, 0.5, 0.6, 0.7, 0.8, \\
& 0.9, 1.0, 1.2, 1.4, 1.6, 1.8, 2.0, 2.5, 3.0, 3.5 \\
$\delta_\mathrm{ISM}$ (power-law slope for the ISM) & $-0.7$ \\
$\delta_\mathrm{BC}$ (power-law slope for birth clouds) & $-1.3$ \\
$A_V^\mathrm{ISM}/(A_V^\mathrm{BC}+A_V^\mathrm{ISM})$ & 0.44
\enddata
\end{deluxetable*}

We derived stellar masses ($M_\ast$) for individual galaxies through SED fitting based on the photometry from the COSMOS2015 catalog together with the precise spectroscopic redshifts. It should be noted that the stellar mass here denotes the mass of living stars at the time of observation, rather than the integral of the star formation rate (SFR).  We will independently perform another fitting to the composite FUV spectra using high-resolution model spectra to estimate the stellar metallicities (see Section \ref{sec:FUV_fit}).

Our SED fitting procedure uses the photometric fluxes measured within 32 broad-, intermediate-, and narrow-band filters from \textit{GALEX} near-UV to \textit{Spitzer}/IRAC ch4, as listed in Table 3 of \citet{2016ApJS..224...24L}.  For CFHT, Subaru, and UltraVISTA photometry, we used the fluxes measured in a $3\arcsec$ diameter aperture and applied offsets provided in the catalog to convert them to the total fluxes.  All the photometric bands whose rest-frame central wavelengths are within $1960\textrm{\AA}\le\lambda_\mathrm{cen}/(1+z)\le2440\textrm{\AA}$ were excluded in order to ensure that the SED fitting is not affected by the possible $2175~\textrm{\AA}$ bump feature in the SED of the galaxies \citep[see][]{2021ApJ...909..213K}.

The stellar masses and SFRs are estimated using a Python code \texttt{CIGALE} \citep{2005MNRAS.360.1413B,2009A&A...507.1793N,2019A&A...622A.103B} for the SED fitting and adopted the stellar population synthesis models of \citet{2003MNRAS.344.1000B} with a \citet{2003PASP..115..763C} IMF.  
We considered delayed star formation histories (SFHs) with an additional recent burst in order to model the long-term star formation that has formed the bulk of the stellar mass and the latest episode of star formation \citep[e.g.,][]{2016A&A...585A..43C,2017A&A...608A..41C,2017A&A...603A.102P},
\begin{equation}
    \SFR(t) = \SFR_\mathrm{delayed} (t) + \SFR_\mathrm{burst} (t)
\end{equation}
where $\SFR_\mathrm{delayed} (t) \propto t e^{-t/\tau_0}$ and $\SFR_\mathrm{burst}(t)\propto e^{-(t-t_1)/\tau_1}$ if $t>t_1$ and $\SFR_\mathrm{burst}(t)=0$ otherwise.  The parameter $t$ denotes the elapsed time since the onset of star formation, $t_1$ the galaxy age when the late episode of star formation onsets, $\tau_0$ and $\tau_1$ the e-folding times of the main stellar population and the late starburst population.  To avoid artificially inferring unrealistically young ages \citep[e.g.,][]{2010MNRAS.407..830M}, we limit the age of the main stellar population to be $\ge1~\mathrm{Gyr}$.

In \texttt{CIGALE}, we also accounted for the effect on the photometry of nebular emission lines assuming a common ionization parameter $\log U=-2.8$ and dust emission based on the templates from \citet{2014ApJ...784...83D}.  The dust attenuation is accounted for using the prescription of \citet{2000ApJ...539..718C}.  The full list of the input parameters is presented in Table \ref{tb:cigale}.

\begin{figure}[tbp]
    \begin{center}
    \includegraphics[width=3.5in]{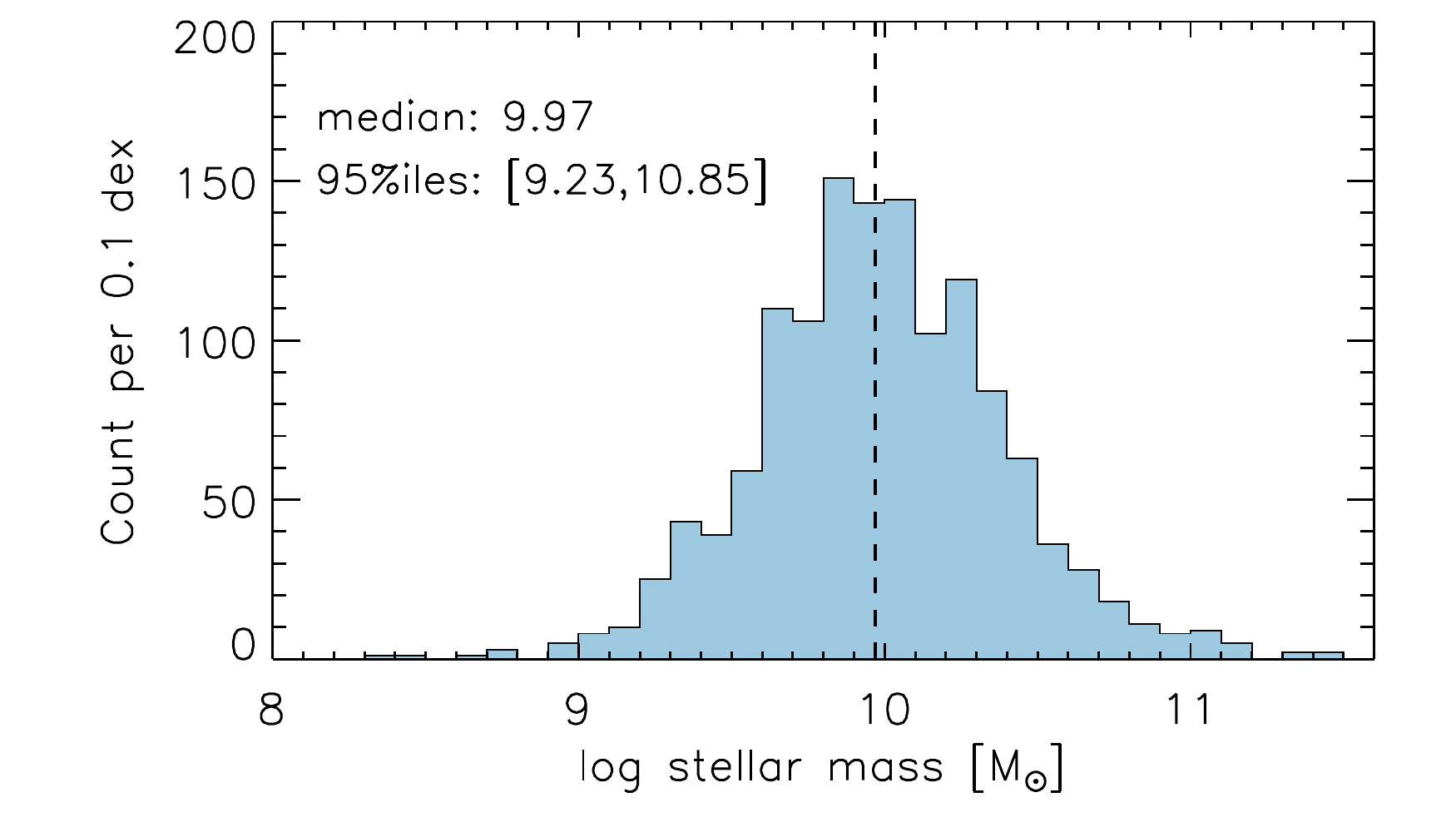}
    \includegraphics[width=3.5in]{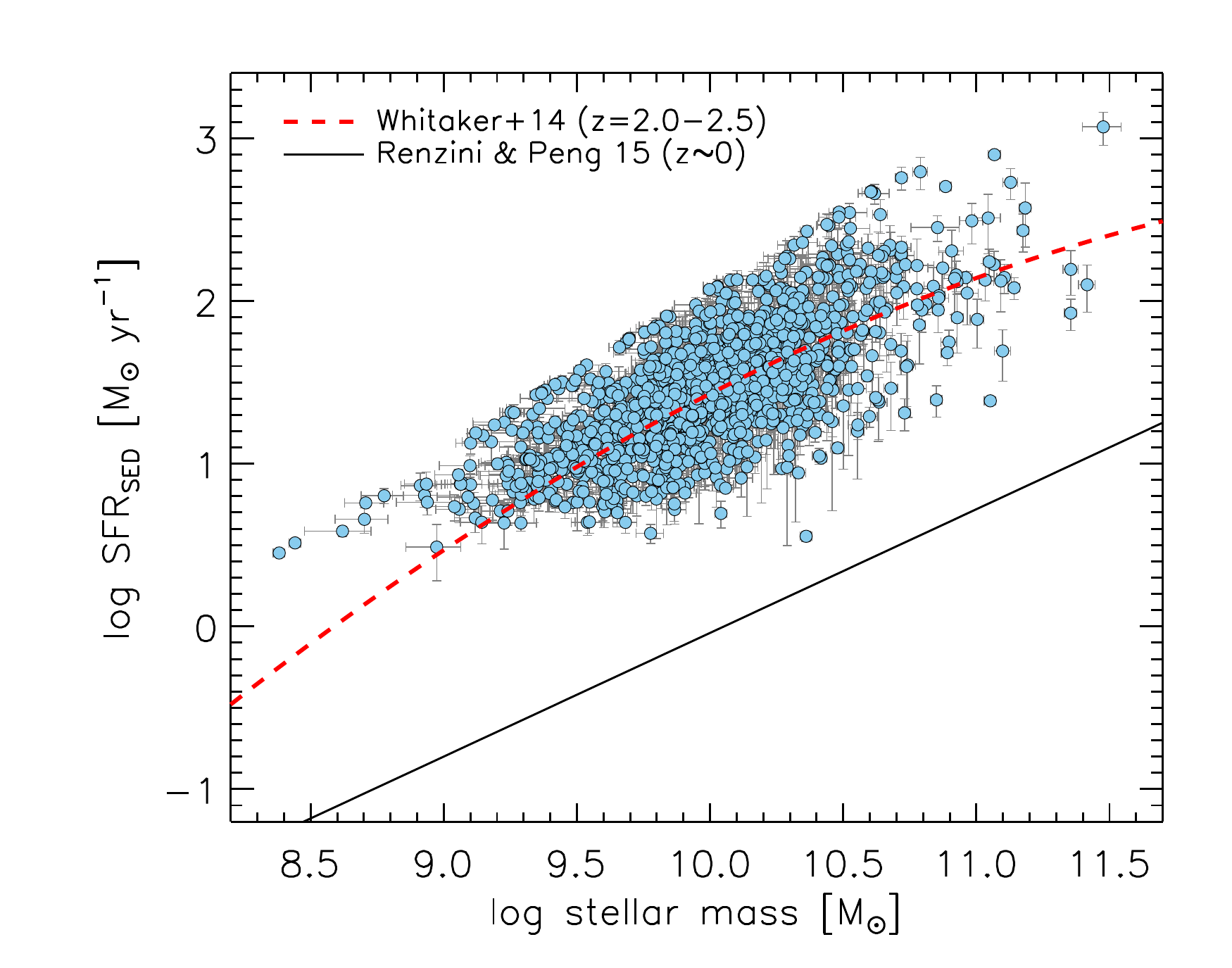}
    \caption{Results of the SED fitting.  Upper panel: distribution of estimated stellar masses.  Lower panel: stellar mass vs. instantaneous SFR for the entire sample of 1336 galaxies.  For reference, the main-sequence relations are shown, taken from \citet[][$z=2.0\textrm{--}2.5$; red dashed line]{2014ApJ...795..104W} and \citet[][$z\approx0.1$; solid line]{2015ApJ...801L..29R}.
    \label{fig:M_vs_SFR}}
    \end{center}
\end{figure}

The estimated stellar masses span the range $9.23\le \log M_\ast/M_\odot \le 10.85$ (the 2.5--97.5th percentiles) with a median value of $\left<\log(M_\ast/M_\odot)\right>_\mathrm{med} = 10.0$.  The reduced $\chi^2$ values ranges mostly between 0.4 and 3.3 with the median value of 0.95.  Figure \ref{fig:M_vs_SFR} shows the distribution of the inferred $M_\ast$ and instantaneous SFR.  For reference, the main-sequence $M_\ast\textrm{--}\SFR$ relations at similar redshifts are taken from \citet[][$z=2.0$--2.5]{2014ApJ...795..104W}.  
We also plot the local main sequence adapted from \citet{2015ApJ...801L..29R}.

Overall, our sample is in good agreement with the epoch's main sequence, indicating that it should be a representative sample at these redshifts.  Note that there is a (spurious) sharp upper boundary of the SFR distribution in Figure \ref{fig:M_vs_SFR}.  This is produced by the limited range of SFR probed by the adopted SFHs. This should be of no consequence as we do not use the SFR values of the individual galaxies in the following analyses.

\section{Stellar metallicity measurement}
\label{sec:measure_metallicities}

In this section, we describe the method to measure the stellar metallicities.  We start with the spectrophotometric calibration of each single spectrum, then improve the accuracy of the spectroscopic redshifts of the sources by fitting a common spectral template to each single spectrum.

\subsection{Flux calibration of the VIMOS spectra }
\label{sec:flux_calibration}

We adopt a spectrophotometric correction to every single spectrum of our sample following the method described in \citet{2021ApJ...909..213K}.  The spectra that are produced through the standard zCOSMOS-deep reduction pipeline were flux-calibrated based on standard star observations.  The nominal flux calibration, however, cannot perfectly correct for the effects of finite slit-width, imperfect slit-centering, and the effects of atmospheric dispersion.  

\begin{figure}[htbp]
\begin{center}
\includegraphics[width=3.3in]{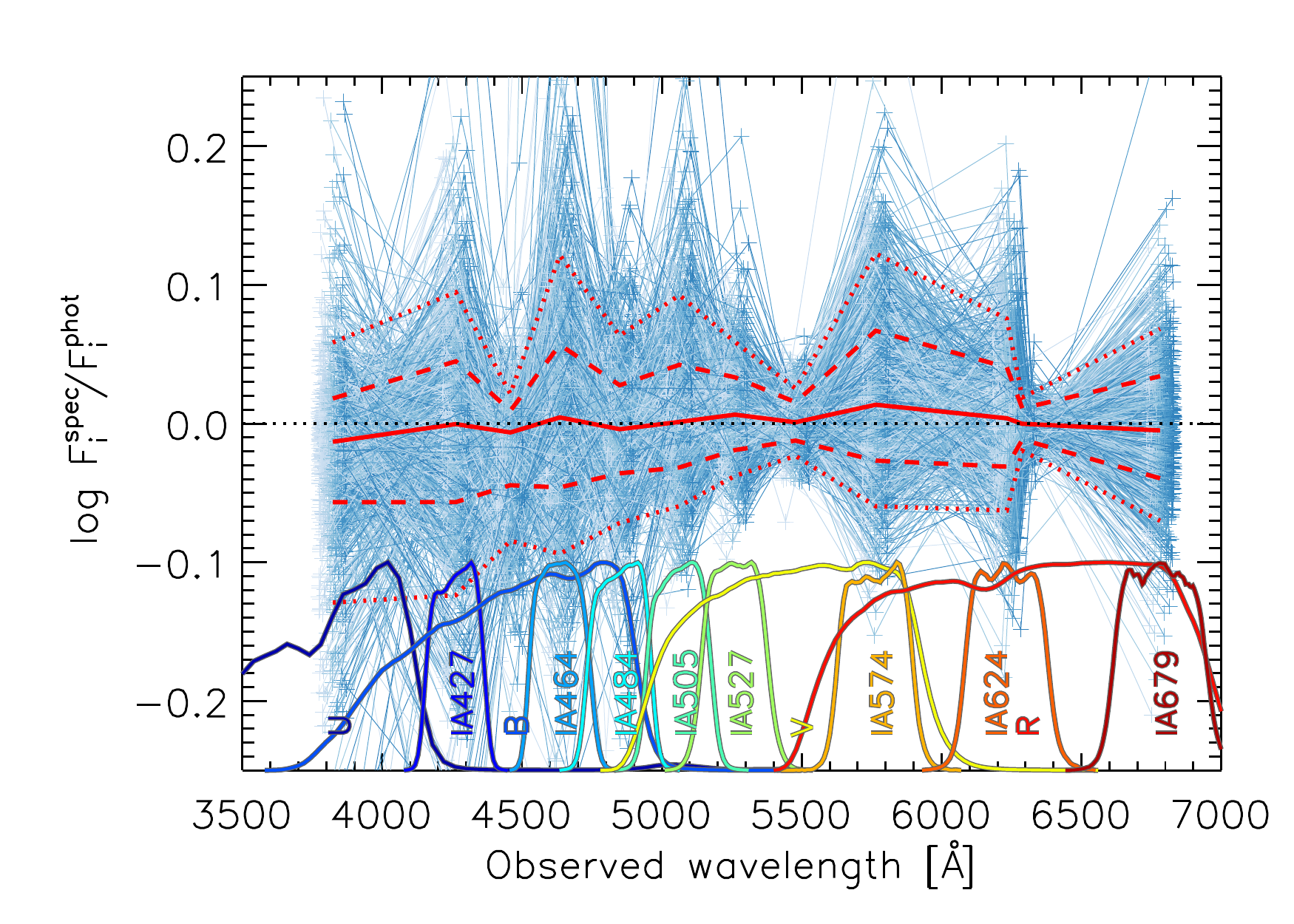}
\includegraphics[width=3.3in]{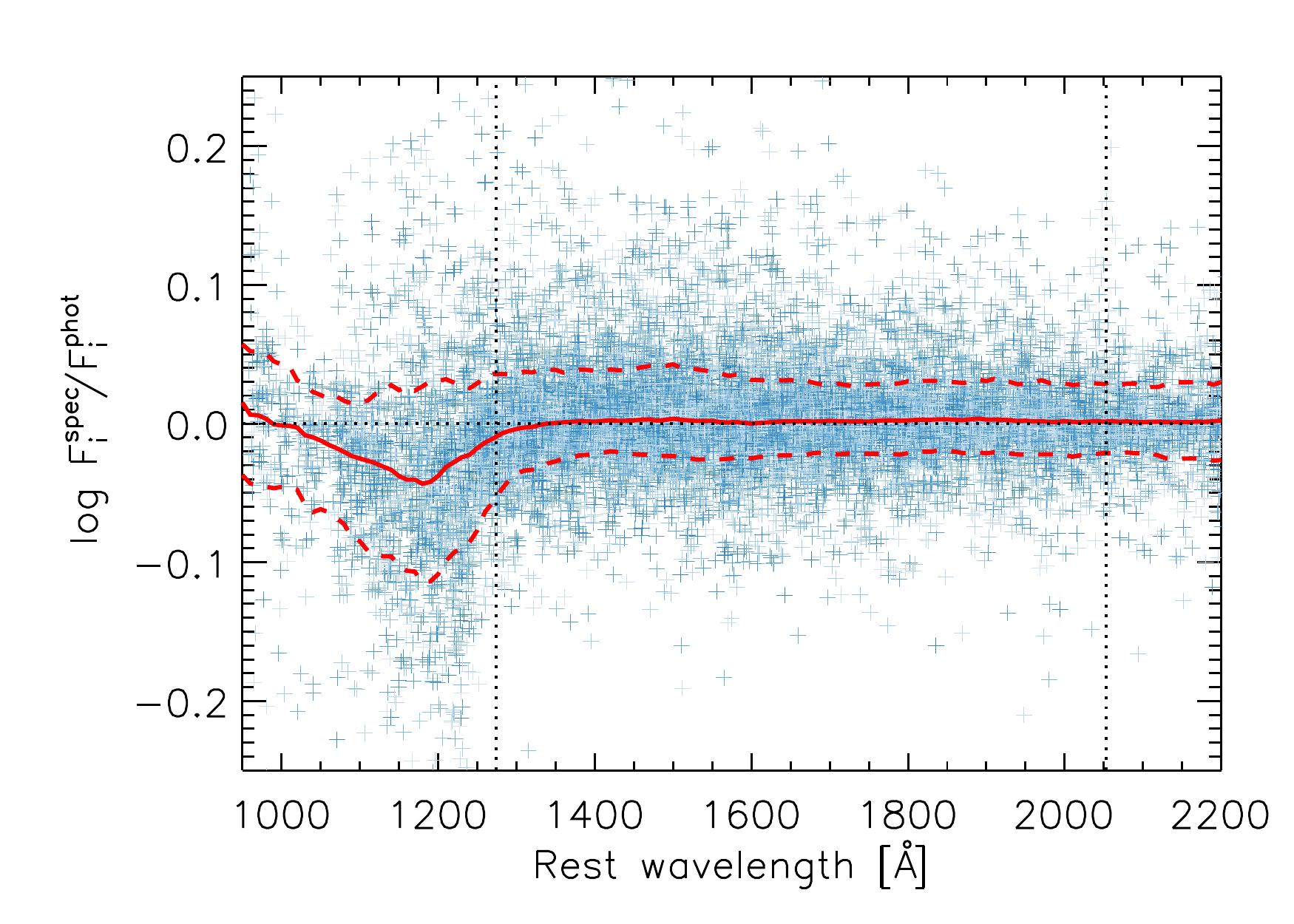}
\caption{
Upper panel: flux ratios, $F^\mathrm{spec}_i/F^\mathrm{phot}_i$, after correction for the sample of 1336 galaxies.  Symbols for each single spectrum are connected by a line.  Different spectra are colored differently for display purposes.  The red solid, dashed, and dotted lines indicate, respectively, the median values, the 16th--84th percentiles, and the 5th--95th percentiles.  The horizontal dotted line indicates $F^\mathrm{spec}_i/F^\mathrm{phot}_i=1$.  The transmission curves of the relevant photometric bands are shown at the bottom.
Lower panel: the same flux ratios, but as a function of rest-frame wavelength.  The red solid and dashed lines indicate running medians and 16th--84th percentiles with a window size of 100~{\AA}. The vertical dotted lines denote the lower and upper limits of the wavelength range used for the stellar metallicity measurement.
\label{fig:flux_ratios_sample}}
\end{center}
\end{figure}

We correct each spectrum with a smooth function of wavelength (see Equation 8 of \citealt{2021ApJ...909..213K}) that is constructed using the differences between the actual photometric fluxes in four broad- and eight intermediate band filters (shown in Figure \ref{fig:flux_ratios_sample}) and the ``pseudo'' fluxes computed from the pipeline-processed spectrum in these same twelve filter bands.
In doing so, we excluded any photometric bands that sampled the rest-frame wavelength of the Ly$\alpha$ emission line.  This is because the flux of the strong Ly$\alpha$ line may be differently affected than the continuum flux due to the possible extended shape of the emission and/or the overestimate of the sky level at these particular wavelengths.

The upper panel of Figure \ref{fig:flux_ratios_sample} shows how this spectrophotometric calibration works in the observed frame by comparing the pseudo broad- and intermediate-band fluxes ($F^\mathrm{spec}_i$ where $i$ denotes a filter) recomputed in the flux-corrected spectra with the photometric fluxes ($F^\mathrm{phot}_i$) from the COSMOS2015 catalog.  
The median values of the corrected $F^\mathrm{spec}_i$-to-$F^\mathrm{phot}_i$ ratios in each band are all within $\pm 0.015~\mathrm{dex}$ and the scatter is $\sim0.02\textrm{--}0.05$~dex, depending on the filters.  
This overall scatter (seen in the rest frame) is quite smaller than before correction ($\sim0.1\textrm{--}0.15$~dex).
Note that there is still scatter for a given filter because the correction used a smooth function of wavelength, so the effects of photometric noise in the different filters is still seen.  

The lower panel of Figure \ref{fig:flux_ratios_sample} shows the same data but now shifted to the rest-frame wavelengths.  There is no systematic trend in the corrected flux ratios across the entire wavelength range for the stellar metallicity measurement ($1274\textrm{--}2053~\textrm{\AA}$).  The running medians with a window size of 100~{\AA} are all within $0.005~\mathrm{dex}$ over this wavelength range of interest.  On the other hand, a significant systematic undercorrection is seen around 1216~{\AA} most likely because of sky-subtraction issues associated with strong Ly$\alpha$ lines mentioned above.  

This flux calibration is important to obtain composite spectra that correctly reflect the average shape of the galaxies' SEDs.  We note, however, that the precision of this calibration is unlikely to be critical for our conclusions, because the stellar metallicities are measured based on the detailed shape of the spectra that results from the blending of numerous narrow stellar absorption lines, whereas the overall shape of the smooth continuum is fit with an arbitrary multiplicative $\lambda$-dependent function (see Section \ref{sec:FUV_fit}).  In the remainder of the paper, the term ``observed VIMOS spectra'' will always refer to these accurately spectrophotometrically recalibrated spectra.

\subsection{Fine adjustment of spectroscopic redshifts}
\label{sec:revise_redshifts}

Precise determination of the spectroscopic redshifts is important to reduce the loss of the potential spectral resolution when spectra are stacked.  The spectroscopic redshifts in the zCOSMOS-deep catalog have been determined by visually inspecting prominent emission and absorption features in each spectrum, but no systematic spectral fitting has been performed.
As a consequence, the spectral redshifts may be more uncertain than the best estimates that can potentially be achieved from the existing spectra; thus some improvements are possible.

We therefore made small adjustments to the spectroscopic redshifts by fitting a common template spectrum to each of the individual spectra.  We constructed this template by stacking the observed spectra of the entire sample of 1336 galaxies used in this work by using the original spectroscopic redshifts from the catalog.  In this template fitting, we applied an arbitrary normalization and a wavelength-dependent multiplicative factors which is intended to mimic dust attenuation.

The differences between the revised and original spectroscopic redshifts, $(z_\mathrm{orig}-z_\mathrm{new})/(1+z_\mathrm{new})$, have a Gaussian-like distribution with a standard deviation of $8.4\times10^{-4}$ ($252~\mathrm{km~s^{-1}}$) and median of $6.4\times10^{-6}$.  
We found that, by using the revised spectroscopic redshifts, the resulting stacked spectra are noticeably improved, showing sharper spectral features both in emission and absorption than those seen in the stacks based on the original redshifts.  However, the adjustment of the spectroscopic redshifts has little affect on the metallicity measurements, and our conclusions do not change if the stacking is done using the original redshift values.

\subsection{Stacking procedure}
\label{sec:stacking}

\begin{figure*}[htbp]
\begin{center}
\includegraphics[width=7in]{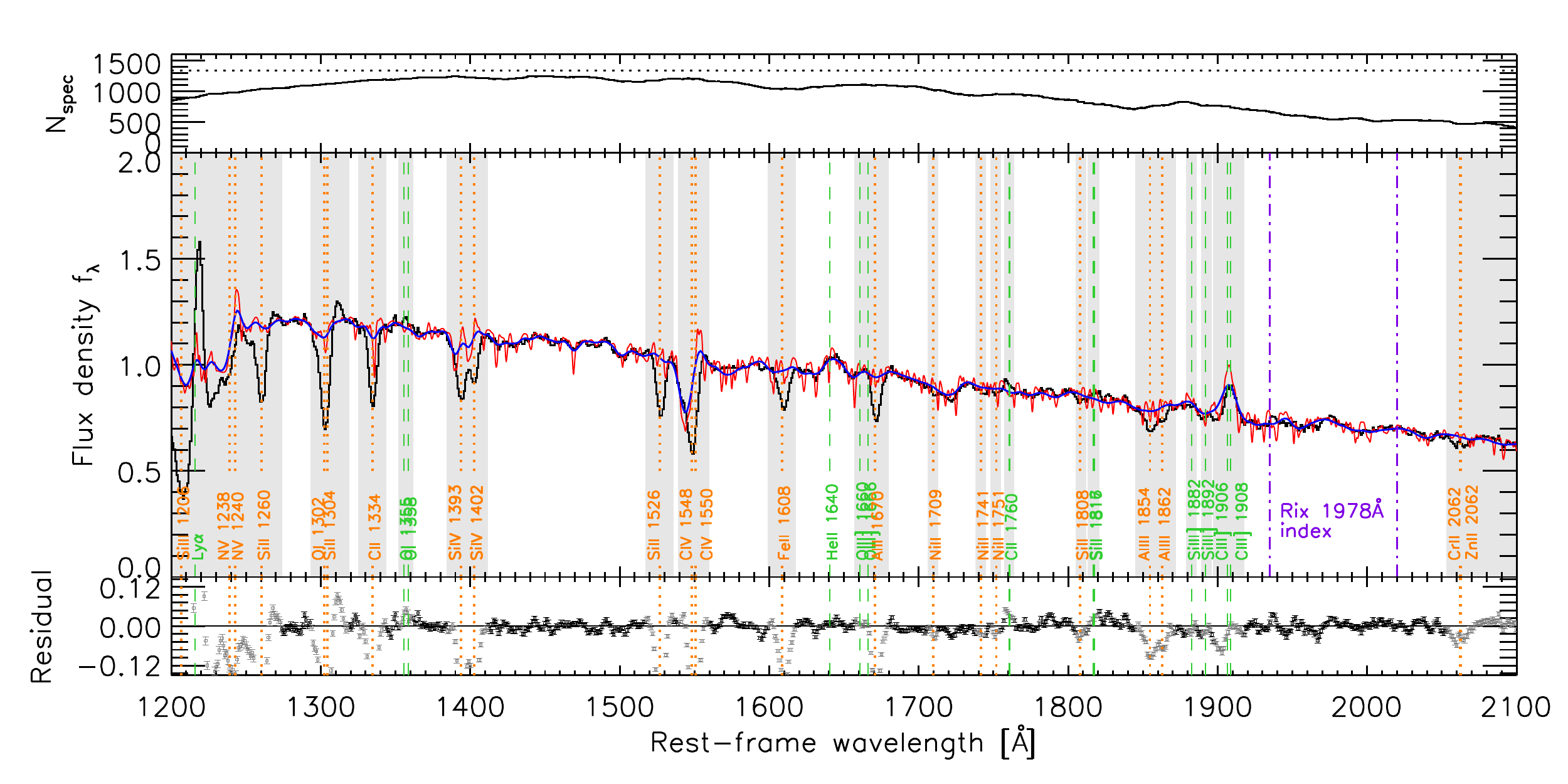}
\caption{
Composite VIMOS spectrum of the entire sample of 1336 galaxies at $1.6\le z\le 3.0$. 
Top panel: the number of spectra that have been stacked at each wavelength grid.  The horizontal dotted line indicates the number of galaxies in the stack.
Middle panel: the stacked spectrum (black line), in which some prominent absorption and emission features are identified as marked by color-coded labels (interstellar absorption features--orange; nebular emission lines--green).  The blue line indicates the best-fit BPASS model obtained from the MCMC analysis and the red line indicates the best-fit model before smoothing.  The gray regions indicate the wavelength ranges that are masked out in the fitting (see Table \ref{tb:mask}).  The purple dot-dashed lines mark the wavelength region that is used for the \citet{2004ApJ...615...98R} 1978~{\AA} index.  
Bottom panel: residuals from the best fit.  The vertical error bars correspond to the associated 1$\sigma$ errors.  The data points that are not included in the fit are shown by gray.  
\label{fig:stack_all}}
\end{center}
\end{figure*}

To infer the stellar metallicity as a function of stellar mass, we rely on stacked spectra from subsamples of galaxies separated by their stellar mass.  The observed spectra are co-added as follows.
We first transform all the individual spectra to the rest-frame wavelength based on their adjusted redshift (see Section \ref{sec:revise_redshifts}) and rebin them to a common wavelength grid with a spacing of $1~\textrm{\AA}$.  

Each spectrum is then normalized by dividing by a fitted continuum of the form $\lambda^\beta \times 10^{-0.4 A_\lambda}$ where $A_\lambda$ is the dust attenuation of the \citet{2000ApJ...539..718C} prescription.  
Here the overall dust attenuation is a free parameter for continuum fitting, independent of the result from the SED fitting in Section \ref{sec:mass_estimation}.  
By doing so, the effects of the variable overall shapes of the spectra are mitigated.  We then take the mean value of the individual continuum-divided spectra at each wavelength grid while ignoring any spectral regions that are missing and/or contaminated, for example, by zeroth order contamination or strong sky lines.  Finally, to recover the global shape of the average SED, the stacked spectrum is multiplied by the mean of the fitted continua.  Our conclusions do not depend on the stacking method; in particular, the results do little change if we take the median values at each wavelength instead of the mean, or if we do not normalize the spectra with the continua before stacking.

Figure \ref{fig:stack_all} shows the composite spectrum (middle panel) of the entire sample.  The number of spectra used at each wavelength is indicated in the top panel. 
Note that, given the redshift range of our sample, the rest-frame wavelength range of 1400--1700~{\AA} is covered by nearly all the input spectra, while shorter and longer wavelengths are less well sampled.  Some prominent spectral features are clearly identified as marked by vertical lines.

\subsection{Stellar population synthesis models for metallicity estimation}
\label{sec:BPASS_model}

To derive stellar metallicities for the galaxies, we compare our observed composite spectra with model spectra, following the approaches described in \citet{2016ApJ...826..159S} and \citet{2019MNRAS.487.2038C}.

We utilize the latest public data release of the population synthesis code ``Binary Population and Spectral Synthesis'' (BPASSv2.2.1; \citealt{2017PASA...34...58E,2018MNRAS.479...75S}).  
The package provides sets of single stellar population synthesis model spectra at a pixel resolution $1~\textrm{\AA}/\mathrm{pixel} $ as a function of stellar ages for different IMFs and for discrete (overall) stellar metallicities ($Z_\ast=10^{-5}, 10^{-4}$, 0.001, 0.002, 0.003, 0.004, 0.006, 0.008, 0.010, 0.014, 0.020, 0.030, and 0.040).  An important note is that, although the BPASS models adopt a fixed abundance ratio based on the solar abundances, the model fit to the rest-frame FUV spectrum is mostly sensitive to the iron abundance of the young stellar component of the galaxies.  We can therefore translate the inferred metallicity $Z_\ast$ into [Fe/H].

\begin{figure}[tbp]
\begin{center}
\includegraphics[width=3.5in]{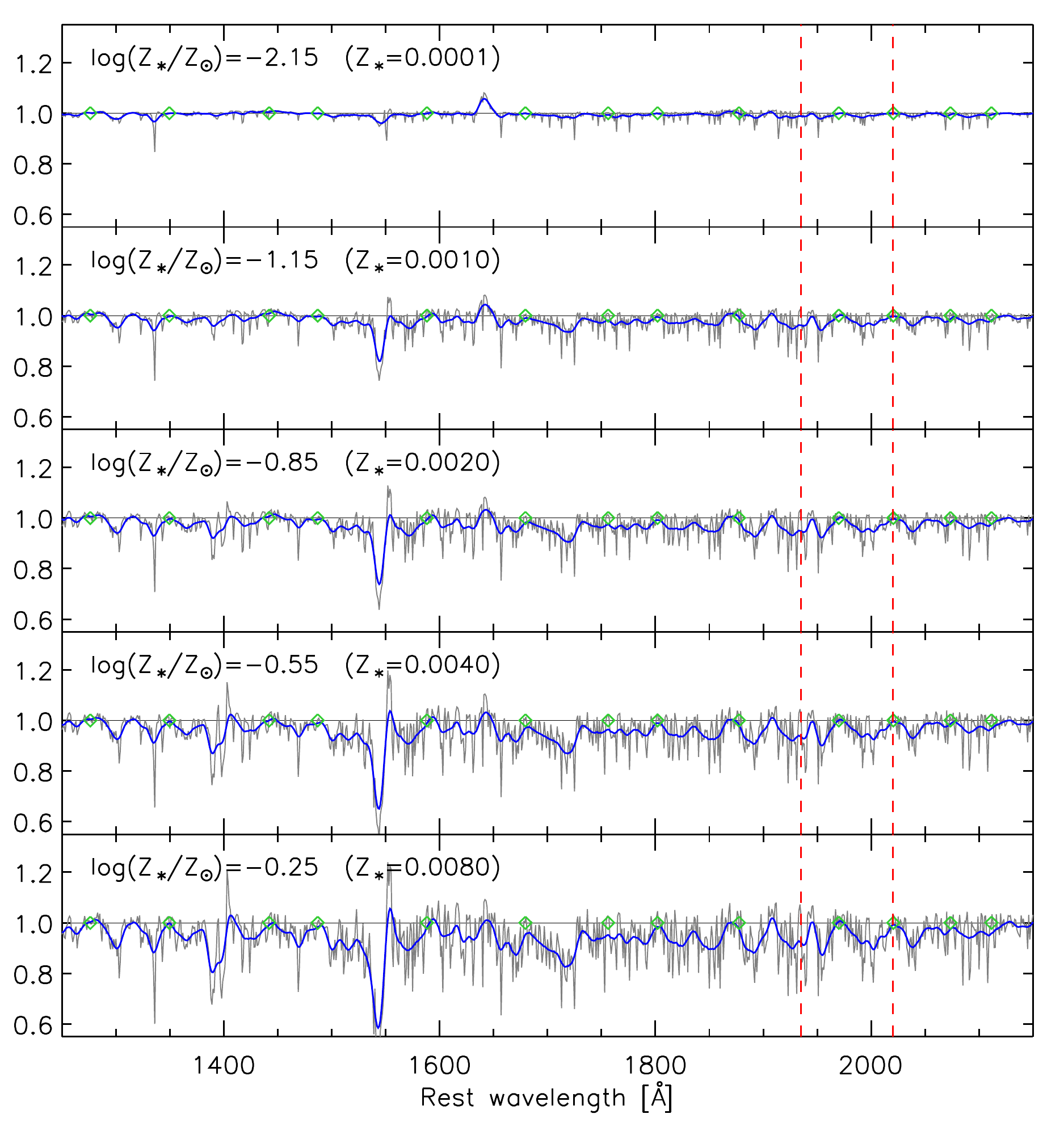}
\caption{The BPASSv2.2.1 stellar population models.  All models assume a constant SFR over 100~Myr.  Each panel shows the model spectrum in the rest FUV window at $\log(Z_\ast/Z_\odot)=-2.15$ to -0.25 from the top to the bottom.  The gray lines show the full resolution model spectra, which are here normalized by the pseudo continuum constructed from a spline fit to the selected points (green diamonds; see Table 3 of \citealt{2004ApJ...615...98R}).  The blue lines indicate the models smoothed to the VIMOS resolution ($\sigma=3.3$~{\AA}, or $2500~\mathrm{km~s^{-1}}$ in FWHM; see Section \ref{sec:results}).
It is clearly seen, even at the VIMOS resolution, the absorption features are stronger with increasing stellar metallicity.  The vertical dashed lines indicate the range of the traditional 1978~{\AA} index \citep{2004ApJ...615...98R} for reference.
\label{fig:templates}}
\end{center}
\end{figure}

We adopted the BPASSv2.2.1 model that used a \citet{2003PASP..115..763C} IMF with a high-mass cutoff of $100~M_\odot$ and included binary star evolution, and considered a continuous SFH with a duration of 100~Myr to construct a set of rest-frame UV template spectra. The results do hardly change for different duration times between 10--1000~Myr.
Figure \ref{fig:templates} shows examples of the model spectra at different metallicities, demonstrating that the detailed shapes of the galaxy spectra across the rest FUV window are sensitive to the stellar metallicity.  A feature of the BPASS binary models is that they model the broad He\,$\lambda$1640 emission line that originates in the winds from very massive stars (see Section \ref{sec:FUV_fit} for a relevant description).

We added nebular continuum and line emission to the templates, although it has only a minor contribution of $\sim5\%$ in the total continuum flux.  The nebular continuum was computed using \textsc{cloudy} v17.00 \citep{2017RMxAA..53..385F} adopting the BPASS spectrum itself as the incident spectrum.  We assumed the electron density to be $n_\mathrm{e}=300~\mathrm{cm^{-3}}$ and the ionization parameter to be $\log U =-2.8$.  These values are consistent with the recent estimates in $z\sim2\textrm{--}3$ star-forming galaxies \citep{2016ApJ...816...23S,2017ApJ...835...88K,2017ApJ...836..164S,2018ApJ...868..117S} and local star-forming galaxies with sSFR as high as our sample ($\sim5~\mathrm{Gyr^{-1}}$; \citealt{2019MNRAS.486.1053K}).  
We change the gas-phase metallicity to calculate nebular emission according to the stellar metallicity, with an offset of $\log(Z_\mathrm{gas}/Z_\ast)=0.42$ as reported for $z\sim2$ galaxies by \citet{2018ApJ...868..117S}.  In other words, we considered an enhanced gas-phase metallicity, $12+\log(\mathrm{O/H}) = 8.69+\log (Z_\ast/Z_\odot)+0.42$, relative to each $Z\ast$ of the BPASS template.  We will indeed measure consistent $\log(Z_\mathrm{gas}/Z_\ast)$, or [O/Fe] values in Section \ref{sec:O/Fe} and discuss the evolution of [O/Fe] in Section \ref{sec:modeling}.


\subsection{Fitting synthesis model spectra}
\label{sec:FUV_fit}

\begin{deluxetable}{ccl}
\tablecaption{Rest-frame wavelength ranges excluded from fitting\tablenotemark{a} \label{tb:mask}}
\tablehead{
    \colhead{$\lambda_\mathrm{min}$}&
    \colhead{$\lambda_\mathrm{max}$}&
    \colhead{Interstellar\tablenotemark{b}}\\
    \colhead{(\AA)}&
    \colhead{(\AA)}&
    \colhead{Spectral features}}
\startdata
 0    &  1273 & Si\,{\sc ii}$\lambda$1260.42, Si\,{\sc ii}$^\ast\lambda$1264.74 \\
 1294 &  1318 & O\,{\sc i}~$\lambda$1302.17, Si\,{\sc ii}~$\lambda$1304.37, Si\,{\sc ii}$^\ast\lambda$1309.28 \\
 1326 &  1343 & C\,{\sc ii}~$\lambda$1334.53 \\
 1353 &  1361 & O\,{\sc i}~$\lambda$1355.60, O\,{\sc i}$\lambda$1358.51 \\
 1385 &  1411 & Si\,{\sc iv}~$\lambda\lambda$1393.76, 1402.77 \\
 1518 &  1535 & Si\,{\sc ii}~$\lambda$1526.71 \\
 1540 &  1559 & C\,{\sc iv}~$\lambda\lambda$1548.19, 1550.77 \\
 1600 &  1617 & Fe\,{\sc ii}~$\lambda$1608.45 \\
 1658 &  1679 & O\,{\sc iii}]~$\lambda\lambda$1660.81, 1666.15, Al\,{\sc ii}~$\lambda$1670.79 \\
 1707 &  1712 & Ni\,{\sc ii}~$\lambda$1709.60 \\
 1739 &  1744 & Ni\,{\sc ii}~$\lambda$1741.55 \\
 1749 &  1754 & Ni\,{\sc ii}~$\lambda$1751.91 \\
 1758 &  1763 & C\,{\sc ii}~$\lambda$1760.5 (blended) \\
 1806 &  1811 & Si\,{\sc ii}~$\lambda$1808.01 \\
 1814 &  1820 & Si\,{\sc ii}~$\lambda\lambda$1816.93, 1817.45 \\
 1846 &  1871 & Al\,{\sc iii}~$\lambda$1854.72, 1862.79 \\
 1880 &  1885 & Si\,{\sc iii}]~$\lambda$1882.71 \\
 1890 &  1895 & Si\,{\sc iii}]~$\lambda$1892.03 \\
 1898 &  1917 & C\,{\sc iii}]~$\lambda$1906.68, 1908.73 \\
 2054 &  $\infty$ & Cr\,{\sc ii}~$\lambda$2062.24, Zn\,{\sc ii}~$\lambda$2062.66
 \enddata
\tablenotetext{a}{The wavelength range used for the fit is limited between $\lambda_\mathrm{rest} = 1274\textrm{--} 2053~\textrm{\AA}$ .}
\tablenotetext{b}{Wavelengths are all given in vacuum.}
\end{deluxetable}

Our goal is to estimate average stellar metallicities for binned subsamples of galaxies by using their composite spectra.  To do so, we limit the wavelength range between $1274~\textrm{\AA}\le \lambda_\mathrm{rest}\le 2053~\textrm{\AA}$ and excluded some narrow wavelength regions where the stellar continuum is impacted by interstellar absorption lines or nebular emission lines.  The wavelength regions excluded from the fitting are summarized in Table \ref{tb:mask}.  We note that the region of the He\,$\lambda$1640 line is not excluded because the BPASS models includes the broad He\,$\lambda$1640 stellar emission line that originates from very hot stars.  In fact, as shown in Section \ref{sec:results}, the observed He\,$\lambda$1640 feature in the composite spectra could be attributed almost entirely to the contribution from stars that is predicted in the BPASS models.

An added advantage of the full spectral fitting over using the traditional indices (e.g., the 1978~{\AA} index) is that the former does not need to identify the pseudo smooth continuum. Contrarily, the latter requires it for measuring equivalent widths (EWs) of particular stellar absorption lines.  The measurements of the EWs of faint features are quite sensitive to the assumed continuum level, whilr the determination of the pseudo continuum has to be based on very limited wavelength regions free from narrow features, thus being easily affected by noise.  The full spectral fitting is free from these difficulties.

We fit the constructed BPASS models to our composite spectra based on the maximum likelihood estimation.  We assume that the associated errors are Gaussian and independent, so the logarithm of the likelihood $\mathcal{L}$ is given by
\begin{equation}
    \ln \mathcal{L}= -\frac{1}{2}\sum_{i} \left[ \frac{(f^\mathrm{obs}_i-f_i^\mathrm{model})^2}{\sigma_i^2} + \ln (2\pi \sigma_i^2)\right]
\end{equation}
where $f_i^\mathrm{obs}$ is the observed composite spectrum at the $i$th wavelength grid, $f_i^\mathrm{model}$ is the model spectrum for a given set of parameters, and $\sigma_i$ is the error on the observed flux.  The summation is over all wavelength grids that are included in the fit.

For our models, we adopted eight free parameters: the logarithm of the overall stellar metallicity ($\log Z_\ast$), the Gaussian smoothing kernel $\sigma_\mathrm{smooth}$ applied to the model spectra, and six parameters for a 5th-order polynomial function that is multiplied to the template spectrum;
\begin{equation}
    f_i^\mathrm{model}(Z_\ast) = f_i^\mathrm{BPASS}(Z_\ast) \times \sum_{k=0}^{5} p_k (\lambda[\mathrm{\AA}]-1650)^k
    \label{eq:polynomial}
\end{equation}
where $f_i^\mathrm{BPASS}(Z_\ast)$ is the template spectrum at given $Z_\ast$ in arbitrary units, in which the nebular emission is accounted for.  The polynomial term is ideally intended to reflect the change of the overall shape of the spectra due to dust attenuation.

To sample the posterior probability distribution of the model parameters, we employed a Markov Chain Monte Carlo (MCMC) technique using the \texttt{emcee} package for python \citep{2013PASP..125..306F}.  In our fitting, we adopted a uniform prior probability function for each parameter.  For stellar metallicity, a flat prior is adopted in log space (i.e., $\log(Z_\ast)$) between $-5\le \log(Z_\ast)\le \log(0.040)$ (which corresponds to $-3.15\le \log(Z_\ast/Z_\odot)\le 0.44$) according to the possible choices in the BPASS models.
Since the models are only provided at 13 discrete metallicity values (see Section \ref{sec:BPASS_model}), we interpolated the flux values in $\log Z_\ast$--$\log(\mathrm{flux})$ space between the models. This enables us to generate a model at any metallicity value within the available range. 

In the fitting, we need to match the spectral resolution of the model templates to that of the observed composite spectra.  However, the spectral resolution of the composite spectra is not precisely known. It may vary from galaxy to galaxy within a subsample because of different velocity dispersions of the galaxies and different slit illumination profiles. 
We thus leave the smoothing scale, $\sigma_\mathrm{smooth}$ in units of {\AA}\footnote{Note that the wavelength sampling is 1~{\AA}/pixel.}, as a free parameter rather than using a pre-determined value.  In the fitting, the model spectra are smoothed by a Gaussian kernel with $\sigma_\mathrm{smooth}$ and the smoothing scale that gives the best fit to the data is constrained along with other parameters.

\subsection{Accuracy of the metallicity measurements}
\label{sec:systematics}

Before applying our procedure to the data, we evaluate the potential systematic uncertainties in the metallicity measurements that may come from the noise and the limited spectral resolution of the spectra.

As we will show below, the width of the smoothing kernel, $\sigma_\mathrm{smooth}$, is estimated from the fits to be $\sim 3.3~\textrm{\AA}$.  This is large enough to wash out individual narrow stellar absorption features in the model spectra. Still, the model spectra with different metallicities will retain their different characteristic shapes that result from blending of the narrow features after smoothing.  Thus the metallicity measurement is indeep possible.  The possible systematic uncertainties, however, should be evaluated.

For this purpose, we constructed a set of artificial spectra using the BPASS models at different metallicities.  We smoothed them with $\sigma_\mathrm{smooth}=3.3~\textrm{\AA}$ and modify the overall shape using the representative polynomial function as in Equation (\ref{eq:polynomial}).  We then added Gaussian noise with the moderate wavelength dependence that replicates the 1$\sigma$ error spectra.
We considered noise with four different levels: the best had the noise seen in the composite spectrum of the entire sample, i.e., the signal-to-noise ratio at $\lambda_\mathrm{rest}\sim1450~\textrm{\AA}$, $\mathrm{S/N}_{1450}\approx 190$ per 1~{\AA}.  The other three are degraded to have twice ($\mathrm{S/N}_{1450}\approx 95$), five times ($\mathrm{S/N}_{1450}\approx 38$), or ten times ($\mathrm{S/N}_{1450}\approx 19$) higher noise than the first one.  The stacked spectra of the subsamples selected by stellar mass have noise levels between these values, depending on the number of galaxies in the bin and their average brightness.  

We then attempted to measure the stellar metallicity in the same way as from the data in order to evaluate how accurately the input metallicities can be recovered.  We repeated each setup 10 times with different random noise realizations. Here only a single fiducial duration of star formation (100~Myr) is considered.

\begin{figure}[tbp]
\begin{center}
\includegraphics[width=3.3in]{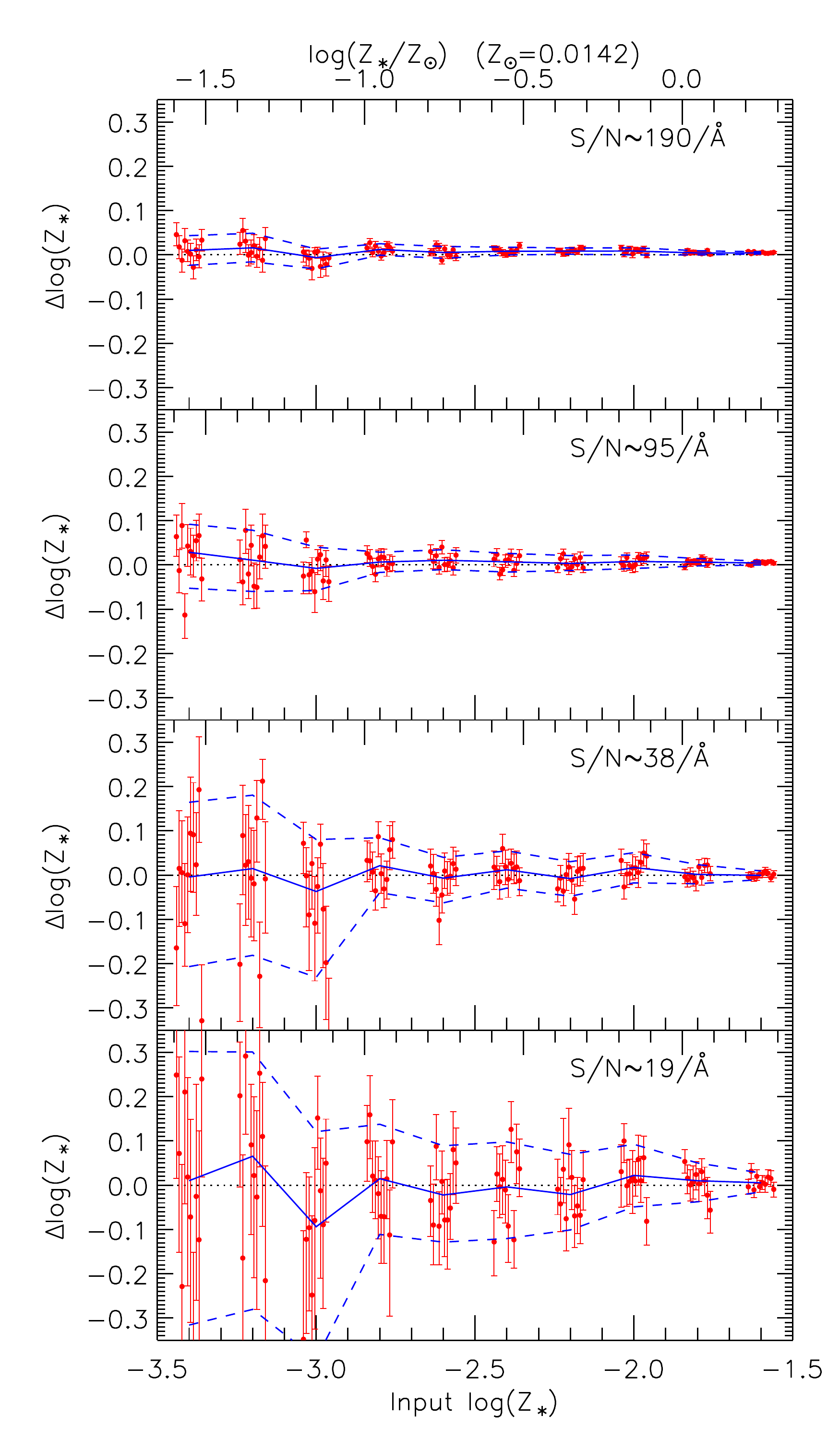}
\caption{
The results of testing how accurately the input metallicities can be recovered with realistic noise and the spectral resolution of the actual data.  The difference between the derived metallicity and the input metallicity is shown as a function of the input value.  The four panels show the results for different levels of random noise, from top to bottom, the same level as the stack of the entire sample ($S/N\approx190~\textrm{\AA}^{-1}$), and twice, five times, and ten times higher than the top one ($S/N\approx95, ~38,~\textrm{and}~19~\textrm{\AA}^{-1}$, respectively).  Red symbols indicate individual results of the MCMC analysis at different input metallicities, each for one of 10 realizations of the random noise.  The results are shown with small offsets along the $x$-axis for display purposes.  Blue solid and dashed lines indicate the median and the 16th--84th percentiles of the posterior probability distribution compiling all ten trials.
\label{fig:test_systematics}}
\end{center}
\end{figure}

Figure \ref{fig:test_systematics} shows the results of this exercise, separately, for four different noise levels.  The difference between the inferred and input metallicities is shown as a function of the input value.  The error bars indicate the 16th--84th percentiles of the inferred posterior for individual measurements.  The results from the ten trials with different random noises are shown separately with slight offsets in the $x$-axis.  We also compile the posterior probability distributions of $\log Z_\ast$ from all the 10 trials and derive the median and the 16th--84th percentiles that are shown by blue solid and dashed lines.  

Generally, the uncertainties reduce with increasing metallicity because the stellar absorption features in the spectrum become more prominent and sensitive to the metallicity value at higher metallicity.  It can be seen that, with all these noise levels, the systematic biases in the metallicity measurements are negligible with no trend across the metallicity range of interest.

\section{Results}
\label{sec:results}

\begin{figure}[htbp]
\begin{center}
\includegraphics[width=3.5in]{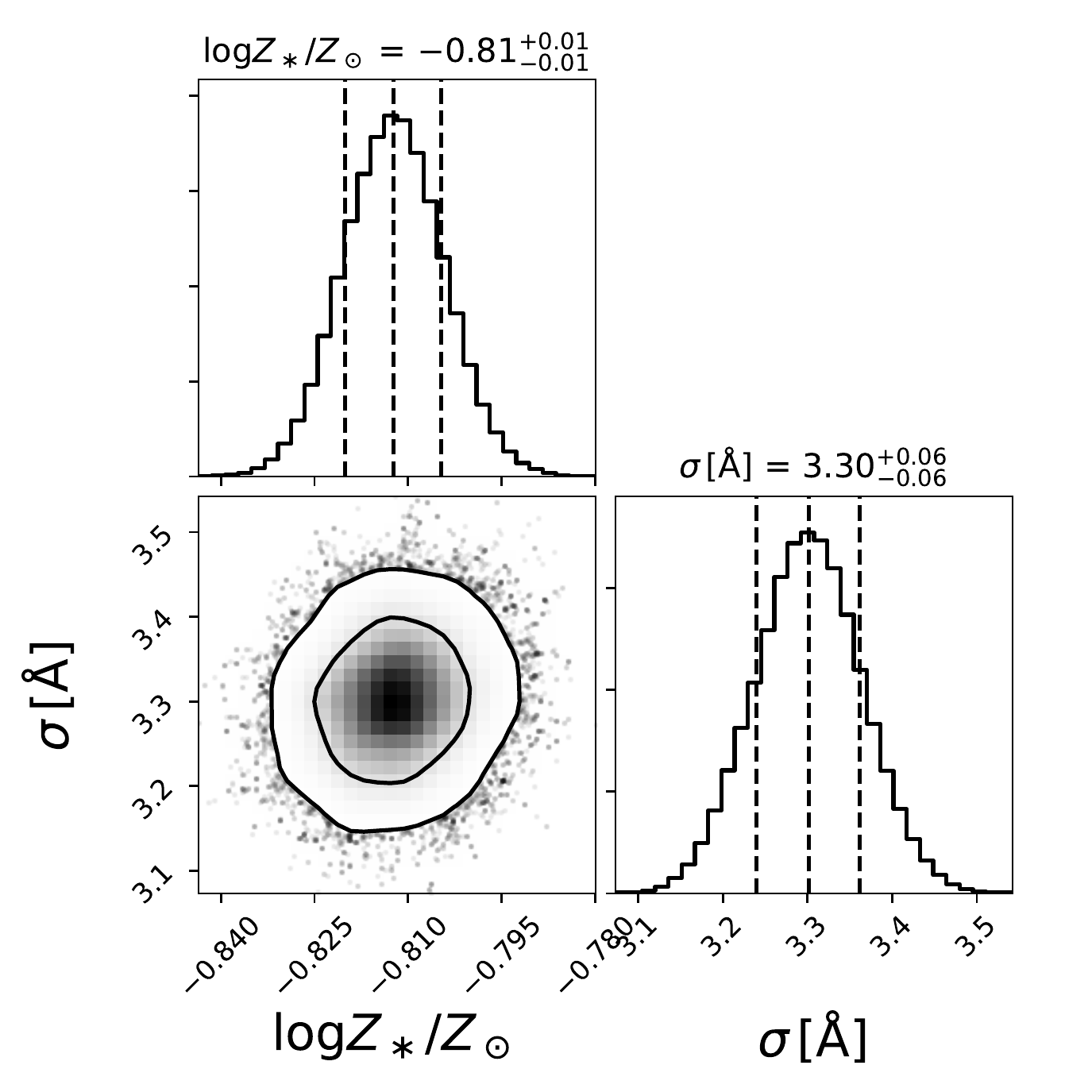}
\caption{
Posterior probability distribution functions for $\log(Z_\ast/Z_\odot)$ and smoothing scale $\sigma_\mathrm{smooth}$.  The contours on the 2D plots correspond to enclosing 68 and 95 per cent of the posterior probability.  The vertical lines in the 1D plots indicate the 16th, 50th (median), and 84th percentiles.  Here we omit the posteriors of the coefficients $p_k$ in Equation \ref{eq:polynomial} that were simultaneously fitted.
\label{fig:corner}}
\end{center}
\end{figure}

Figure \ref{fig:corner} shows the posterior distributions of the parameters for the composite stack of the entire sample of 1336 galaxies.\footnote{This figure was created using the python module \texttt{corner.py} \citep{2016JOSS....1...24F}.}  The stellar metallicity, $Z_\ast$, and smoothing scale, $\sigma_\mathrm{smooth}$, are constrained with a single preferred solution ($\log(Z_\ast/Z_\odot)=-0.812\pm0.008$ and $\sigma_\mathrm{smooth}=3.30\pm0.07$~{\AA}). The constraints on the coefficients $p_k$ in Equation \ref{eq:polynomial} also show single peak posterior distributions.  This holds when the sample is divided into stellar mass bins.

Figure \ref{fig:stack_all} shows the best-fit model for the stack of the entire sample.  The detailed shape of the FUV continuum is well reproduced after smoothing the original BPASS model spectrum with $\sigma_\mathrm{smooth}=3.3$~{\AA}.  The recovered smoothing width $\sigma_\mathrm{smooth}$ is marginally less than the nominal resolution of the VIMOS spectrograph with the LR blue grism and $1\arcsec\!.0$ arcsec slits of $\approx3.4$~{\AA}, possibly due to non-uniform illumination of the slit.\footnote{The resolution $R=200$ in FWHM at the center of the spectral window ($\sim5150$~{\AA}) corresponds to $\sigma\approx3.4$~{\AA} in the rest-frame for the median redshift of 2.2.}

For comparison to a standardized metallicity indicator, we also measured the ``1978~{\AA}-index'' of \citet{2004ApJ...615...98R}, i.e., the EW across 1935--2020~{\AA}, to be $2.62\pm0.08$~{\AA}.  The calibration of \citet[][see Equation 8]{2004ApJ...615...98R} converts this into $\log(Z_\ast/Z_\odot)=-0.93\pm0.05$, which is consistent within $\sim0.1$~dex with the fiducial result.\footnote{Here we adopted $Z_{\odot}=0.0142$ while \citet{2004ApJ...615...98R} assumed $Z_\odot=0.020$.}  Note that the statistical uncertainty here is about six times larger than that of the fiducial result because the spectral regions used for the 1978~{\AA}-index is limited.

In the following, we present the measurement of the stellar MZR based on the stacks in stellar mass bins. We here recall that the metallicity estimated from the photospheric absorption in the FUV spectra reflects the iron abundance \citep[e.g.,][]{2016ApJ...826..159S} in short-lived, recently-formed stars, and thus can be regarded as being almost equivalent to the gas-phase value.

\subsection{Stellar mass versus stellar metallicity}
\label{sec:MZR}

\begin{deluxetable}{cccccc}
\tablecaption{Stack statistics and metallicity estimates \tablenotemark{a} \label{tb:result} }
\tablehead{
    \multicolumn{2}{c}{$\log \left(M_\ast/M_\odot\right)$}&
    \colhead{$N$}&
    \colhead{$\mathrm{S/N}_{1450}$\tablenotemark{b}}&
    \colhead{$\log \left(Z_\ast/Z_\odot\right)$\tablenotemark{c}}
    \\
    \colhead{Median}&
    \colhead{min--max}&
    \colhead{}&
    \colhead{}&
    \colhead{([Fe/H])}}
\startdata
 9.97 &  8.38--11.48 & 1336 &  189 & $-0.812^{+0.008}_{-0.008} $  \\
 \hline 
 \multicolumn{5}{l}{Binned equally into six subsamples in $M_\ast$} \\
 9.45 &  8.38-- 9.62 &  222 &   71 & $-1.045^{+0.029}_{-0.029} $  \\
 9.74 &  9.63-- 9.83 &  222 &   72 & $-0.901^{+0.029}_{-0.030} $  \\
 9.89 &  9.83-- 9.97 &  223 &   81 & $-0.848^{+0.017}_{-0.018} $  \\
10.06 &  9.97--10.14 &  223 &   86 & $-0.785^{+0.017}_{-0.018} $  \\
10.24 & 10.14--10.34 &  223 &   82 & $-0.717^{+0.017}_{-0.017} $  \\
10.52 & 10.34--11.48 &  223 &   73 & $-0.662^{+0.016}_{-0.012} $  \\
\hline
 \multicolumn{5}{l}{Binned into the 0.2-dex-width intervals in $M_\ast$} \\
 9.06 &  8.91-- 9.10 &   13 &   13 & $-1.281^{+0.159}_{-0.230} $  \\
 9.24 &  9.10-- 9.30 &   35 &   28 & $-1.273^{+0.143}_{-0.180} $  \\
 9.40 &  9.30-- 9.50 &   82 &   46 & $-1.087^{+0.095}_{-0.412} $  \\
 9.62 &  9.50-- 9.70 &  169 &   67 & $-0.970^{+0.031}_{-0.031} $  \\
 9.82 &  9.70-- 9.90 &  257 &   83 & $-0.837^{+0.017}_{-0.017} $  \\
10.00 &  9.90--10.10 &  287 &   98 & $-0.792^{+0.016}_{-0.016} $  \\
10.20 & 10.10--10.30 &  221 &   86 & $-0.711^{+0.017}_{-0.017} $  \\
10.37 & 10.30--10.50 &  147 &   62 & $-0.684^{+0.011}_{-0.017} $  \\
10.59 & 10.50--10.70 &   64 &   42 & $-0.672^{+0.025}_{-0.021} $  \\
10.78 & 10.70--10.90 &   29 &   27 & $-0.611^{+0.037}_{-0.037} $  \\
11.00 & 10.91--11.10 &   17 &   20 & $-0.393^{+0.072}_{-0.078} $  
\enddata
\tablenotetext{a}{The top row is for the stack of the entire sample.  The middle set of rows is for six equally populated bins of $M_\ast$, and the bottom set of rows is for stacks in 0.2~dex fixed-width $M_\ast$ bins.}
\tablenotetext{b}{The S/N per unit pixel (1~{\AA}) of the stacked spectra around 1450~{\AA}, represented by the median S/N within $\lambda_\mathrm{rest}=1430\textrm{--}1470~\textrm{\AA}$.}
\tablenotetext{c}{Metallicity estimates, represented by the median value of the posterior probability distribution function obtained from the MCMC analysis.  The associated errors corresponds to the 16th-84th percentiles.  Because the fit is sensitive to the iron abundance, this can be translated into [Fe/H].}
\end{deluxetable}

To investigate the relation between stellar mass and stellar metallicity, we divided the sample of 1336 star-forming galaxies into bins of stellar mass in two different binning schemes: first we equally split the sample into six mass bins, and second we divided the sample into bins with a constant width of $\Delta \log M_\ast=0.2~\mathrm{dex}$.
After stacking, we performed the MCMC analysis to fit the BPASS template to the stacked spectra as described in Section \ref{sec:FUV_fit}.

\begin{figure*}[tbp]
\begin{center}
\includegraphics[width=3.5in]{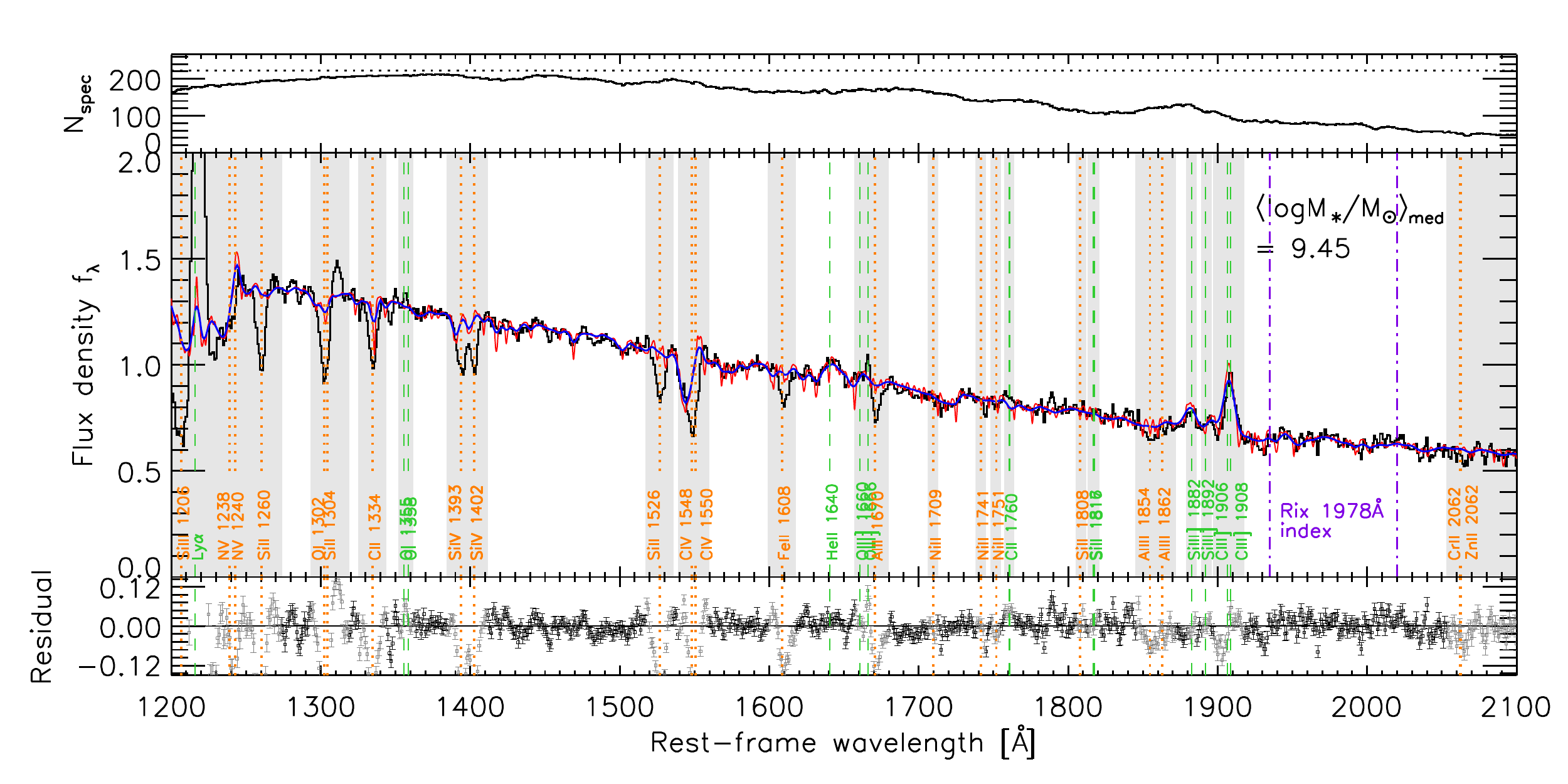}
\includegraphics[width=3.5in]{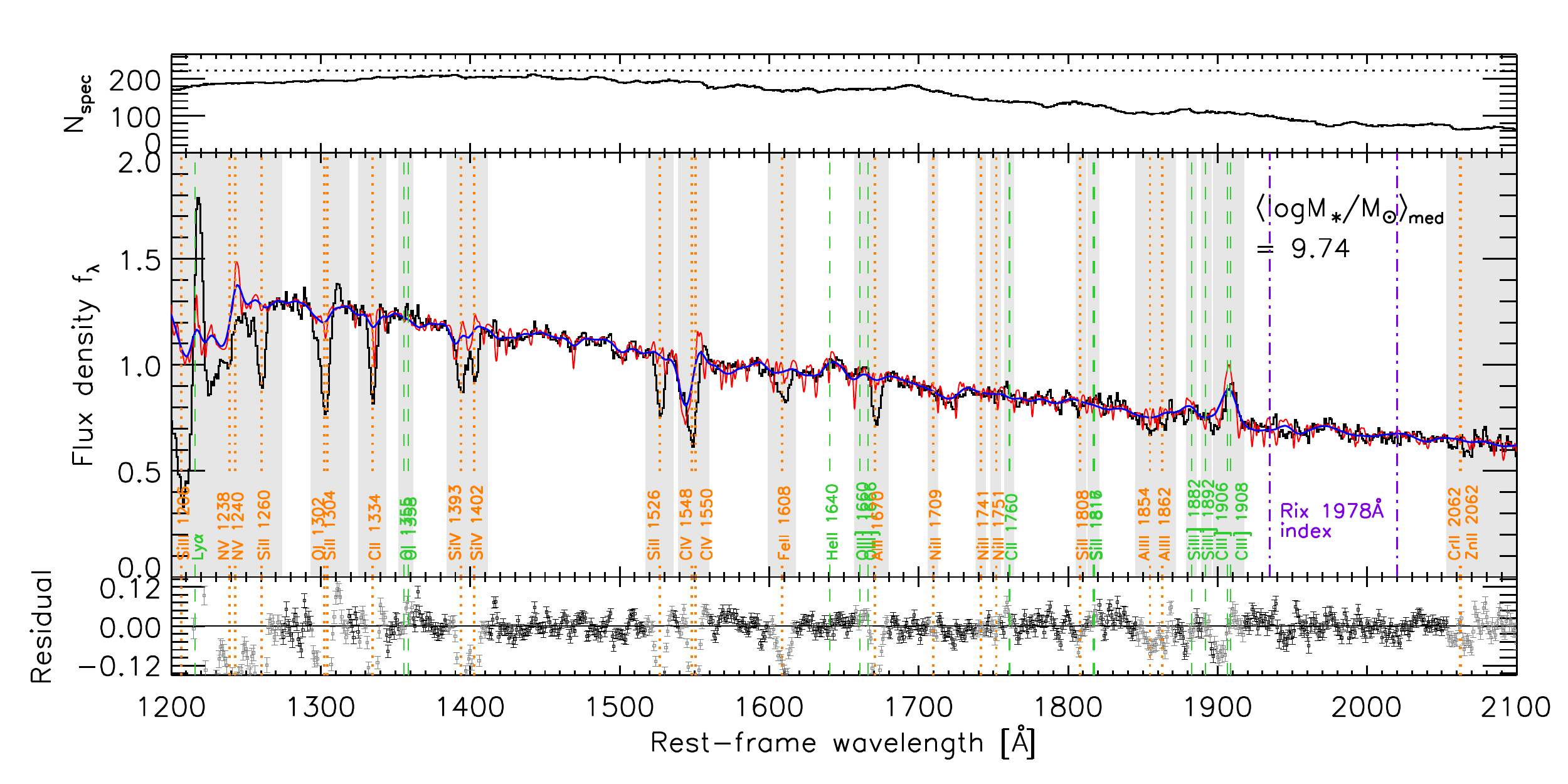}
\includegraphics[width=3.5in]{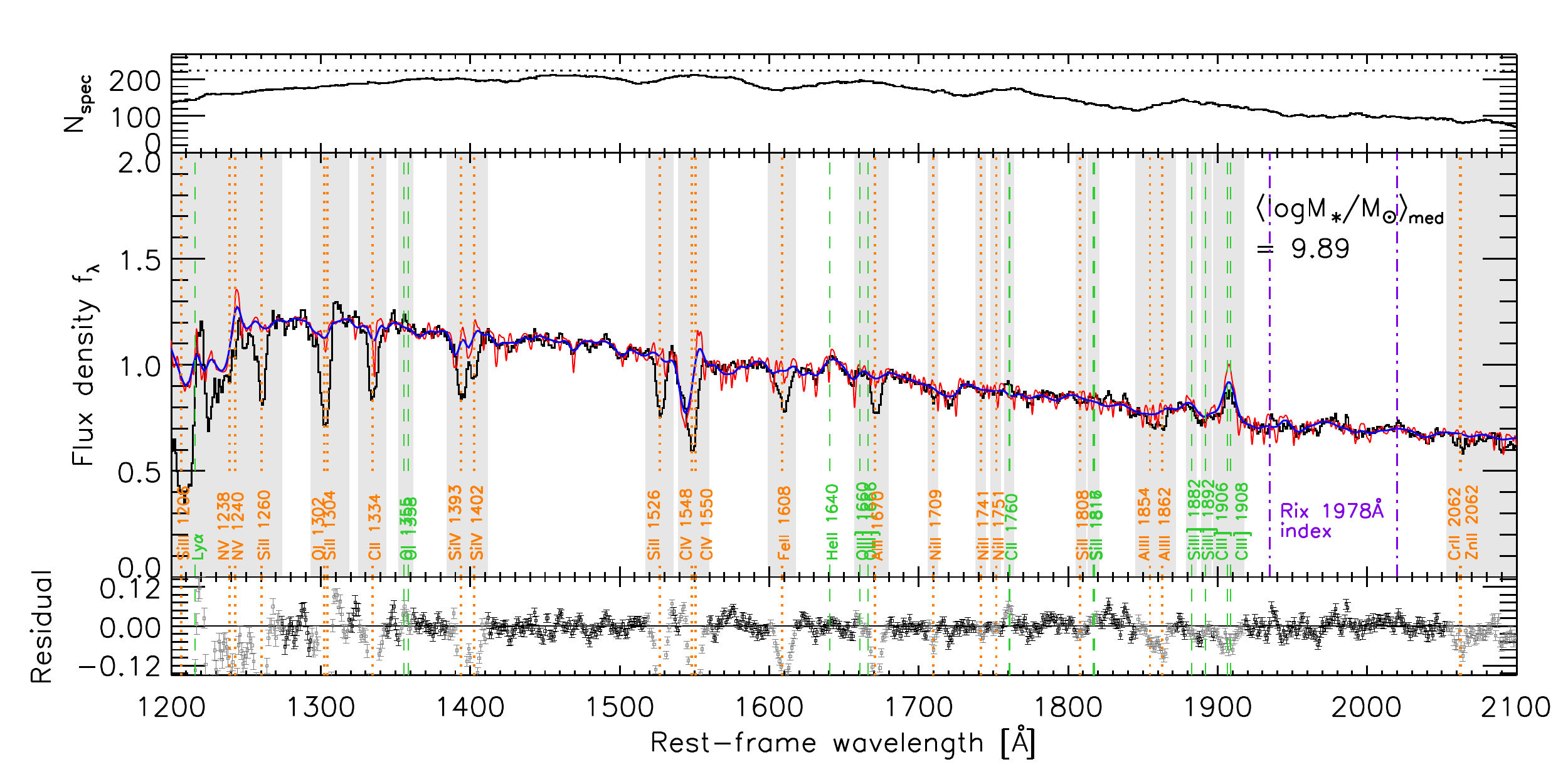}
\includegraphics[width=3.5in]{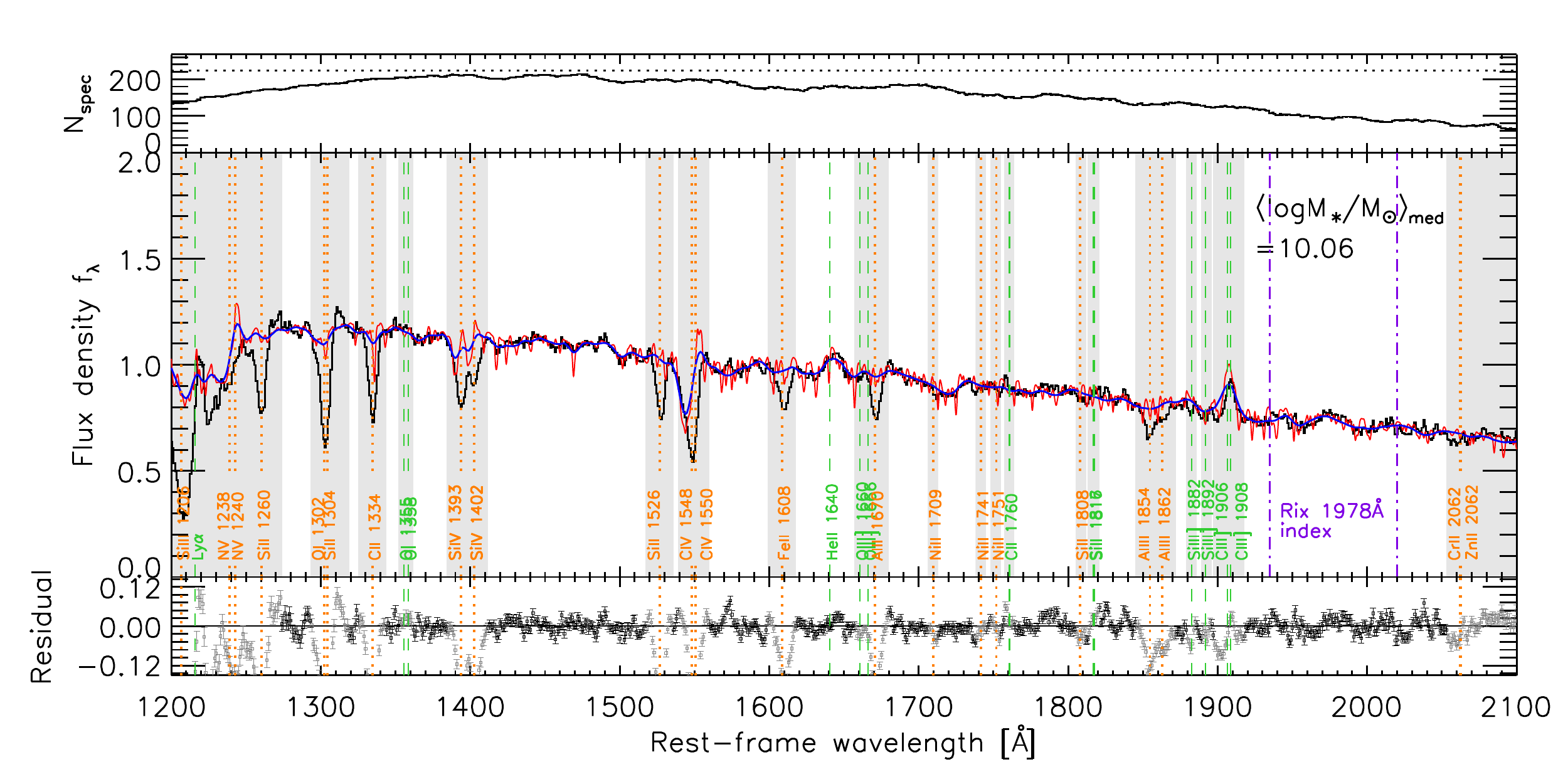}
\includegraphics[width=3.5in]{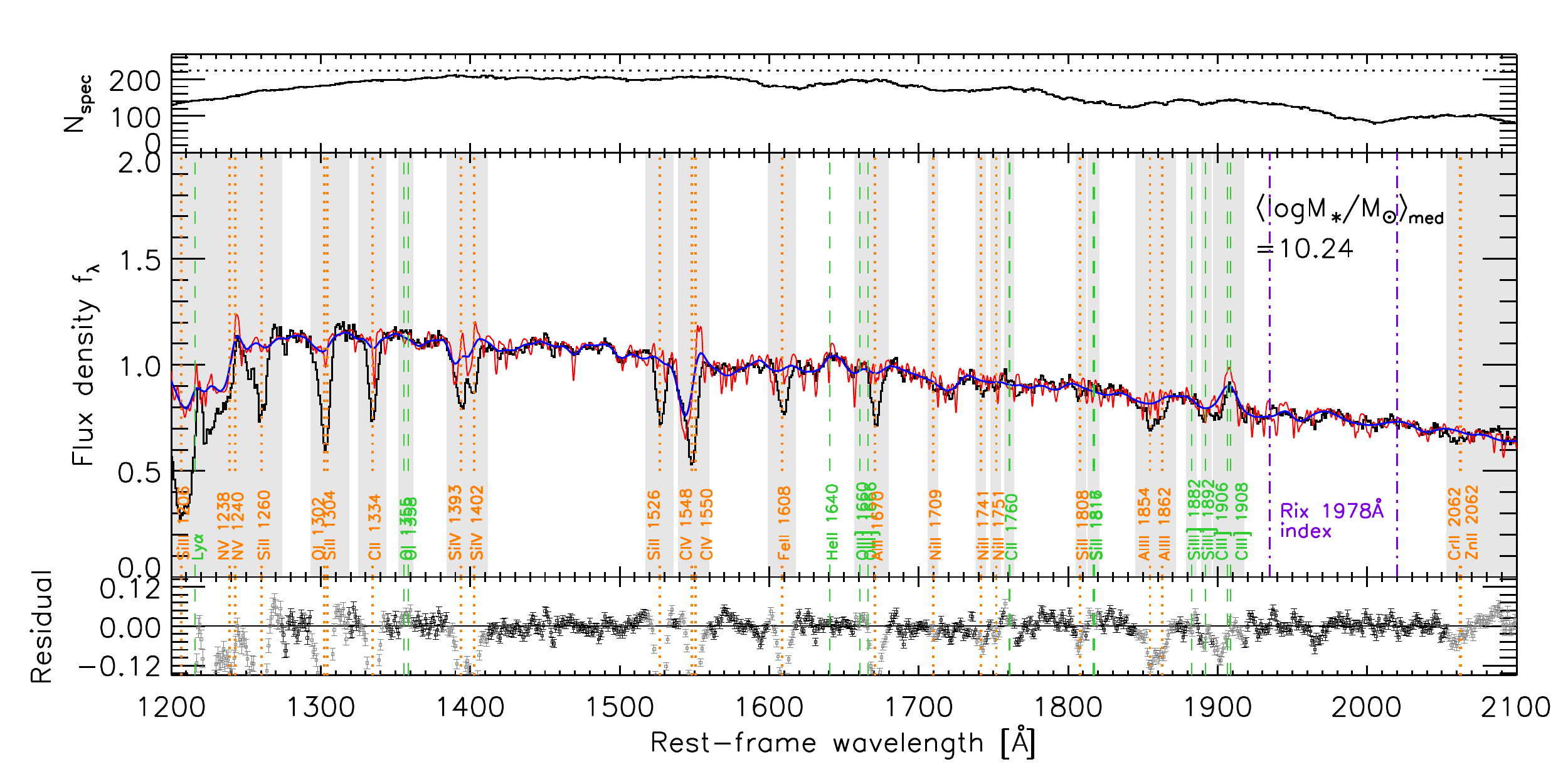}
\includegraphics[width=3.5in]{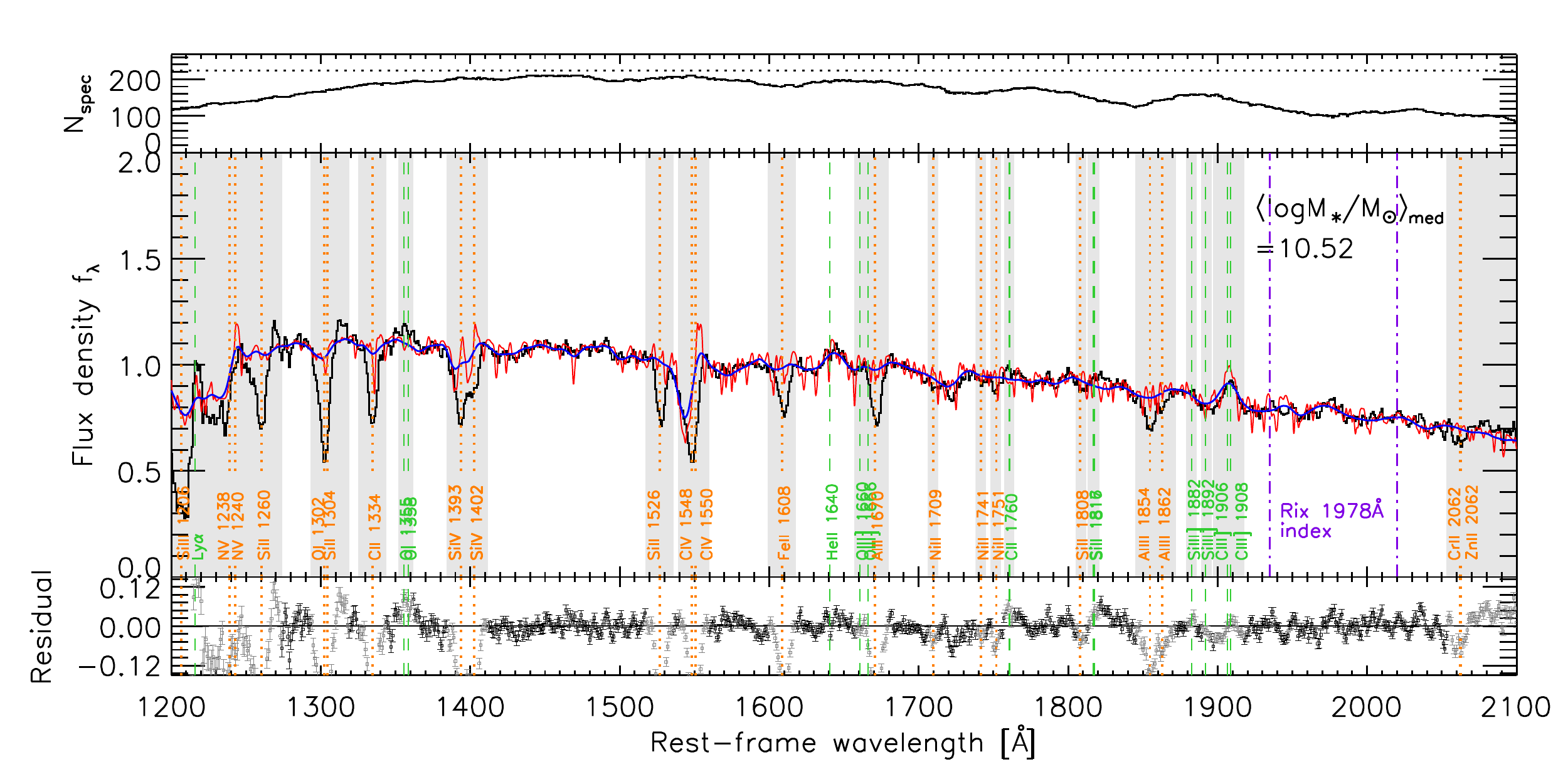}
\caption{Composite VIMOS spectra and the best-fit BPASS models are shown, separately for the six subsamples equally separated by stellar mass.  The median masses are indicated in each panel.  Each panel is in the same format as Figure \ref{fig:stack_all}.
\label{fig:stack_Mbins}}
\end{center}
\end{figure*}

We successfully determined a best-fit BPASS model for all of these bins.  Figure \ref{fig:stack_Mbins} shows the composite spectra and the corresponding best-fit models in the six equally-populated subsamples. It is noticeable that the overall slope of the spectra becomes shallower for higher $M_\ast$ due to the increasing average dust attenuation.  The overall shape and detailed features are both well reproduced across the entire wavelength range of interest by the model spectra.

The results in different bins are summarised in Table \ref{tb:result}.  The stellar metallicities are measured to range over $-1.3\lesssim \log (Z_\ast/Z_\odot) \lesssim-0.4$, increasing with $M_\ast$.  
Note that the BPASS models cover this observed range of metallicity with a sufficiently large margin at either end and thus our MCMC analysis should not be affected by the artificial limit of the explored metallicity range.
We obtained consistent results from the fixed-width binning scheme.

\begin{figure}[tbp]
\begin{center}
\includegraphics[width=3.5in]{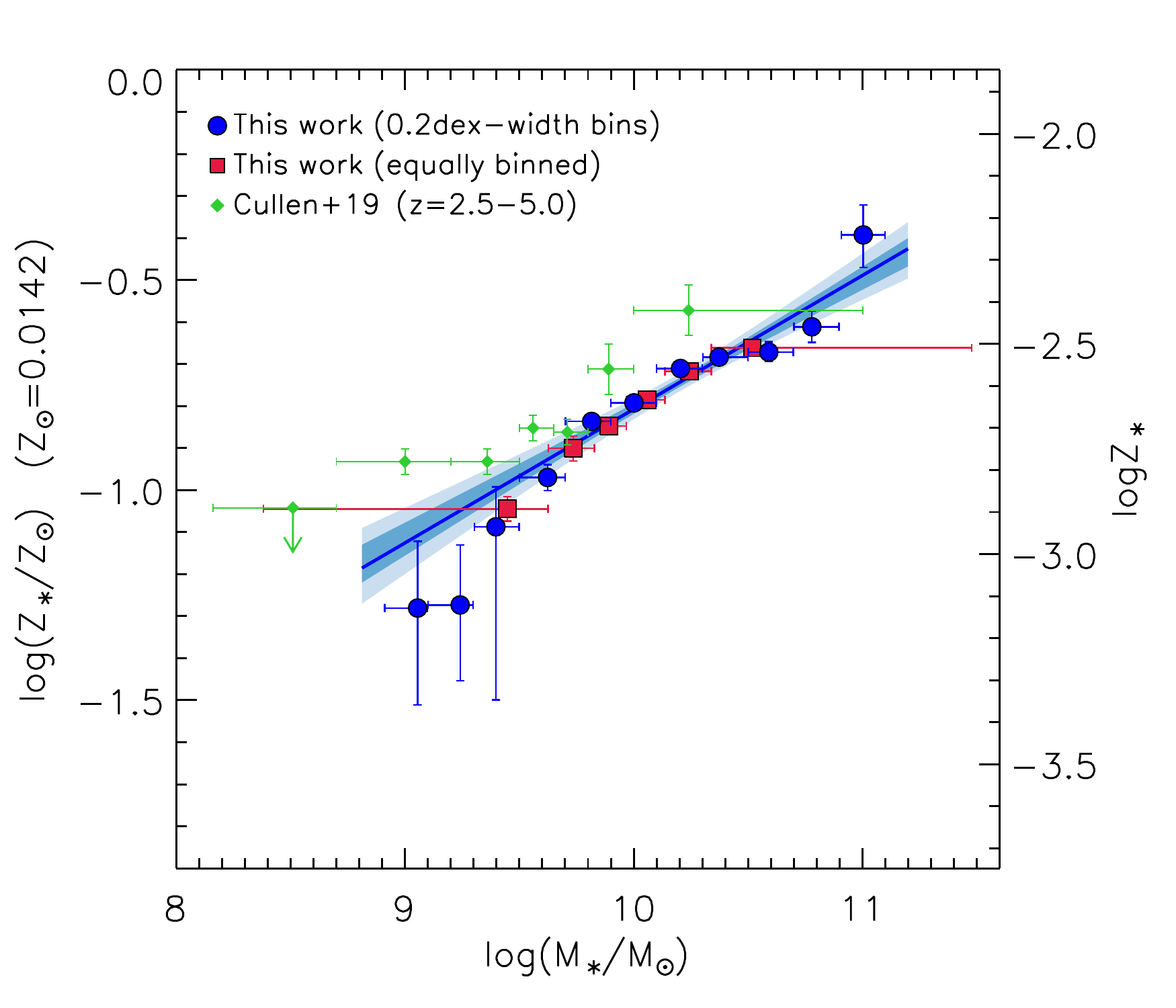}
\caption{
Stellar metallicity as a function of stellar mass, i.e., the stellar MZR, for our sample of star-forming galaxies at $1.6\le z \le 3.0$ (median $\left< z\right>=2.22$).  Here the stellar metallicity mostly reflects the iron abundance ([Fe/H]).
The Red squares and blue circles show the results based on two different binning schemes (see text).  The blue solid line indicates the best linear fit to our data (Equation \ref{eq:MZ_linear}) with the 1$\sigma$ (dark blue) and 2$\sigma$ (light blue region) confidence limits.  
For comparison, the relation for a galaxy sample at $2.5\le z \le 5.0$ ($\left< z\right>=3.5$; green diamonds) is taken from \citet{2019MNRAS.487.2038C}.
\label{fig:M_vs_Zstar}}
\end{center}
\end{figure}

Figure \ref{fig:M_vs_Zstar} shows the relationship between stellar mass and stellar metallicity.  
The $x$-axis values of the points correspond to the median stellar masses within the bin, and the horizontal error bars indicate the minimum and maximum $M_\ast$ values in each bin.  The $y$-axis values and the error bars indicate the median and the 16th--84th percentiles of the posterior distribution of $\log (Z_\ast/Z_\odot)$.  There is a tight correlation between these quantities, though the $\log (Z_\ast/Z_\odot)$ values in the lowest mass bins are relatively insecure.  A linear fit to the results in log-log space, obtained from binning the sample into 0.2~dex-width bins, yields
\begin{eqnarray}
   \lefteqn{ \log (Z_\ast/Z_\odot) }\nonumber \\
  && = -(0.81\pm0.01) + (0.32\pm0.03)\log [M_\ast/(10^{10}M_\odot)],
    \label{eq:MZ_linear}
\end{eqnarray}
where the errors in the parentheses denote the nominal $1\sigma$ errors.  This is shown as the blue line in Figure \ref{fig:M_vs_Zstar}.

This result is generally in good agreement with that of \citet{2019MNRAS.487.2038C}, who used stacks of in total 681 star-forming galaxies at $z=2.5\textrm{--}5$ ($\left< z\right>\approx3.5$).  There is a difference at the low-mass end, where the \citet{2019MNRAS.487.2038C} data show a tentative sign of flattening at $8.8\lesssim \log M_\ast/M_\odot \lesssim 9.5$, which, however, is not seen in our data (it should be noted that their lowest mass point is only an upper limit).  Overall, our metallicities are slightly lower at given $M_\ast$, which appears to be counter to the expected redshift evolution since the \citet{2019MNRAS.487.2038C} sample is generally at somewhat higher redshifts.  This apparent discrepancy may be accounted for by a systematic effect from the methodology that we now explain.

The method in \citet{2019MNRAS.487.2038C} is very similar to ours, and was based on stacks of rest-frame FUV spectra and model templates with various metallicities.  The authors, however, used the Starburst99 high-resolution WM-basic stellar population models for their fiducial results.  They compared their fiducial estimates to the ones derived using the BPASSv2.1 models and found that the use of the BPASS models leads to systematically lower metallicities by $\sim0.1~\mathrm{dex}$.  Accounting for this systematic offset will then bring the two results into better agreement.  Note that the lower metallicity limit of the Starburst99 model is $\log(Z_\ast/Z_\odot)=-1.15$ which is not sufficiently low for unbiased fitting for our sample, and thus we adopted the BPASS model in this work.  

Our results also appear to be lower by $\sim0.4~\mathrm{dex}$ than the average measurement of $\log (Z_\ast/Z_\odot) = -0.425\pm0.159$\footnote{This value is converted to $Z_\odot=0.0142$.} at $M_\ast\sim10^{10}~M_\odot$ obtained by \citet{2008A&A...479..417H} using the \citet{2004ApJ...615...98R} 1978~{\AA} index.  This offset suggests that there possibly remain substantial systematic uncertainties in the stellar metallicity measurements for different samples and methodologies. 

It should also be noted that the stellar metallicities may be underestimated because the integrated stellar emission would be biased towards stars formed in the less-obscured, and thus lower-metallicity environments.  Such systematic biases in the stellar metallicity measurements will be explored in future papers.

\subsection{Comparison with the local stellar MZR}
\label{sec:vs_local_MZR}

\begin{figure*}[tbp]
\begin{center}
\includegraphics[width=5in]{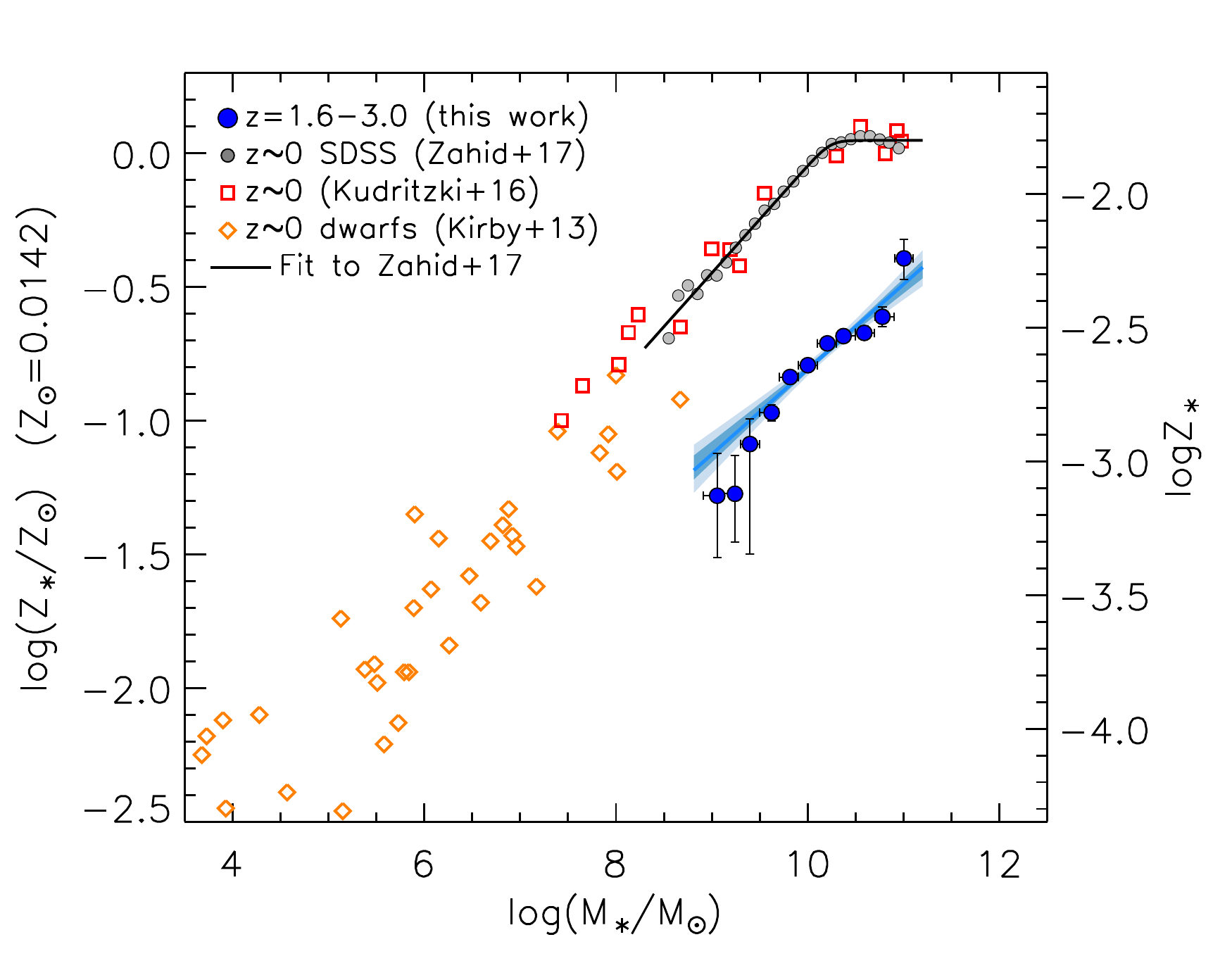}
\caption{
The stellar MZR for our sample (blue circles) at $\left<z\right>=2.22$ in comparison with those for $z\sim0$ galaxies: SDSS galaxies (gray circles; \citealt{2017ApJ...847...18Z}), individual stars in nearby galaxies (red open squares; \citealt{2016ApJ...829...70K}), and dwarf galaxies (orange open diamonds; \citealt{2013ApJ...779..102K}).  These stellar MZRs refer to iron abundances.
The blue solid line indicates the best linear fit to our data (Equation \ref{eq:MZ_linear}) with the 1$\sigma$ (dark blue) and 2$\sigma$ (light blue region) confidence limits.  The black solid line indicates the best-fit relation of Equation (\ref{eq:MZ_Curti20_Eq2}) to the local sample.  The stellar metallicities of $z\sim2.2$ galaxies are lower by $\sim0.8~\mathrm{dex}$ than the local galaxies at a given stellar mass.  Note that the local stellar metallicities are not corrected for the reduction caused by the measurements based on the rest-frame optical spectra (see Section \ref{sec:O/Fe}).
\label{fig:M_vs_Zstar_comp}
}
\end{center}
\end{figure*}

We now compare our result with other work at low redshifts to see how the stellar MZR evolves through cosmic time. \citet{2017ApJ...847...18Z} established the stellar MZR in the range $10^{8.5}\lesssim M_\ast/M_\odot \lesssim 10^{11}$ using $\sim2\times10^5$ star-forming galaxies at $z<0.25$ (median $\left<z\right>=0.08$) from SDSS.  
The authors employed a full spectral fitting approach using the observed composite spectra in the rest-frame optical.  Their result is in agreement with the measurements obtained from spectroscopy of individual stars in nearby galaxies compiled by \citet{2016ApJ...829...70K}.  Furthermore, comparing to the stellar metallicities of dwarf galaxies, measured by \citet{2013ApJ...779..102K}, revealed that the $M_\ast$--$Z_\ast$ correlation holds smoothly down to $M_\ast\sim10^4~ M_\odot$.  

We note that the metallicities from \citet{2013ApJ...779..102K} are intended to purely reflect the iron abundance (Fe/H).  The other local metallicity measurements are also assumed to reflect the iron abundance, though they might be slightly affected by other elements, including $\alpha$-elements such as magnesium.

Figure \ref{fig:M_vs_Zstar_comp} compares these low-redshift measurements with our results at high redshift.  Our high-redshift stellar MZR is clearly offset below the local relation (at our sampled masses) with $\Delta \log Z\approx 0.8~\mathrm{dex}$ at a given mass.  We should note that, the local MZR is not yet corrected for the underestimation that arises from the measurements based on the rest-frame optical integrated light of the galaxies.  We will take account of this correction, which is about $+0.1$~dex (see Appendix \ref{sec:offset_Fe_H}), in the subsequent sections where we will compare high- and low-redshift MZRs more precisely.

We employed the empirical parameterization introduced by \citet[][Equation 2]{2020MNRAS.491..944C} to express the local stellar MZR:
\begin{equation}
\log \left( \frac{Z}{Z_\odot}\right) = \mathcal{Z}_0 - \frac{\gamma}{\beta} \log\left(1+\left(\frac{M_\ast}{M_0}\right)^{-\beta}\right),
\label{eq:MZ_Curti20_Eq2}
\end{equation}  
where $Z_0$ is the asymptotic metallicity at the massive end, $M_0$ is a characteristic mass where the relation begins to flatten, $\gamma$ is the power-law slope of the relation at $M_\ast \ll M_0$, and $\beta$ determines the width of the transition region between the two extremes.  Fitting the measurements from \citet{2017ApJ...847...18Z} yields $(\mathcal{Z}_0,~M_0,~\gamma,~\beta)_{z\sim0} = (0.049,~10.24,~0.40, ~6.36)$.  In Figure \ref{fig:M_vs_Zstar_comp}, this best-fit relation clearly represents the local measurements.  

Comparing with the local relation, our result at $z=1.6\textrm{--}3.0$ is found to have a slightly shallower slope, $0.32\pm0.03$ (see above Equation \ref{eq:MZ_linear}).  As noted above, the high redshift relation is offset to lower metallicities and, perhaps for this reason, does not show evidence of a saturation at the high mass end.

\subsection{The [O/Fe]--metallicity relations in low and high redshift galaxies}
\label{sec:O/Fe}

As already noted, the gas-phase and stellar ``metallicities'' can be more or less translated, at both high and low redshifts, into oxygen and iron abundances, respectively.  Comparison of the two may therefore give some insight into the dependence and evolution of the relative abundance ratio, O/Fe, often used as a proxy of the $\alpha$-enhancement.  

The gas-phase metallicity reflects the instantaneous oxygen abundance at the time of observation.  Measuring the gas-phase iron abundance is, however, quite challenging because of the faintness of iron emission lines as well as the uncertainties in dust depletion factors; we therefore need to rely on the stellar metallicities.
The stellar metallicities of our high-$z$ galaxies estimated from the rest-frame FUV spectra, that are dominated by short-lived, recently formed stars, offers a more ``instantaneous'' measurement, similar to the gas-phase O/H.  As we discuss in Appendix \ref{sec:offset_Fe_H}, we find that the FUV-weighted [Fe/H] is consistent within $\sim0.02$~dex with the gas-phase values.  We can thus adopt the ratio of the gas-phase oxygen abundance to the stellar iron abundance of $z\sim2.2$ galaxies, [O/Fe], as a proxy of the instantaneous $\alpha$-enhancement of the gas in the galaxies for the remainder of the paper.

In contrast, the iron abundance based on the rest-frame optical spectra could be lower than the instantaneous value because of the substantial contribution to the integrated light from older stars.  This is the case of the stellar MZR at $z\sim0$ from \citet{2017ApJ...847...18Z}.  Therefore, the comparison between the two [Fe/H] measurements is not straightforward.  As shown in Appendix \ref{sec:offset_Fe_H}, the offset with respect to the instantaneous [Fe/H] is probably around $\approx -0.1~\mathrm{dex}$ for low-redshift galaxies.  In the subsequent sections, we thus shift the local stellar MZR by $\Delta\log(Z_\ast)=+0.1~\mathrm{dex}$ so that it better reflects the instantaneous iron metallicity.  This correction, however, does not have any significant impact on our conclusions.

We take the gas-phase [O/H] MZRs at $z\sim0$ and $z\sim2.2$, respectively, from \citet[][Equation 2]{2020MNRAS.491..944C} and \citet{2020MNRAS.491.1427S}.  The former uses the latest accurate calibration between the optical strong-line ratios and the metallicity determined from the direct method.  The latter is purely based on the direct method O/H measure in individual 18 galaxies at $z=1.7\textrm{--}3.6$ (median $\left<z\right>=2.17$).  An important caveat is that the sample of \citet{2020MNRAS.491.1427S} is not representative of our zCOSMOS-deep sample, but rather a compilation from the literature of different surveys.  The following analysis is thus based on an assumption that the result of \citet{2020MNRAS.491.1427S} and ours both independently represent the same, typical star-forming galaxy population at these epochs.

\citet{2020MNRAS.491.1427S} corrected their O/H measurements for the residuals around the epoch's main sequence, $\Delta_\mathrm{MS}\log(\mathrm{SFR})$, of their sample galaxies to obtain more representative metallicities at given redshift.  The corrections achieve $\sim0.2$~dex on average.  These corrections are, however, questionable at some level due to the systematic uncertainties in determining the shape of the main sequence and the $\Delta_\mathrm{MS}\log(\mathrm{SFR})$-dependence of the metallicity, both certainly being a function of redshift. Indeed, this adopted dependence is approximately twice as strong as that inferred from the strong-line method locally.  We therefore adopt here their direct measurements of O/H (Equation 7 of \citealt{2020MNRAS.491.1427S}) with a moderate constant correction of $\Delta \log(\mathrm{O/H})=+0.1$.  Although this choice is more or less arbitrary, it is indeed within the statistical error of the direct measurement.

A similar bias might also be expected in our own sample.  However, we do not find a significant offset from the main sequence at this epoch given by \citet{2014ApJ...795..104W} in the $M_\ast$--SFR diagram of Figure \ref{fig:M_vs_SFR}.  We therefore assume that the observed galaxies, and their stellar metallicities, are reasonably representative of galaxies of the given $M_\ast$ at these epochs.

\begin{figure}[tbp]
\begin{center}
\includegraphics[width=3.5in]{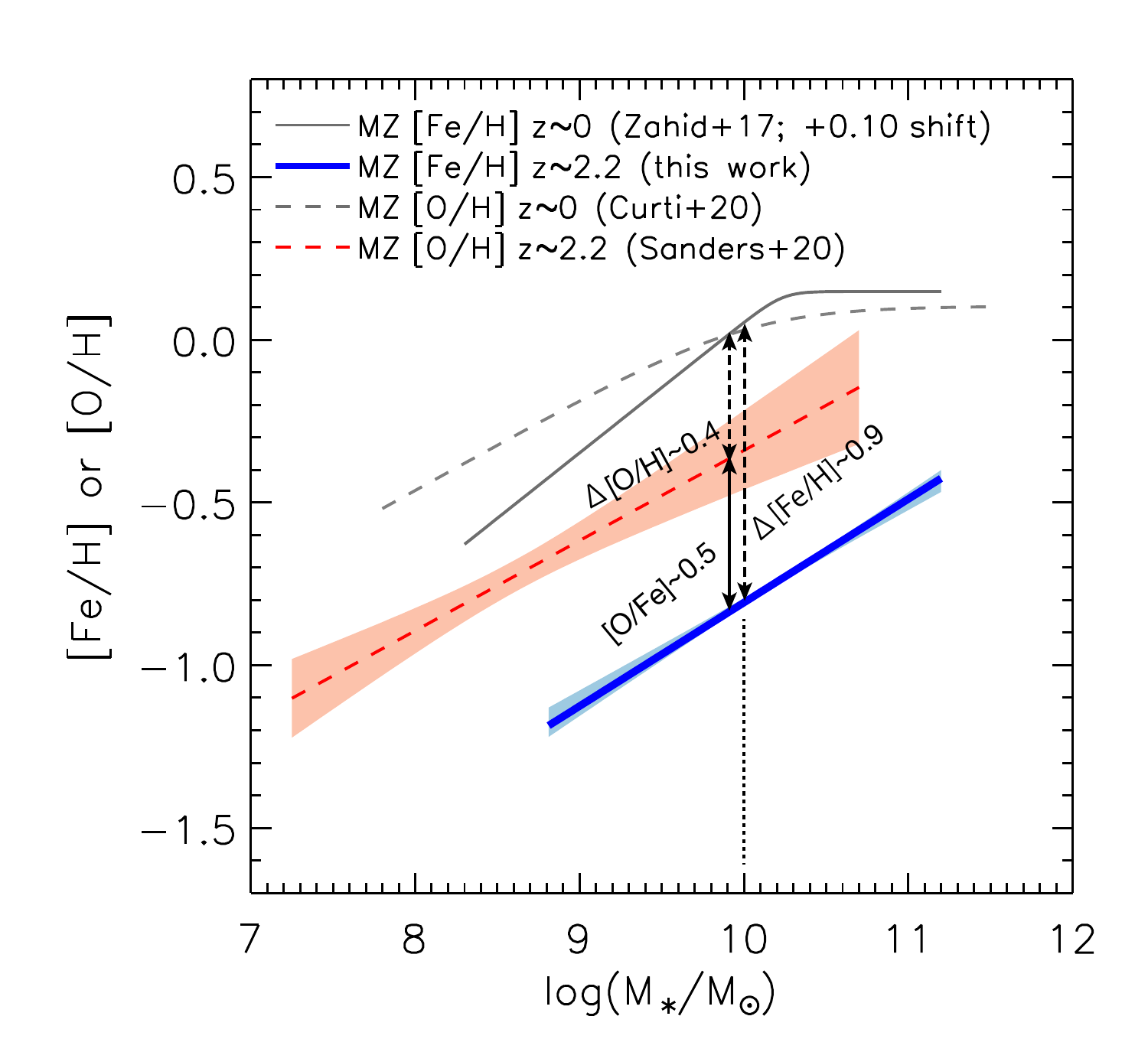}
\caption{
Comparison between the stellar ([Fe/H]) and gas-phase ([O/H]) MZRs at low and high redshifts: stellar MZR at $z\sim0$ (gray solid line; \citealt{2017ApJ...847...18Z} fitted with Equation \ref{eq:MZ_Curti20_Eq2}), stellar MZR at $z\sim2.2$ (thick solid blue line; our data, Equation \ref{eq:MZ_linear}), gas MZR at $z\sim0$ (gray dashed line; \citealt{2020MNRAS.491..944C}), gas MZR at $z\sim2.2$ (red dashed line; \citealt{2020MNRAS.491.1427S}).  Here the local stellar MZR is shifted by $\log Z_\ast = +0.1$~dex so that it reflects better the instantaneous iron abundance in the gas phase.  No correction is applied for the $z\sim2.2$ stellar MZR as it is based on the rest-frame FUV spectra.
The light blue and red shaded regions indicate, respectively, the 68\% confidence limits of the fits.
\label{fig:Zstar_vs_Zgas}}
\end{center}
\end{figure}

In Figure \ref{fig:Zstar_vs_Zgas}, we compare the empirical fits to the four MZRs (stellar and gas-phase; $z\sim0$ and $z\sim2.2$).  All the metallicities are here normalised to the solar values.
It can be seen that the evolution in [Fe/H] between $z\sim0$ and $z\sim2.2$ is larger than that seen in [O/H]. At $M_\ast\sim10^{10}~M_\odot$, the [Fe/H] metallicities at $z \sim 2.2$ are about 0.9~dex lower than locally (including the $+0.1$~dex shift in the local stellar MZR; see above), whereas the [O/H] metallicities are only about 0.4~dex lower than that at $z\sim0$.
In other words, the offset between the instantaneous [O/H] and [Fe/H] metallicities evidently increases with redshift; put another way, at $M_\ast=10^{10}~M_\odot$, $\mathrm{[O/Fe]} \approx 0$ is indicated locally, but $\mathrm{[O/Fe]} = 0.47\pm0.12$  at $z\sim2.2$. Interestingly, the implied value of [O/Fe] at $z\sim2.2$ is close to the average $\textrm{[O/Fe]}=0.42$ reported by \citet{2018ApJ...868..117S} for star-forming galaxies at $z\sim2.3$.  It is also approaching the predicted maximum value ($\sim0.6$) that is predicted for metal enrichment from core-collapse supernovae (CCSNe; \citealt{2006NuPhA.777..424N,2017ApJ...835..224A}) alone, suggesting that the production of iron by SNe~Ia had not progressed far at these redshifts.

\begin{figure}[tbp]
\begin{center}
\includegraphics[width=3.5in]{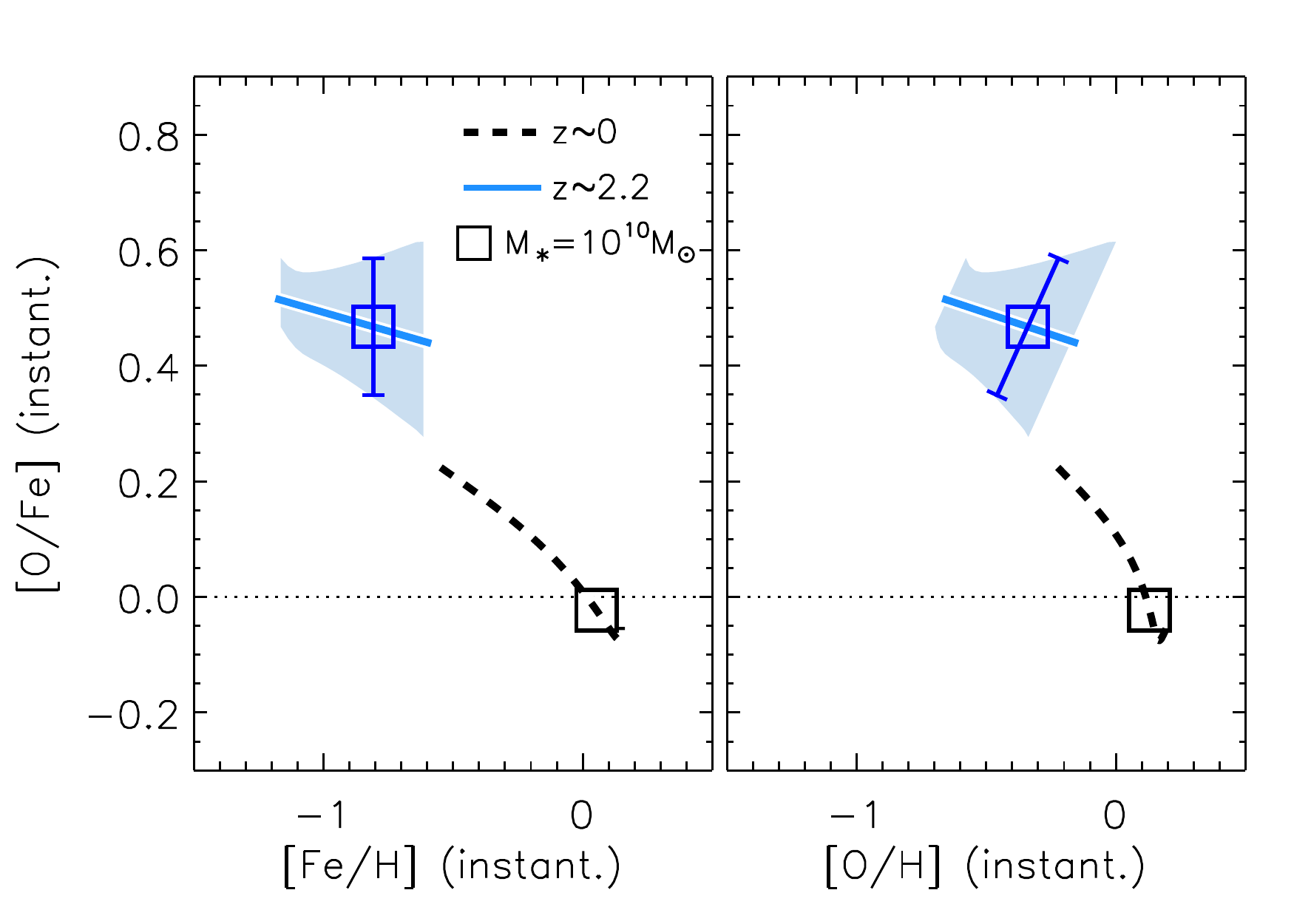}
\caption{
The implied [O/Fe] as a function of the iron abundance [Fe/H] (left panel) and the oxygen abundance [O/H] (right panel) obtained by eliminating $M_\ast$ from the $M_\ast$--[Fe/H] and $M_\ast$--[O/H] relations shown in Figure \ref{fig:Zstar_vs_Zgas}.  The blue solid and black dashed lines indicate the inferred relations at $z\sim2.2$ and $z\sim0$, respectively.  The mass ranges are limited to $\log(M_\ast/M_\odot)=8.5\textrm{--}11.0$ ($z\sim 0$) and $8.8\textrm{--}10.7$ ($z\sim2.2$), respectively.  
For the $z\sim2.2$ relations, the shaded regions indicate the 68\% confidence limit of each relation.  The squares indicate the values at $M_\ast = 10^{10}~M_\odot$ at each redshift.  The error bar correspond to the 1$\sigma$ error.  Note that the error in [O/Fe] is correlated with that in the $x$-axis values; thus the error bar is nearly vertical in the left panel while being tilted in the right panel because it is dominated by the error in [O/H] while the error in [Fe/H] is much smaller than the symbol size.
\label{fig:Z_vs_OFe}}
\end{center}
\end{figure}

We can in principle eliminate $M_\ast$ from the gas and stellar MZRs (taken from Figure \ref{fig:Zstar_vs_Zgas}) to yield the relations between [O/Fe] and either [O/H] or [Fe/H].  This is done in Figure \ref{fig:Z_vs_OFe} which shows the [O/Fe] as a function of [Fe/H] (left panel) and of [O/H] (right panel).  The mass ranges of the $z\sim 0$ and $z\sim 2.2$ relations are limited to $\log(M_\ast/M_\odot)=8.5\textrm{--}11.0$ and $8.8\textrm{--}10.7$, respectively.

Of course, this procedure of eliminating $M_\ast$ from the MZR fits may not produce the same [O/Fe] vs. metallicity relations as would be obtained by considering individual galaxies, because it does not consider the scatter in the observed quantities at a given $M_\ast$.  In other words, this procedure is tantamount to assuming that galactic mass is the primary driver of the variations in the [Fe/H] and [O/H] within the observed sample.  In the current study, we have no alternative to this procedure, because we relied on stacked spectra (in stellar mass bins) for our metallicity measurements.  

The results show that these relations have negative slopes in both panels and at both redshifts.  Interestingly, the low-redshift [O/Fe]--[Fe/H] (and [O/Fe]--[O/H]) relations in Figure \ref{fig:Z_vs_OFe} appear to broadly line up with the overall evolutionary vectors from $z\sim2.2$ to $z\sim0$.  The high-redshift [O/Fe]--[Fe/H] (and [O/Fe]--[O/H]) relation is also consistent with having the same slope, but this slope is quite uncertain.  

Uncertainties in the slopes of the input MZR can propagate to have a large effect on the resultant slopes of the inferred [O/Fe] vs. metallicity relations.  We constructed confidence intervals on these slopes by considering the range of possible linear relations that are obtained using random combinations of fits (within the uncertainties) to the input $M_\ast$--[O/H] and $M_\ast$--[Fe/H] relations.  These are shown in Figure \ref{fig:Z_vs_OFe}.  Flat or even positive correlations between [O/Fe] and overall metallicities are evidently allowed by our available high-redshift data and analysis methods.  We do not try to evaluate the range of possible slopes at low redshift because, while the statistical errors in the fits are very small, it is difficult for us to assess any systematic uncertainties in those taken from the independent studies.  

Given the large uncertainties in the slopes of these relations, we focus instead on the [O/Fe] and [Fe/H] (or [O/H]) values at a single fiducial mass of $M_\ast=10^{10}~M_\odot$ (close to the middle of our high-$z$ mass range) as being representative of each redshift.  These representative values for $M_\ast=10^{10}~M_\odot$ at high and low redshifts are shown by the two blue and black squares in each of the panels of Figure \ref{fig:Z_vs_OFe}.

\section{Modeling of iron and oxygen abundances}
\label{sec:modeling}

In this section, we explore the observed evolution in both [O/Fe] and [Fe/H] (and [O/H]) in the context of ``flow-through'' gas-regulated models of galaxies.  The goal is to demonstrate that the simple chemical evolution model explains well all the observed changes in these quantities from $z\sim2$ to $z\sim0$ while assuming the galaxies have followed the evolving main sequence through cosmic time.  In particular, we will derive the evolutionary tracks in the [O/Fe]--[Fe/H] (and [O/H]) planes to show that all galaxies must follow limited evolutionary paths in these diagrams and the location of the galaxies are determined almost entirely by sSFR alone.

All the metallicities, [Fe/H] and [O/H] (and thus [O/Fe]) refer to the instantaneous (gas-phase) values if not specified throughout the section.

\subsection{Model framework}
\label{sec:model_framework}

In what follows, we adopt the gas-regulator model of \citet{2013ApJ...772..119L} in which the SFR is instantaneously regulated by the mass of gas ($M_\mathrm{gas}$) present in some reservoir, via the star formation efficiency ($\mathrm{SFE} = \SFR/M_\mathrm{gas}$), and with a wind-driven mass loss that scales with the SFR via a ``mass-loading'' factor $\eta$.  Mass conservation then straightforwardly gives (see Equation 9 of \citealt{2013ApJ...772..119L})
\begin{equation}
    \Phi = (1 - r+\eta)\cdot\SFR + \dot{M}_\mathrm{gas}
    \label{eq:mass_conservation}
\end{equation}
where $\Phi$ is the mass inflow rate and $r$ is the recycling factor (or called the return fraction; hereafter fixed to 0.4), i.e., the fraction of mass that is formed into stars then at later times returned to the ISM.    

Obviously, once the two parameters, SFE and mass-loading $\eta$, are specified (possibly as a function of mass and/or redshift), the gas accretion history (and gas-content history ${M}_\mathrm{gas}$(t)) of a given system follows {\it completely} from its SFH, since $\dot{M}_\mathrm{gas}$ will also be given by the change in $\mathrm{SFR}(t)$.  Note that it is the changing gas reservoir that distinguishes this ``gas-regulator'' model from the ``bathtub'' models of \citet{2010ApJ...718.1001B} and \citet{2012MNRAS.421...98D}, in which  $\dot{M}_\mathrm{gas}$ is set to be zero.  

This means that, within the context of the gas-regulator model, the instantaneous metallicity of the gas reservoir will also be completely determined by the SFH once the (possibly mass- and/or epoch-dependent) SFE and $\eta$ are specified, along with an assumption of the metallicity of the inflowing gas; we will here assume for simplicity that this is zero.

In the following, we consider the instantaneous gas-phase metallicity as the ratio of metal mass in the gas phase and gas mass at given time.  We do not consider metal depletion onto dust grains, and thus the gas-phase metal mass represents all metals except locked in surviving stars and those metals gone in the wind.

We can assume that the oxygen is produced only by core-collapse supernovae (CCSNe) that occur ``promptly'' (with zero time-delay) after the birth of the progenitor stars.  The change of oxygen mass $\MO$ in the gas phase is therefore expressed as 
\begin{equation}
    \frac{d \MO}{d t} = \yOcc \cdot \SFR - \ZO (1-r+\eta)\cdot \SFR + \Phi \ZOinf
    \label{eq:dMO/dt}
\end{equation}
where $\yOcc$ is the IMF-weighted oxygen yield defined as the oxygen mass synthesized then returned into the ISM per unit mass formed\footnote{This definition of the yield is different from that in some other literature where the yield is denoted in units of mass that is locked up into long-lived stars and remnants.  The difference between these definition is thus a factor of (1-r).}, $\ZO (=\MO/\Mgas)$ is the oxygen abundance, and $\ZOinf$ is the oxygen abundance of the infalling gas.

We use Equation (\ref{eq:dMO/dt} to numerically track the chemical evolution.  However, it is also useful to formalize the metallicity too.
The change in $\ZO$ is then obtained by eliminating $\Phi$ using Equation (\ref{eq:mass_conservation}) as
\begin{equation}
    \frac{d\ZO}{dt}=\yOcc-(\ZO-\ZOinf)(1-r+\eta) \mathrm{SFE} - (\ZO-\ZOinf)\frac{\dot{M}_\mathrm{gas}}{\Mgas}.
\end{equation}
The metallicity in an equilibrium condition is thus derived by setting $d\ZO/dt$ to zero, i.e., 
\begin{eqnarray}
    \ZOeq &=& \ZOinf +  \frac{\yOcc}{1-r+\eta + \mathrm{SFE}^{-1}\left( (1-r) \mathrm{sSFR} + \dot{M}_\mathrm{gas}/\Mgas\right)}.
    \label{eq:ZOeq}
\end{eqnarray}
\cite{2013ApJ...772..119L} showed that the timescale for driving $\ZO$ toward $\ZOeq$ is shorter than the timescale on which the equilibrium conditions are varying.  Assuming equilibrium is therefore a good approximation. 

\begin{figure}[tbp]
\begin{center}
\includegraphics[width=3.5in]{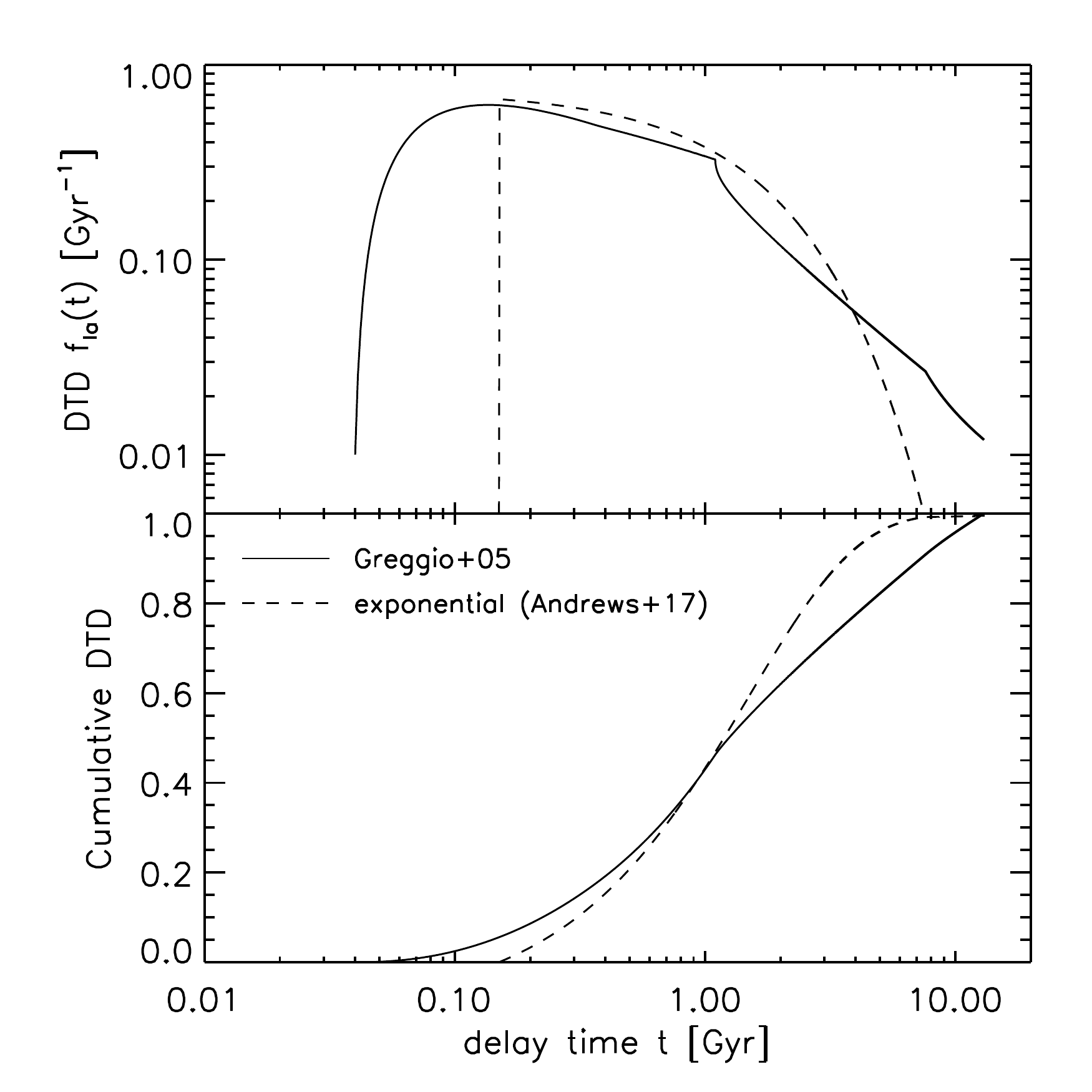}
\caption{The adopted SN~Ia DTD adapted from \citet{2005A&A...441.1055G} is compared to the exponential DTD adopted in \citet{2017ApJ...835..224A}.  Both are normalized so that the total number of events equals 1.  The lower panel shows the cumulative DTDs.
\label{fig:dtd}}
\end{center}
\end{figure}

For iron, we must consider the substantial amount of iron that is produced by Type-Ia SNe (SNe Ia), which occur with some considerable delay after the birth of their stellar progenitors.  In this analysis, we adopt the expression for the distribution of the SN~Ia delay time, $\fIa (t)$, that was formalized by \citet{2005A&A...441.1055G} for a single stellar population (see also \citealt{2008MNRAS.388..829G,2010MNRAS.406...22G}). We consider contributions from the single-degenerate and double-degenerate channels.  Note that $\fIa (t)$ is normalized so that the time integration equals 1.
Figure \ref{fig:dtd} shows the adopted delay time distribution (DTD), compared with a simple exponential parameterization used in \citet{2017ApJ...835..224A} and \citet{2017ApJ...837..183W}.  The adopted one is more sensitive to SFR at earlier times in the past.

Using $\fIa$, the change of the gas-phase iron mass $\MFe$ is written as 
\begin{eqnarray}
    \frac{d \MFe}{d t} &=& \yFecc \cdot \SFR - \ZFe (1 - r + \eta)\cdot \SFR + \Phi \ZFeinf \nonumber \\
    && + \yFeIa \int_0^{t} \mathrm{SFR}(t^\prime) \fIa(t-t^\prime) dt^\prime.
    \label{eq:dMFe/dt}
\end{eqnarray}
where $\yFecc$ is the CCSN iron yield and $\ZFeinf$ is the iron abundance of the infalling gas. The last term denotes the contribution from SNe~Ia, where $\yFeIa$ is the time-integrated SN~Ia yield of iron for unit mass formed\footnote{The SNe~Ia yield $\yFeIa$ is usually expressed as the product $\KFeIa R_0$ where $R_0$($\sim10^{-3}~M_\odot^{-1}$; \citealt{2008MNRAS.388..829G,2012PASA...29..447M}) is the time-integrated number of SNe~Ia per unit stellar mass formed and $\KFeIa$ is the average mass of iron from an individual SN~Ia.}.

The steady-state iron metallicity is then written as
\begin{eqnarray}
    \ZFeeq &=& \ZFeinf + \nonumber \\
    && \frac{\yFecc + (\yFeIa/\SFR) \int_0^{t} \SFR(t^\prime) \fIa(t-t^\prime) dt^\prime}
    {1-r+\eta + \mathrm{SFE}^{-1}\left( (1-r) \mathrm{sSFR} + \dot{M}_\mathrm{gas}/\Mgas\right)}.
    \label{eq:ZFeeq}
\end{eqnarray}
Assuming $\ZOinf = \ZFeinf=0$, Equations (\ref{eq:ZOeq}) and (\ref{eq:ZFeeq}) yield
\begin{eqnarray}
    \frac{\ZOeq}{\ZFeeq}&=&\frac{\yOcc}{\yFecc+(\yFeIa/\SFR)\int_0^{t} \SFR(t^\prime) \fIa(t-t^\prime) dt^\prime} \\
    &\approx&\frac{\yOcc \SFR}{\yFeIa\int_0^{t} \SFR(t^\prime) \fIa(t-t^\prime) dt^\prime}
    \label{eq:ZOeq/ZFeeq}
\end{eqnarray}
where the approximation holds when the SNe~Ia dominates the iron production.   This indicates that the [O/Fe] is approximately proportional to the number ratio of the CCSNe and SNe~Ia at any time.  In other words, as the denominator is  some kind of average SFR in the past, the [O/Fe] would be tightly correlated with the sSFR.  We will see this in Section \ref{sec:model_alpha-metallicity}.

The parameters in these equations are not very well constrained from observations.  We therefore basically follow the ``fiducial'' choice of \citet{2017ApJ...835..224A} and \citet{2017ApJ...837..183W}; $r=0.4$ and $\yOcc=0.017$ (see also \citealt{2016MNRAS.455.4183V} for the IMF-weighted yield and return mass fraction).  As discussed below (Section \ref{sec:model_params}), we adjusted the values of the CCSN and SN~Ia iron yields, taking $\yFecc=0.00081$ (instead of 0.0012), $\yFeIa = 0.0022$ (instead of 0.0017), together with the parameters determining the SFE and $\eta$.  As noted above, for simplicity we set both $\ZOinf$ and $\ZFeinf$ to zero.

\subsection{Calculating chemical evolutionary tracks}
\label{sec:model_calc}

\subsubsection{Choice of star formation histories}
\label{sec:model_SFHs}

Again for simplicity, we construct a representative set of SFHs by integrating the evolving main-sequence of star-forming galaxies across cosmic time.  We adopted the local $M_\ast$--SFR relation ($z\approx0.08$) derived by \citet{2015ApJ...801L..29R} and the redshift evolution at fixed $M_\ast$ as follows:
\begin{eqnarray}
    \log \mathrm{SFR}~(M_\odot~\mathrm{yr^{-1}}) &= 0.76 \times \log M_\ast (M_\odot) - 7.55 \nonumber \\
    & + 3.0\times\log(1+z)
    \label{eq:sSFR_evol_lowz}
\end{eqnarray}
for $z \le 2.4$ and 
\begin{eqnarray}
    \log \mathrm{SFR}~(M_\odot~\mathrm{yr^{-1}}) &= \log \mathrm{SFR}(z=2.4) \nonumber \\
    & + 1.2\times\log\left(\frac{1+z}{1+2.4}\right)
    \label{eq:sSFR_evol_highz}
\end{eqnarray}
for $z>2.4$ (see e.g., \citealt{2015A&A...577A.112L,2015A&A...581A..54T}).

\begin{figure*}[tbp]
\begin{center}
\includegraphics[width=3.5in]{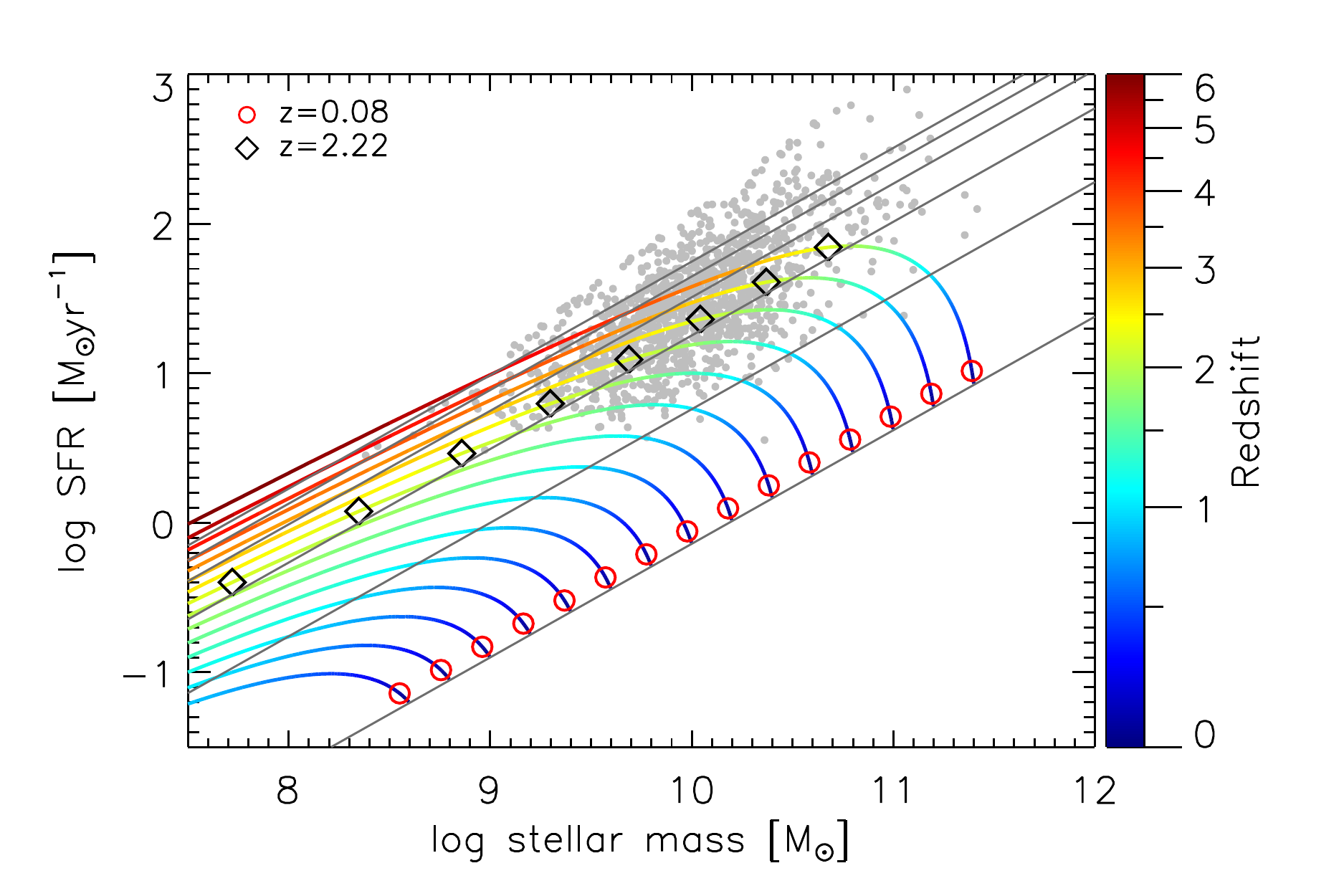}
\includegraphics[width=3.5in]{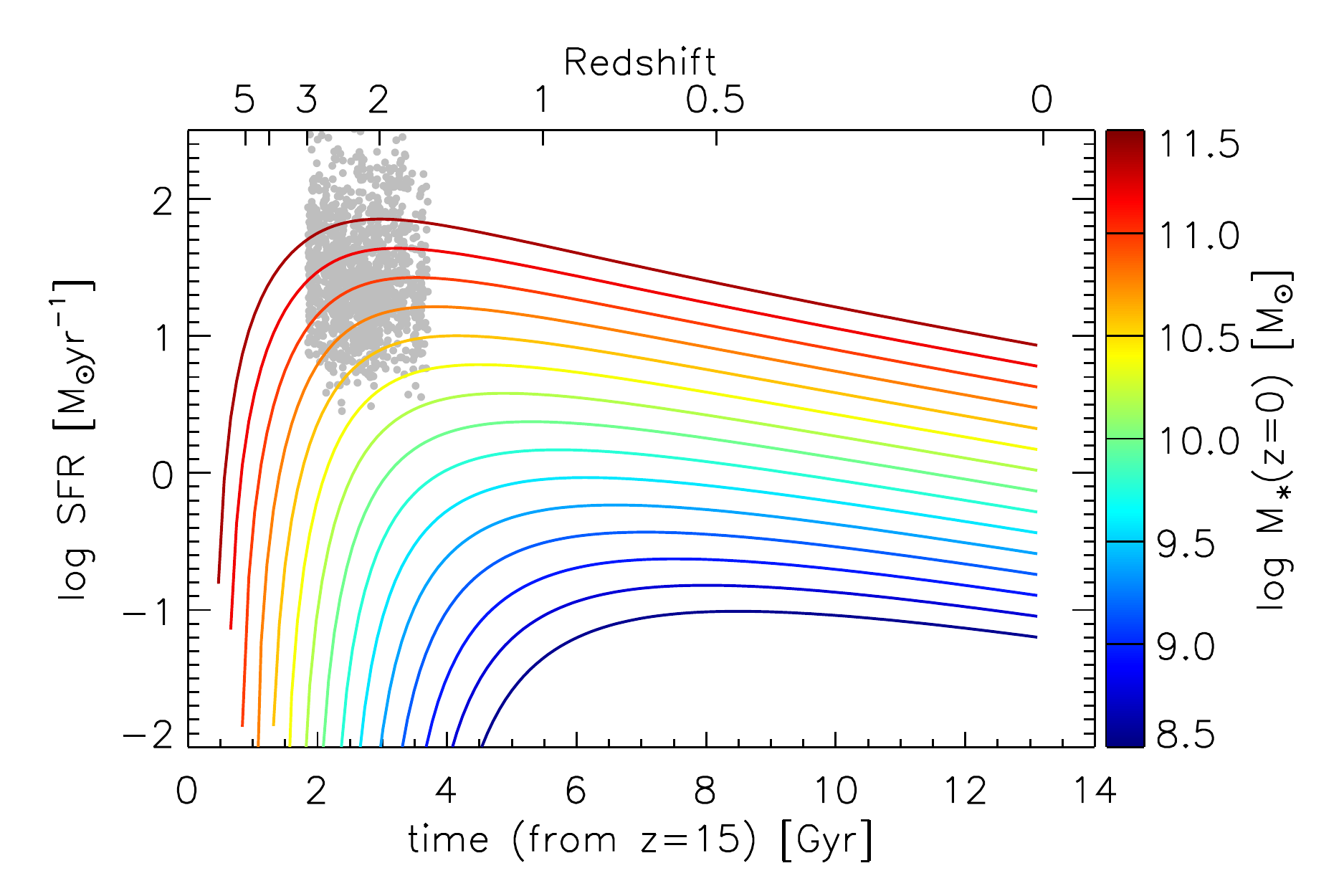}
\caption{Left panel: the evolving main-sequence of star-forming galaxies shown by gray lines at $z=1,2,3,4$ and 5 from bottom to top.  The curves show the evolutionary tracks of galaxies at given stellar masses in steps of 0.2~dex at $z=0$, color-coded by redshift.  The black diamonds and red circles mark the time steps of $z=2.22$ and $z=0.08$, respectively, for all shown tracks.  Gray dots indicate our sample galaxies at $1.6 \le z_\mathrm{spec} \le 3.0$.  Right panel: the corresponding SFHs.  Each line is color-coded by stellar mass at $z=0$.  The gray dots indicate SFR vs. $z_\mathrm{spec}$ for our sample galaxies.
\label{fig:models_SFR}}
\end{center}
\end{figure*}

The inferred evolutionary tracks in the $M_\ast$--SFR plane and SFHs are shown in Figure \ref{fig:models_SFR}.  In the former, we mark the values of our representative galaxies at $z=0.08$ and $z=2.22$.  These simulated galaxies at $z=2.22$ are in broad agreement with our sample galaxies, as shown by gray dots.

\subsubsection{Parameters of the regulator systems and iron yields}
\label{sec:model_params}

In order to derive the time-dependent gas content of each model galaxy from their individual SFHs, and thereby compute the corresponding chemical evolution, we need to define the two parameters of the gas-regulator system, the mass-loading factor $\eta$ and SFE.  We assume that these values scale with the instantaneous $M_\ast$ of the system, but that they are not redshift-dependent, i.e.: 
\begin{eqnarray}
    \eta &&= \mathrm{max}\left\{\eta_{10} \times (M_\ast/(10^{10}~M_\odot))^a, \eta_\mathrm{min}\right\}, \\
    \label{eq:eta}
    \mathrm{SFE} &&= \mathrm{SFE}_{10} \times (M_\ast/(10^{10}~M_\odot))^b.
    \label{eq:SFE}
\end{eqnarray}
Here we consider the minimum value for $\eta$ at high masses (assuming a negative $a$) for better representation of the saturation feature of the local MZR at the high-mass end.  Althoguh this is an arbitrary treatment, the bending in the average $\eta$ toward high masses is seen due to effects of active galactic nuclei (AGN) in simulations \citep{2019MNRAS.490.3234N}.

Now Equations (\ref{eq:dMO/dt}) and (\ref{eq:dMFe/dt}) can be used to compute the evolution with time of [O/H] and [Fe/H] for {\it any} arbitrary $\mathrm{SFR}(t)$, i.e., for each of the representative SFH identified above.  However, using the fiducial iron yields given by \citet{2017ApJ...835..224A} and \citet{2017ApJ...837..183W} gives an [Fe/H] MZR that is slightly higher at $z\sim2.2$ but lower locally than those observed at each redshift shown in Figure \ref{fig:Zstar_vs_Zgas}.  We also notice that the value of $\yFecc=0.0012$ adopted in the above papers gives [O/Fe] ($=0.5$ with $\yOcc=0.017$) from pure CCSNe which is smaller than the values seen in the literature ($\sim0.6$; e.g., \citealt{2006NuPhA.777..424N}).  We therefore allowed the CCSN and SN~Ia iron yields to vary in order to better reproduce the observed mass--metallicity relations.

We used an MCMC algorithm to determine values of the six parameters to be $(\eta_{10}, a, \eta_\mathrm{min}, \mathrm{SFE}_{10}(\mathrm{Gyr^{-1}}), b, \yFecc, \yFeIa)$
$= (2.58, -0.267, 2.34, 0.310, 0.043, 0.00083, 0.00215)$
that, using our representative SFHs, well reproduce the $M_\ast$--[O/H] and $M_\ast$--[Fe/H] relations at $z \sim 0$ and $z \sim 2.2$ shown in Figure \ref{fig:M_vs_Zstar_comp}.  

We note that these values are not far from those suggested from observations and/or simulations.  
The mass loading factor and its scaling is similar to what is found in \citet[][see their Equation 8]{2015MNRAS.454.2691M}.  The typical (total gas) depletion timescales ($t_\mathrm{dep}=1/\mathrm{SFE}$) of several Gyr have been observed \citep{2008AJ....136.2846B} and reproduced in simulations \citep{2017ApJ...845..133S}.  From Equations (\ref{eq:sSFR_evol_lowz}) and (\ref{eq:SFE}), we obtain $\Mgas \propto M_\ast^{0.72}$, which is in broad agreement with the scaling relation between the molecular gas mass and stellar mass ($\sim M_\ast^{0.59}$; \citealt{2020ARA&A..58..157T}).
The corrections of the iron yields are also small ($\sim30\%$) with respect to the fiducial values.

\subsubsection{Calculated chemical evolutionary tracks}

\begin{figure*}[tbp]
\begin{center}
\includegraphics[width=3.3in]{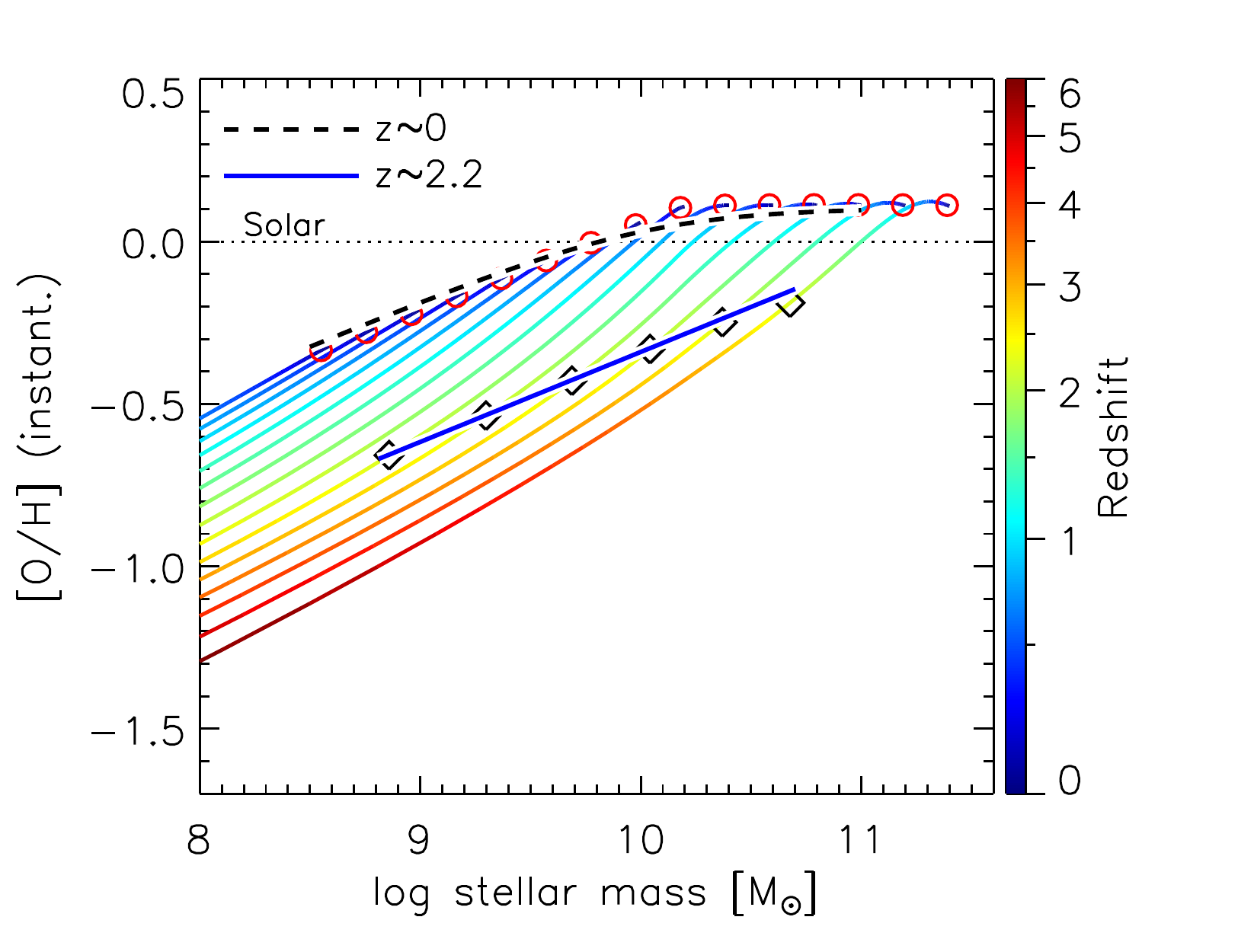}
\includegraphics[width=3.3in]{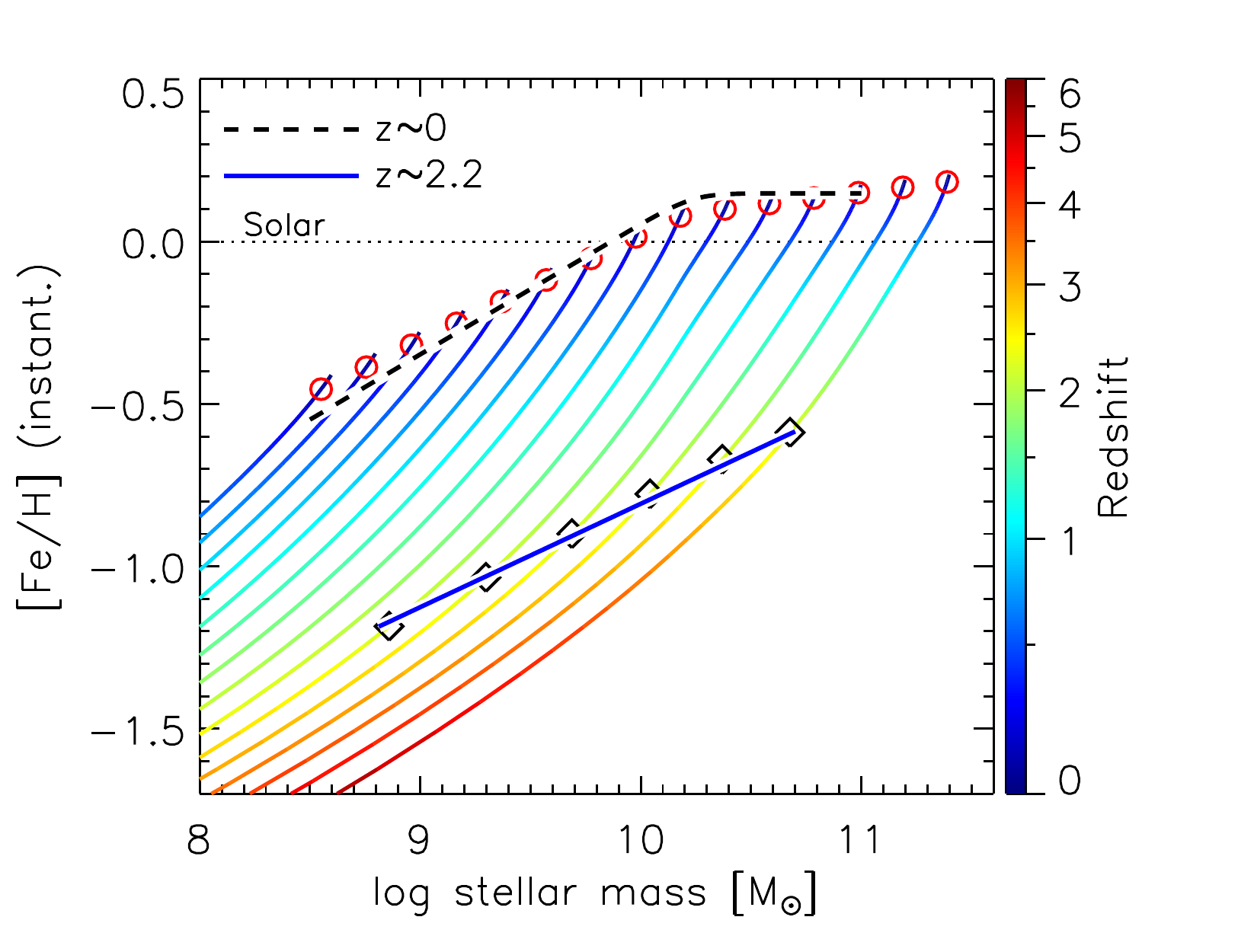}
\caption{Left panel: the calculated evolutionary tracks in the $M_\ast$--[O/H] diagram in comparison with the observations; \citet{2020MNRAS.491..944C} for $z\sim0$ (black dashed line) and \citet{2020MNRAS.491.1427S} for $z \sim 2.2$ (blue solid line; shifted by $+0.1$~dex as mentioned in Section \ref{sec:O/Fe}).  The color-coding is according to the redshift.  Black diamonds mark those with $M_\ast \gtrsim 10^{8.8}~M_\odot$ at $z=2.22$ and red circles mark those at $z=0.08$.  Right panel: same as the left panel but in the $M_\ast$--[Fe/H] diagram.  The observed relations come from \citet{2017ApJ...847...18Z} at $z\sim0$ (shifted by $-0.1$~dex as mentioned in Section \ref{sec:O/Fe}) and our result (Equation \ref{eq:MZ_linear}).
\label{fig:models_MZ}}
\end{center}
\end{figure*}

In Figure \ref{fig:models_MZ}, we show the chemical evolution tracks for the input SFHs in the iron and oxygen MZR diagrams.  
The individual lines correspond to each of SFHs that were shown in Figure \ref{fig:models_SFR}. As there, the positions of the galaxies at $z = 2.22$ and at $z = 0.08$ are marked (again, black diamonds and red circles respectively). At $z=2.22$ we limit those to having $M_\ast$ higher than $\approx 10^{8.8}~M_\odot$, which is the lower mass limit of our high-$z$ sample.

This figure illustrates how the models well reproduce the observed MZRs both in oxygen and iron, including their slopes, at both $z\sim0$ and $z\sim2.2$.  This was of course to be expected since both the $\mathrm{SFE}(M_{\ast})$ and $\eta(M_{\ast})$ functions of the regulator model and the iron yields have been adjusted so as to match these overall MZRs at both $z \sim 2$ and $z \sim 0$, although we stress that the adopted parameters are in our view completely reasonable.

Rather, our interest in constructing these models is in examining and understanding the {\it expected} relations between the overall metallicity and the $\alpha$-enhancement at these two redshifts, especially in comparison with the evolutionary offset between them.  We address this in the following sections of the paper.

\subsection{The expected [O/Fe]--metallicity relation}
\label{sec:model_alpha-metallicity}

\begin{figure*}[tbp]
\begin{center}
\includegraphics[width=3.3in]{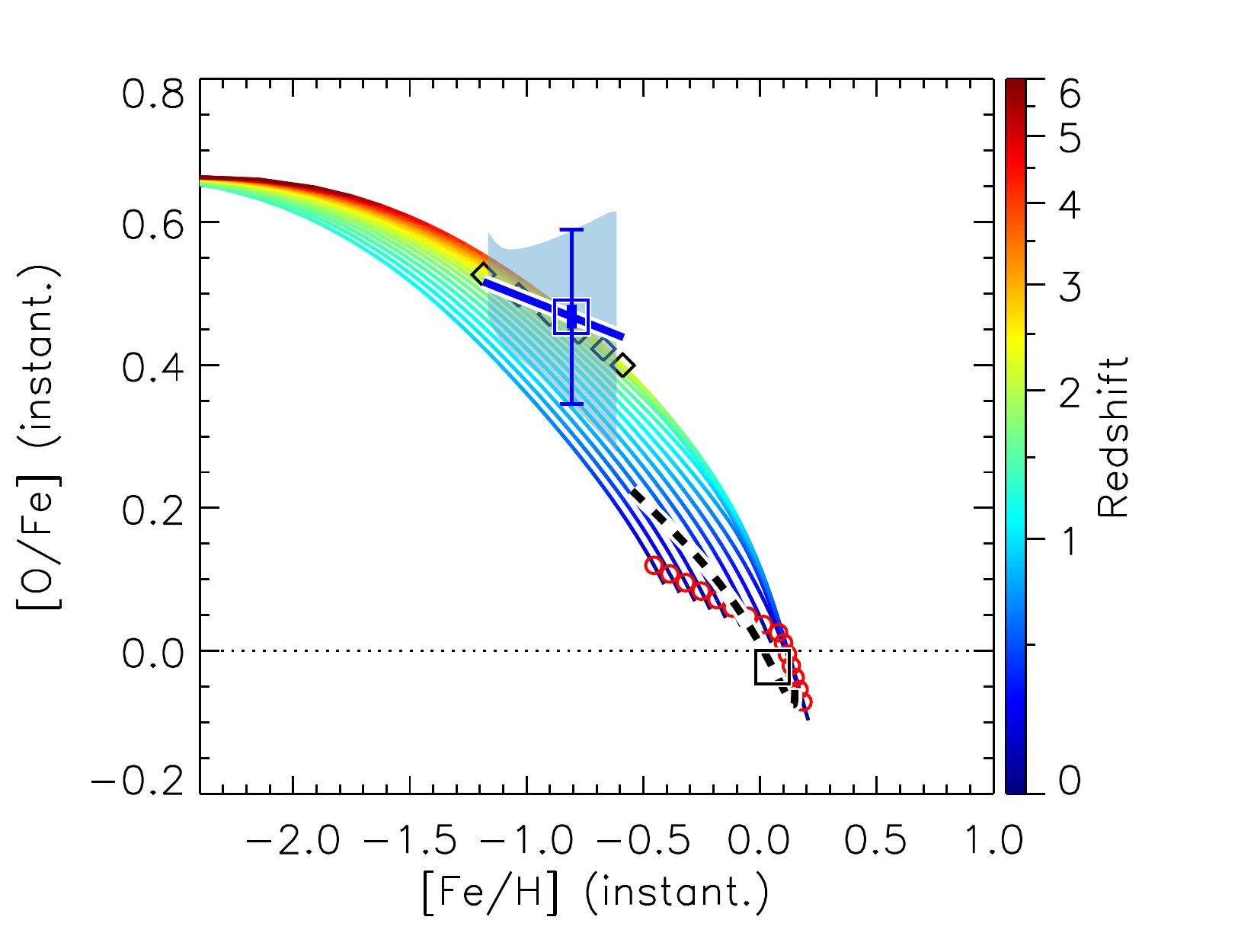}
\includegraphics[width=3.3in]{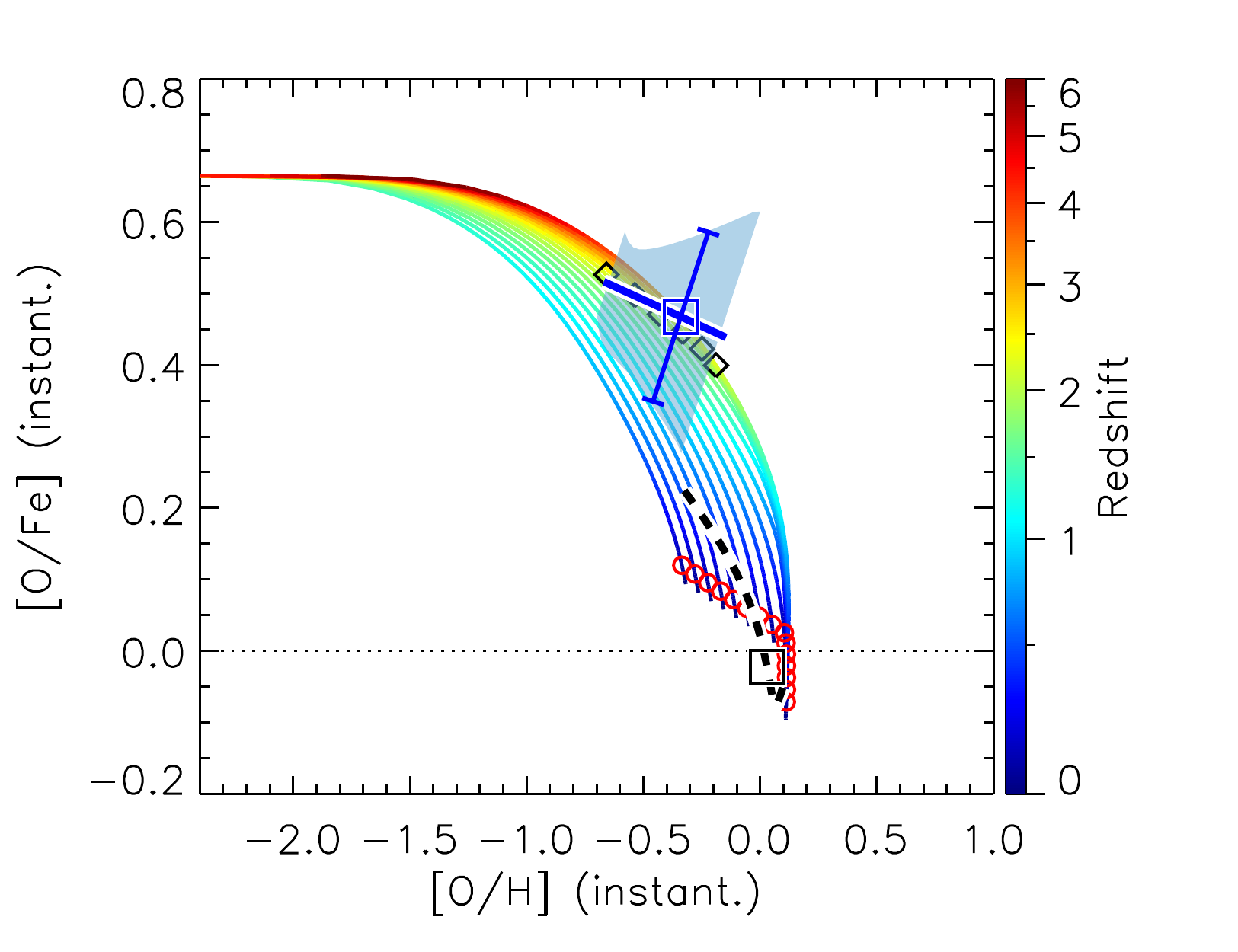}
\caption{
The calculated evolutionary tracks in the [Fe/H]--[O/Fe] (left) and [O/H]--[O/Fe] (right) diagrams in comparison with the observations (same as Figure \ref{fig:Z_vs_OFe}).  Visualization of the evolutionary tracks are the same as in Figure \ref{fig:models_MZ}.
The horizontal dotted line marks [O/Fe]$=0$ (i.e., the solar value) in both panels.
\label{fig:models_Z_vs_OFe}}
\end{center}
\end{figure*}

In Figure \ref{fig:models_Z_vs_OFe}, we show the $\alpha$-enhancement versus overall metallicity of these same models, i.e., [O/Fe] as a function of either [Fe/H] (left panel) or [O/H] (right panel).  As in Figure \ref{fig:models_MZ}, the model galaxies with $M_\ast \gtrsim10^9~M_\odot$ are marked at $z=2.2$ and at $z=0$ in order to compare with the observations (as in Figure \ref{fig:Z_vs_OFe}).  Both the evolutionary tracks in this diagram and the variations within the population at fixed epoch (locus of black and red points) may be read from this diagram. 

It can be seen that the modeled evolutionary tracks are in good agreement with the overall change in both metallicity and [O/Fe] between $z\sim 2.2$ and the present epoch.  This is not surprising because the model parameters were tuned (within reasonable ranges) to fit the oxygen and iron MZRs, and thus implicitly the [O/Fe], at both redshifts.

The models present variations across mass that show a negative slope between these quantities.  The observed slopes are similar to what are observed at these redshifts.  However, as mentioned in Section \ref{sec:O/Fe}, large uncertainties in the slopes do not enable us to make any robust statement.
We thus focus on the representative [O/Fe] and [Fe/H] (or [O/H]) values at our single fiducial mass of $M_\ast=10^{10}~M_\odot$ at each redshift (squares in Figure \ref{fig:models_Z_vs_OFe}).

It is noticeable that, at least for the range of representative SFHs considered, the model tracks all follow a relatively narrow path in the $\alpha$/Fe--metallicity planes. The tightness of this path is enhanced because the locus of $\alpha$/Fe and metallicity (at different masses) at a given epoch is evidently quite close to being parallel to the individual evolutionary tracks.  

\begin{figure*}[tbp]
\begin{center}
\includegraphics[width=3.2in]{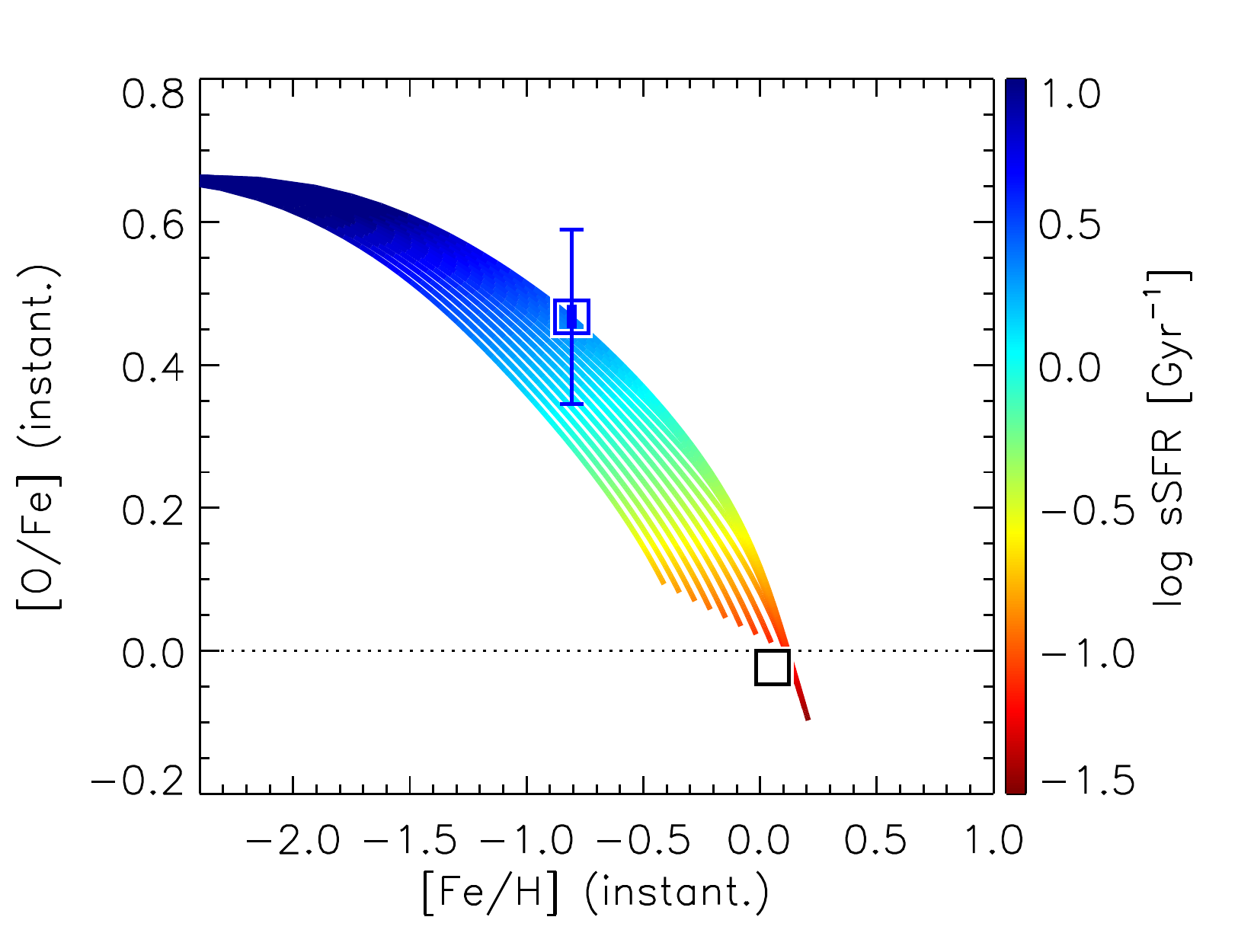}
\includegraphics[width=3.2in]{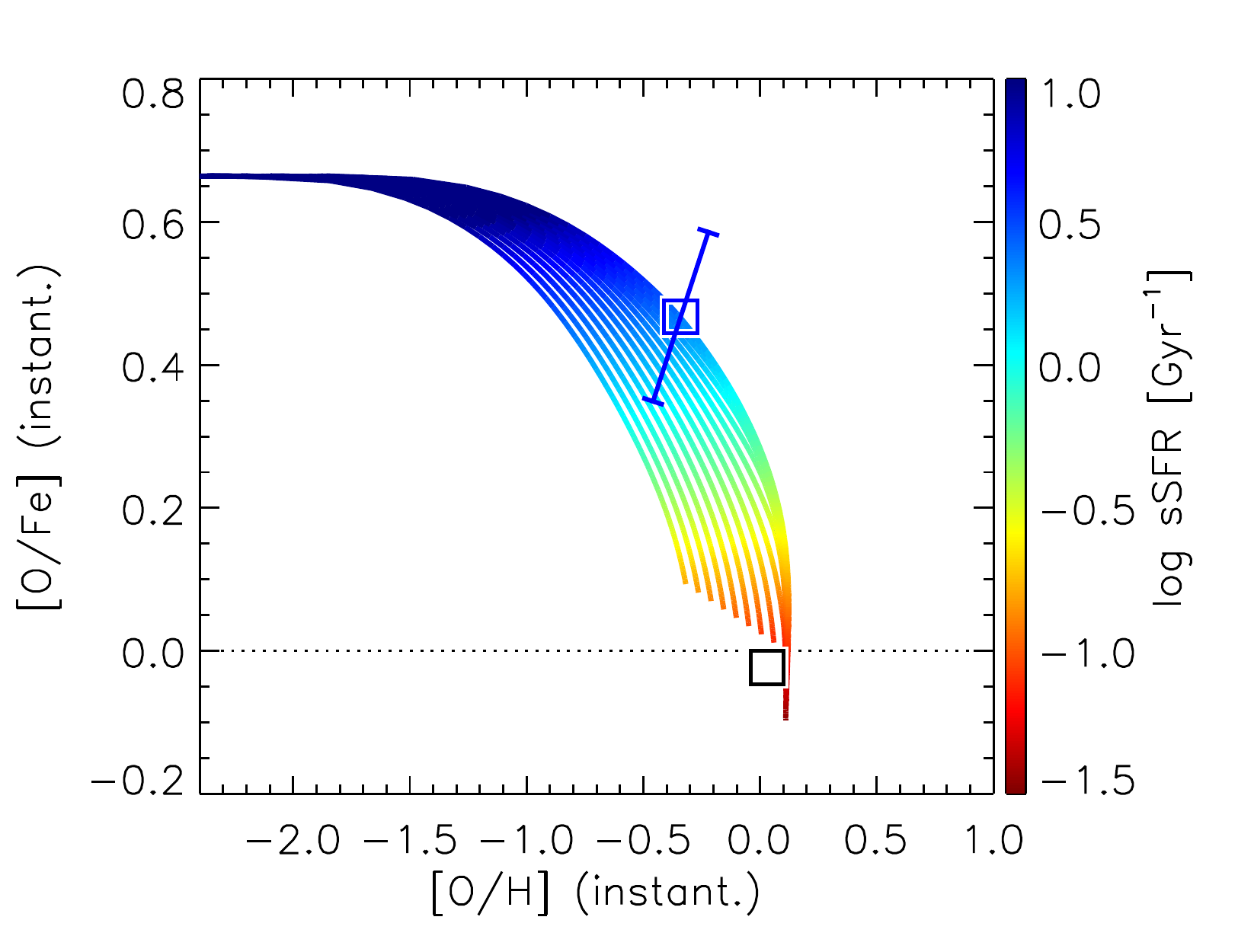}
\includegraphics[width=3.2in]{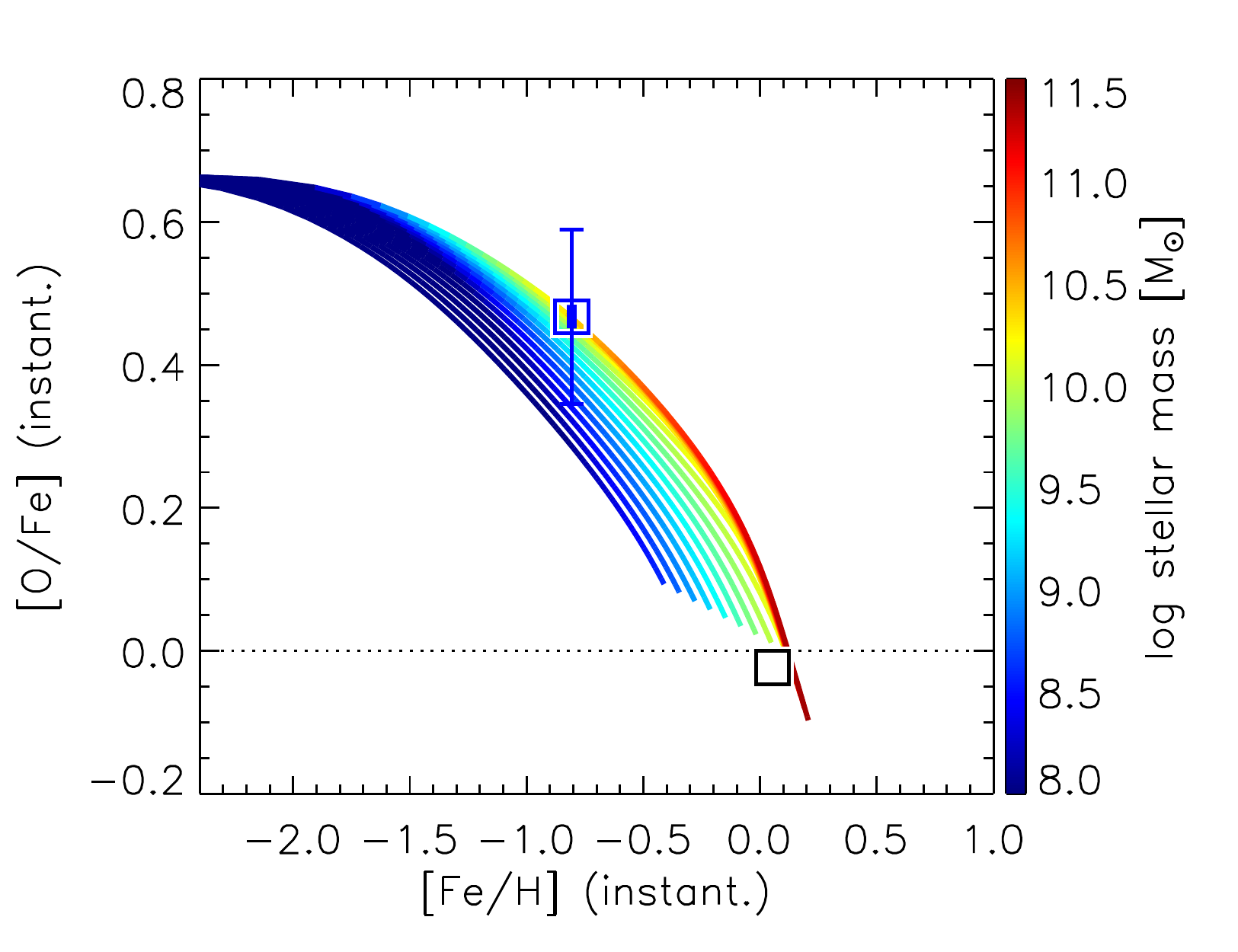}
\includegraphics[width=3.2in]{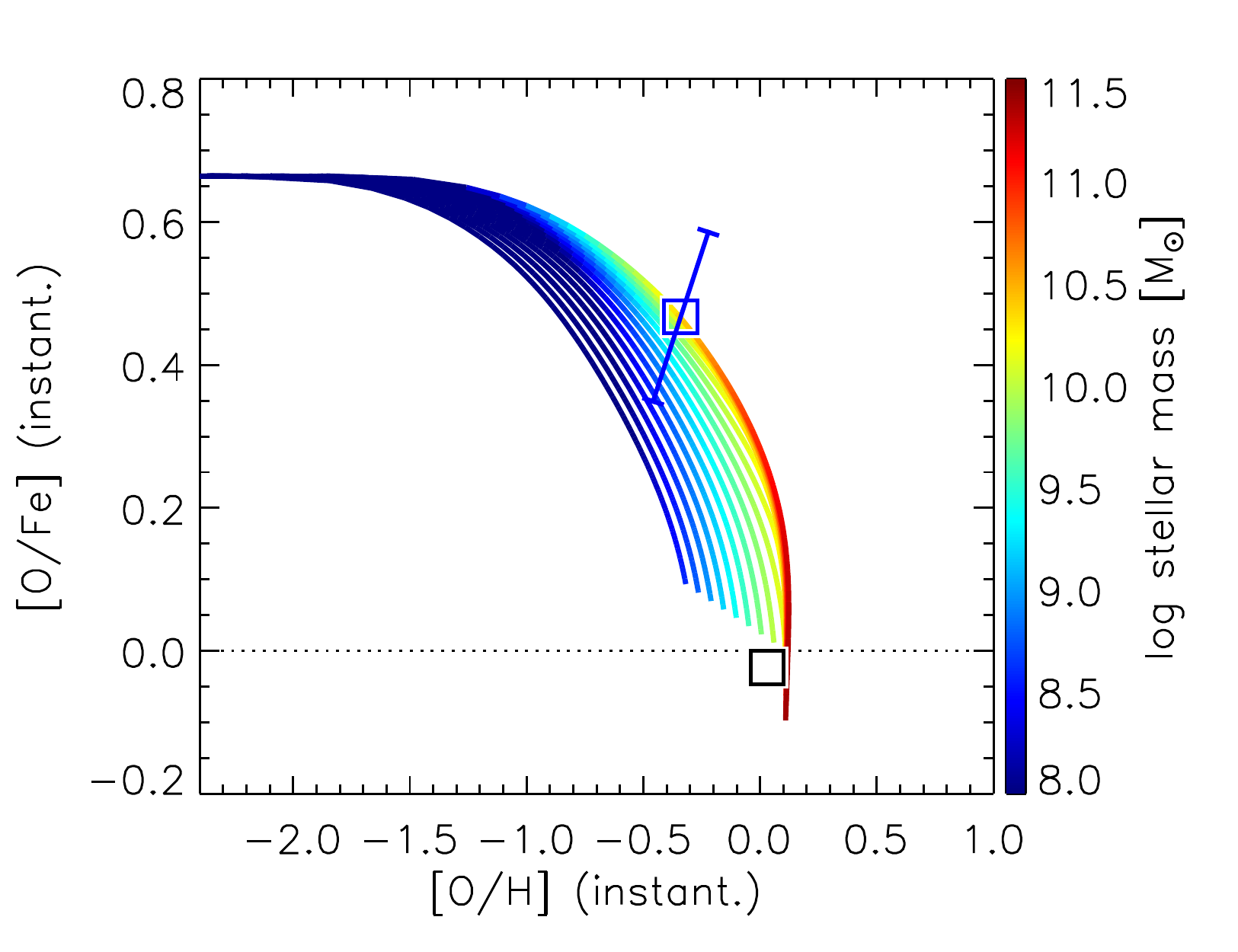}
\caption{
Same as Figure \ref{fig:models_Z_vs_OFe} but the evolutionary tracks are here color-coded by the instantaneous sSFR (upper panel) and stellar mass (lower panel).  We omit all the symbols except the representative measurements at two redshifts.
\label{fig:models_Z_vs_OFe_M_sSFR}}
\end{center}
\end{figure*}

\begin{figure}[tbp]
\begin{center}
\includegraphics[width=3.3in]{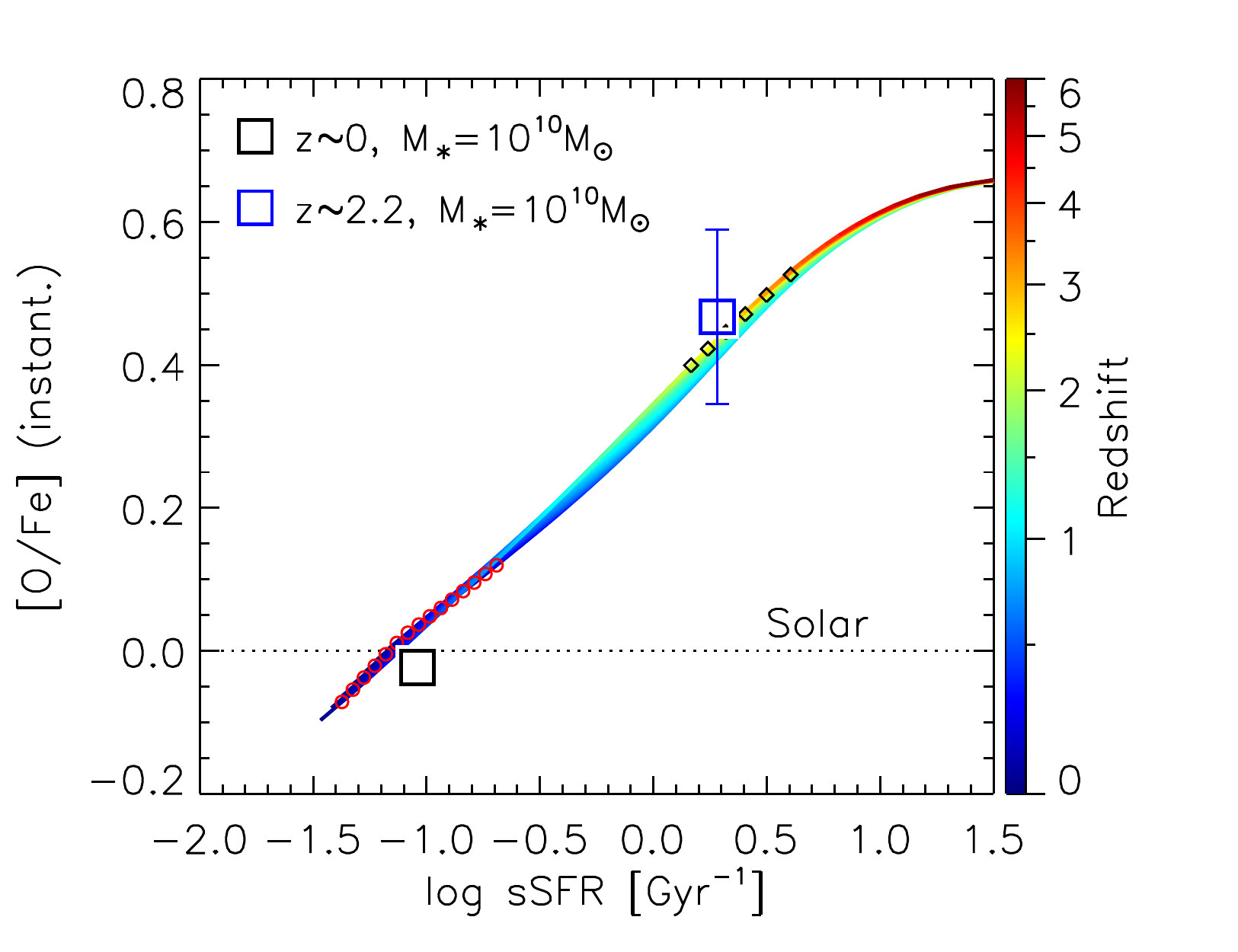}
\caption{[O/Fe] as a function of sSFR for all the model tracks.  The squares indicate the representative measurements at $M_\ast=10^{10}~M_\odot$ at the two redshifts.  Visualization of the evolutionary tracks are the same as in Figure \ref{fig:models_MZ}.
\label{fig:models_sSFR_vs_OFe}}
\end{center}
\end{figure}

We now turn to investigating whether the observed offset between $z\sim0$ and $z\sim2.2$ (i.e., the ``evolution vector'') in this figure can be explained by the change with redshift in some other parameter, and whether that parameter, if present, might also be responsible for producing the variation within the population at a single epoch.  
While the [O/Fe] (or $\alpha$/Fe) is often loosely considered as being indicative of the ``age'' of a stellar system, because of the time delay in producing much of the iron via SNe~Ia, it should really reflect, especially in a continuous ``flow-through'' scenario, the sSFR.  The sSFR reflects the ratio of the current SFR to some average SFR in the past.  This ratio will thus determine the relative number of CCSNe and SFe~Ia at any point in time, which will then determine the ``instantaneous'' $\alpha$-enhancement of the gas (see Equation \ref{eq:ZOeq/ZFeeq}).

To explore this, we show in Figure \ref{fig:models_Z_vs_OFe_M_sSFR} the same evolutionary tracks (as in Figure \ref{fig:models_Z_vs_OFe}) but now color-coded by the instantaneous sSFR (upper panels) and $M_\ast$ (lower panels) of the model galaxy.  Now the different dependences of the evolutionary tracks on these quantities are quite obvious: it is clear that the [O/Fe] is very tightly correlated with the sSFR across the whole range of these evolutionary tracks, and is largely unaffected by either the metallicity or the mass of the system. This is indicated by the striking horizontal banding of the sSFR-coded colors in the upper panels.  In contrast, the inclined $M_\ast$-coded color banding in the lower panels shows that the metallicity (whether [O/H] or [Fe/H]) depends on both the stellar mass and on [O/Fe].  In other words, using the tight relation between [O/Fe] and sSFR, the metallicity depends on the {\it mass} and {\it sSFR} together. 

Figure \ref{fig:models_sSFR_vs_OFe} shows the same data in a different way, demonstrating that, at least in the gas-regulated models and representative SFHs considered in this paper, there is remarkably little scatter between [O/Fe] and the instantaneous sSFR of the model galaxies.  The [O/Fe] scales as $\approx \mathrm{sSFR}^{0.34}$ at $\log(\mathrm{sSFR}~[\mathrm{Gyr^{-1}}]) < 0.5$. 

These results support the idea that the tight relation between [O/Fe] and sSFR is quite fundamental, being almost independent of the epoch, the shape of the SFHs, and thus, in our modeling, the present-day mass and thus the values of SFE and $\eta$.

Lastly we compare these results to the concept of the so-called ``fundamental metallicity relation'' (FMR; e.g., \citealt{2010MNRAS.408.2115M,2010A&A...521L..53L}; see also \citealt{2013ApJ...765..140A,2013MNRAS.434..451L}).  The FMR established that the SFR appears to be a second parameter in the gas-phase [O/H] MZR at low redshift, in the sense that higher SFR galaxies at a given mass have lower oxygen metallicities.  Further, it was then shown that the observed evolutionary change in [O/H] to high redshift (at a given mass) was the same as that obtained by simply extrapolating the trend with SFR established at $z \sim 0$ to the much higher SFR seen at high redshift.

An important insight into the FMR came from the introduction of the gas-regulated model of galaxies \citep{2013ApJ...772..119L}.  Assuming a quasi-equilibrium state, the (gas-phase) metallicity of the galaxy is set ``instantaneously'' by two considerations. The first is the specific rate at which the system is being fed by gas ($d\Mgas/dt/\Mgas$), which itself can be inferred from the sSFR. The second is the values of the two regulator parameters: the star-formation efficiency (SFE) and mass-loading $\eta$.  These latter parameters determine the gas content of the system that is necessary to achieve the required sSFR (see \citealt{2013ApJ...772..119L} for details and discussion).
The evolving sSFR of main-sequence galaxies can be viewed in this framework as the primary driver of the observed redshift evolution of the gas metallicity, but the positive MZR at a given epoch is the consequence of the change in sSFR along the main sequence together with the mass-dependences of the SFE and mass-loading $\eta$.

It is important to clarify that the dependence of [O/H] on the mass-dependent SFE and $\eta$ (see \citealt{2013ApJ...772..119L}) contrasts with the fact that [O/Fe] is determined almost entirely by the sSFR alone, acting as a proxy of the number ratio of CCSNe and SNe~Ia at a given time. The variations in SFE and $\eta$ (with mass) cause the variations in O/H at fixed sSFR (thus at fixed O/Fe) and thus the range in the evolutionary tracks in the [O/Fe]--metallicity diagrams (Figure \ref{fig:models_Z_vs_OFe}).  Note that, in the context of the gas-regulator model, if SFE and $\eta$ are constant, the gas-phase metallicity is also determined by sSFR alone.  The [O/Fe]--metallicity tracks in Figure \ref{fig:models_Z_vs_OFe} will thus be completely independent of SFHs, having no scatter, while, at a single epoch, the values of [O/Fe] and metallicities vary along this single path if the sSFR changes with mass along the main sequence, or a population has a catter in SFR at fixed mass as expected in the real universe.

\section{Comparison with Galactic stars}
\label{sec:Galactis_stars}

An important and revealing comparison may be made between the location in the [O/Fe]--metallicity planes of our high-redshift galaxies and those of individual Galactic stars.  Are we seeing at high redshift the formation of stars of the same type as seen today in the Galaxy?  This enables a direct confrontation between ``Galactic archaeology'' and observations of galaxies at high redshifts.  In this section, we will compare our results to data of the Milky Way (MW) stars from the literature.

\subsection{Thick disk stars and high-$z$ galaxies}
\label{sec:thick_disk_stars}

\begin{figure*}[tbp]
\begin{center}
\includegraphics[width=3.5in]{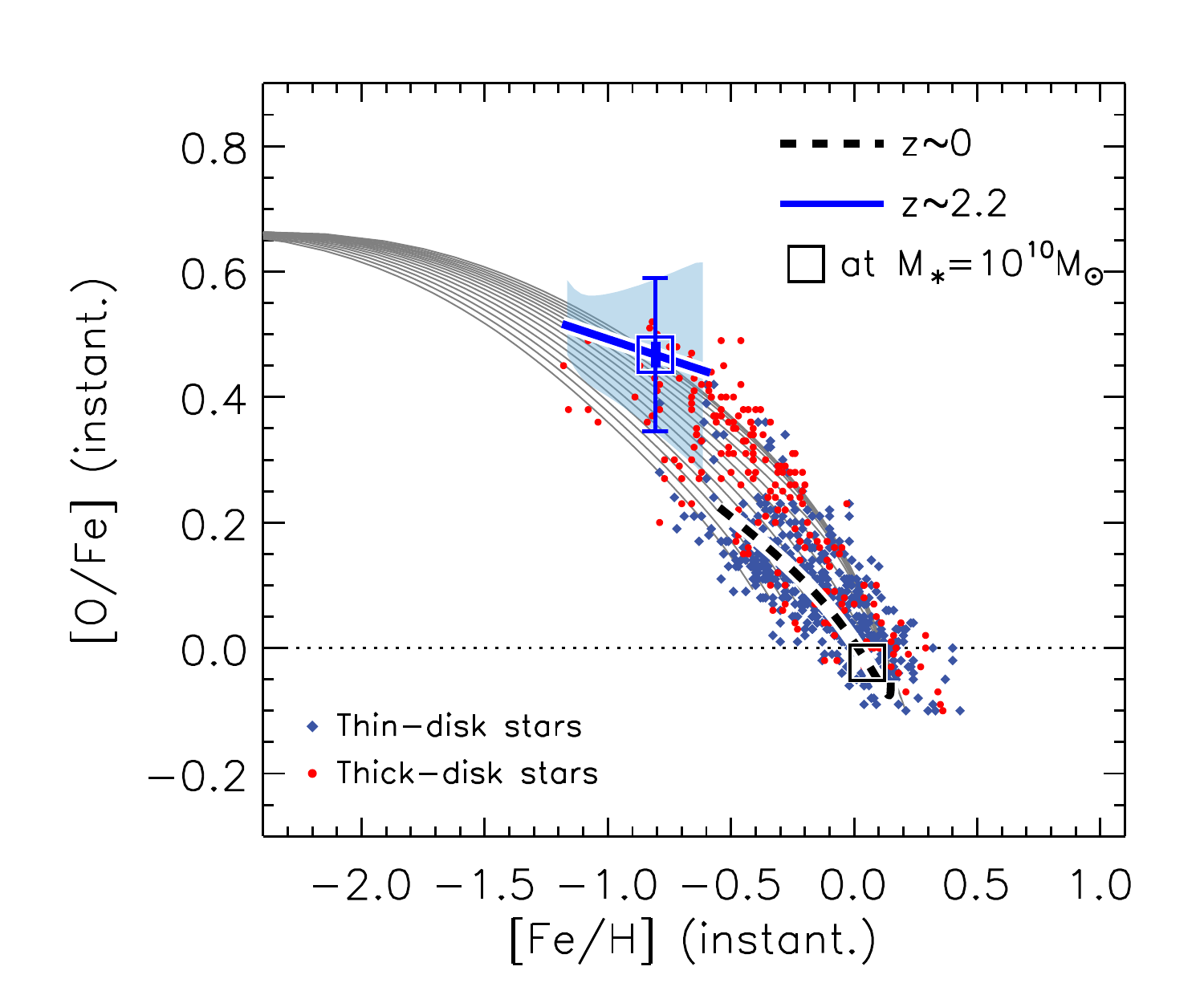}
\includegraphics[width=3.5in]{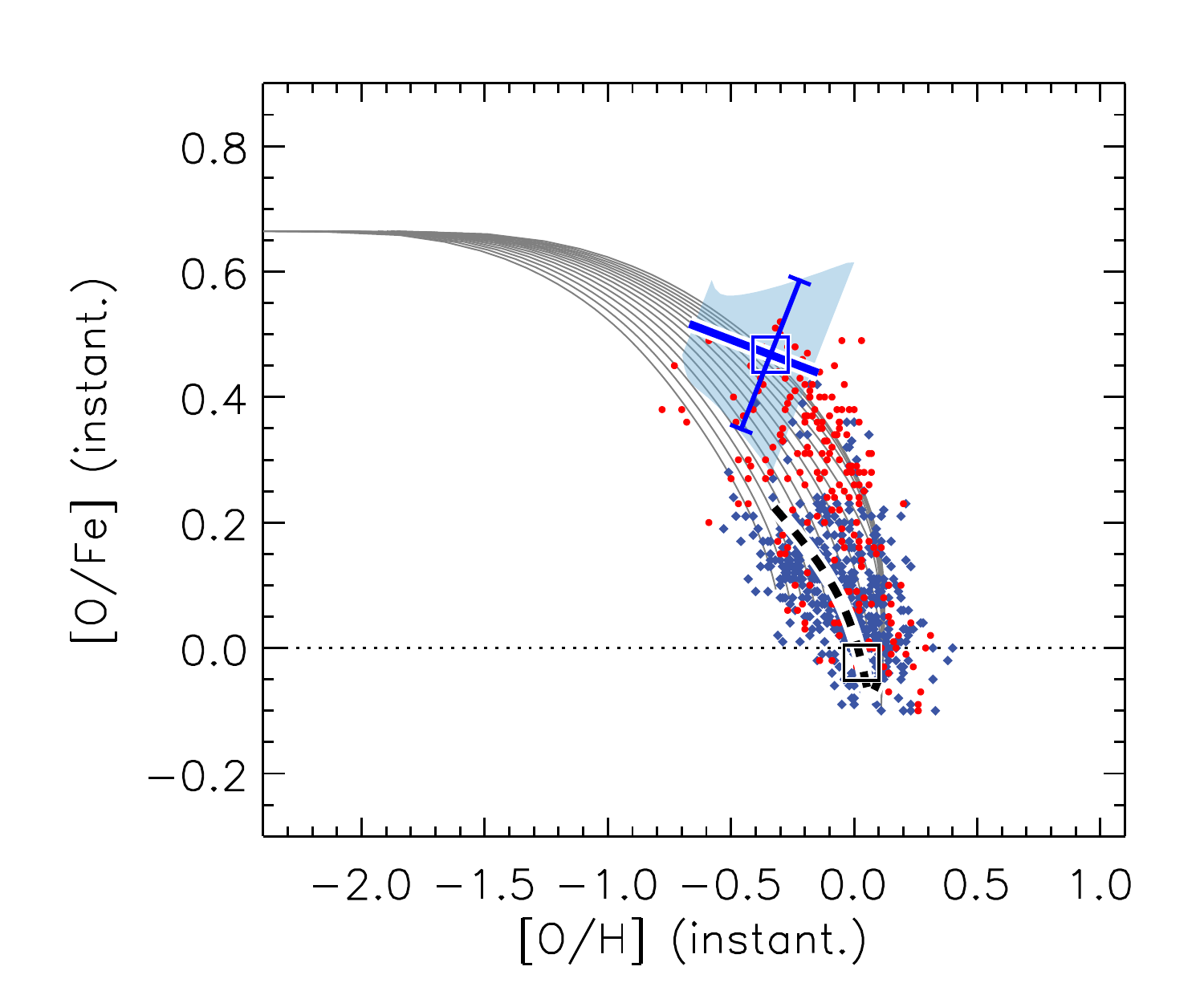}
\caption{The observed data for $z\sim2$ and $z\sim0$ galaxies (same as in Figure \ref{fig:models_Z_vs_OFe}) are compared to the data for the individual Galactic stars taken from \citet{2013ApJ...764...78R} in the [Fe/O]-metallicity diagrams.  The stars are separated into either thin disk stars (dark blue diamonds) or thick disk stars (red circles).  The gray curves show the model evolutionary tracks that are shown in Figure \ref{fig:models_Z_vs_OFe}.
\label{fig:Z_vs_OFe_MWstars}}
\end{center}
\end{figure*}

Figure \ref{fig:Z_vs_OFe_MWstars} shows data for nearby ($\lesssim 200~\mathrm{pc}$) FGK stars in the MW adapted from \citet{2013ApJ...764...78R}.  These stars are separated into either the thin-disk ($P_1>0.5$) or thick-disk ($P_2>0.5$) populations using the membership probability $P_i$ determined from their kinematics (see \citet{2013ApJ...764...78R} for details).  It is commonly known that the stars in the so-called ``thick disk'' are typically older, have lower-metallicity, and are $\alpha$-enhanced (i.e., have higher [O/Fe]) than the stars in the ``thin disk''.  It is indeed clear that the thin and thick disk stars form distinct sequences in the [O/Fe]--metallicity planes, although the classification based on kinematics is not perfect, as noted by the authors.

Galactic stars, especially the thick-disk stars, clearly lie along the locus extending between the two $M_\ast\sim10^{10}~M_\odot$ squares in Figure \ref{fig:Z_vs_OFe_MWstars}. 
In particular, it can be seen that our best-estimate locations of intermediate mass ($M_\ast=10^{10}~M_\odot$) galaxies at $z \sim 2$ in the two [O/Fe]-metallicity diagrams lie at the top of the sequence of thick disk stars in our own Galaxy.  

Figure \ref{fig:Z_vs_OFe_MWstars} also shows that the simple models constructed in Section \ref{sec:modeling} also successfully reproduce the locus of these Galactic stars. This gives added confidence that they are reasonable.

\citet{2016ApJ...831..139M} found the median ages of the thick disk population of stars to be $\approx9$--5~Gyr, decreasing from the inner to the outer disk.  \citet{2017ApJ...837..162K} also estimated the average age of the thick disk stars to be $\approx9$~Gyr.  It is thus quite plausible that those Galactic thick-disk stars located around our high-$z$ measurement (i.e., the relatively metal-poor $\alpha$-enhanced stars) formed around $z\sim2\textrm{--}3$, i.e., some 10 Gyr ago and some 3.5 Gyr after the Big Bang.  We have already discussed how, in terms of the simple flow-through scenario of chemical evolution discussed in Section \ref{sec:modeling}, it is natural to get these high $\alpha$-enhancements $\sim10$~Gyr ago.
Note that, in our modeling, present-day galaxies with the same mass as the Galaxy ($6.4\times10^{10}~M_\odot$; \citealt{2011MNRAS.414.2446M}) had a mass around this fiducial value of $M_\ast=10^{10}~M_\odot$ at $z\sim2$.

The evident agreement of [O/Fe] and [Fe/H] (or of [O/Fe] and [O/H]) between the old stars in the Galactic thick disk and the high-redshift galaxies with masses comparable to those expected for MW progenitors provides, in our opinion, a beautifully direct link between the results of Galactic archaeology and observations of galaxies in the high-$z$ universe that has long been assumed but rarely, if ever, established directly.

\subsection{Comparison with the MW bulge-like stars}
\label{sec:metal-rich_bulge}

In the subsection above, we found that the agreement between the chemical abundances seen in high-$z$ ($z \sim 2$) galaxies and the MW thick disk population, plus the estimated ages of the latter, support the idea that, when observing $z\sim2$ galaxies, we are witnessing the formation of stars that will constitute the thick disk population in MW-like galaxies.  However, as we will see below, some caveats and questions are posed when a detailed comparison is made with bulge stars in the Galaxy.

Recently, \citet{2020A&A...638A..76Q} have presented a detailed study of chemical abundances from APOGEE data \citep{2018ApJS..235...42A} combined with stellar distances from Gaia \citep{2018A&A...616A...1G}.  They constructed detailed distributions in [$\alpha$/Fe] versus [Fe/H] plots at several radial distances from the center of the Galaxy and at three vertical distances from the Galactic plane (see their Figure 6).  
Note that they measure [$\alpha$/Fe] using different multiple $\alpha$-elements, and the comparison to [O/Fe] probably needs some offset (order 0.1~dex; \citealp{2008AJ....136..367M,2020AJ....160..120J}).
They showed that there are basically two disjoint populations in the [$\alpha$/Fe]--[Fe/H] plot producing, at each spatial location, a bimodal distribution of stars.  Broadly speaking, these two distinct populations comprise a metal rich (solar or more) component with [$\alpha$/Fe] close to solar and a lower metallicity component with high [$\alpha$/Fe] ($>0$).  Globally, $\sim 50\%$ of bulge stars belong to the metal-rich component \citep{2017A&A...599A..12Z}

In the bulge region ($R<2~\mathrm{kpc}$), the supersolar metal rich component with [$\alpha$/Fe] spread around zero dominates.  Moving radially outwards in the plane of the disk, the metal-rich component shifts slightly to lower (i.e., subsolar) metallicities without significant change in [$\alpha$/Fe], while the metal-poor component progressively vanishes.  Instead, moving vertically away from the plane, it is the metal-rich component that tends to vanish leaving the metal-poor component dominant, without much change in its abundances.  This variation in the spatial manifestation of the bimodality may therefore be related to the ``thin'' and ``thick'' disk components, at least at intermediate radii.  However, the dominant metal-rich component in the central bulge region challenges this simple picture and poses questions for the beautiful concordance between $z\sim2$ galaxies and the thick disk stars shown in Section \ref{sec:thick_disk_stars}.

It is clear that we find no evidence in our zCOSMOS-deep sample at $z \sim 2.2$ for star-formation with abundances comparable to the metal-rich component in the \citet{2020A&A...638A..76Q} distributions.
Some studies, however, have indicated that this metal-rich component is dominated by stars of age $\sim 10$~Gyr or older, as demonstrated by multi-band Hubble Space Telescope (HST) photometry extending from the near UV to the near-IR \citep{2009AJ....137.3172B,2018ApJ...863...16R}.\footnote{In this respect we note that \citet{2017A&A...605A..89B} derive a broad age distribution for the supersolar low-$\alpha$ stars in the bulge, but still a fraction of them are given ages older than 8--10~Gyr.}  Based on these indications, this component seems to have formed at $z\sim 2$. The obvious question is why are such abundances not seen in our $z\sim 2$ zCOSMOS-deep galaxies?  There are some possible explanations for this observational discrepancy.

The first possibility is that the central metal-rich component in the MW had not in fact formed by $z\sim2$, i.e., that the ages of these stars have been over-estimated.  This possibility would be consistent with our observational results and also with our models, since these suggest that galaxies with SFH that follow the evolution of the main sequence cannot have $\mathrm{[\alpha/Fe]}\sim0$ at $z\sim2$.  The main problem with this first possibility is the observational evidence for old ages for these metal-rich low-$\alpha$ stars.  
One way out could come from assuming that the central regions of the MW are unusual and not representative of the general galaxy population seen in the high redshift universe.

A second possibility is that metal-rich stars are indeed forming at $z\sim2$, as indicated by their ages, but that they these metal-rich stars will not have contributed to the rest-frame FUV spectra for being heavily obscured by dust, which may be common in a supersolar metallicity environment.

To test the hypothesis of obscured star-formation {\it within} the zCOSMOS-deep galaxies, we examined the far-infrared emission of our sample using the public super-deblended catalog adapted from \citet{2018ApJ...864...56J}.  The total FIR luminosity and FIR-based SFR are estimated in the same way as in \citet{2021ApJ...909..213K}.  We found only 54 sources out of the entire sample of 1336 galaxies for which the FIR luminosity is detected at $>3\sigma$.  Although the nominal FIR-based SFRs (median $\sim140~M_\odot~\mathrm{yr^{-1}}$) are slightly higher than the SED-based SFR estimates ($\sim100~M_\odot~\mathrm{yr^{-1}}$) in this FIR-detected subset, there is no compelling evidence of substantial star formation hidden across the entire sample within the available data.

It is harder to exclude the possibility that the required obscured star-formation is concentrated at any particular time in a small subset of the population, i.e. in high-luminosity infrared sources. However this would almost inevitably require starburst situations, with very high sSFR.  Such starbursts can produce high metallicities but are very unlikely environments to produce the low (solar) $\alpha$-enhancements that are seen in the metal-rich Galactic population that we are trying to account for.

The third possibility is that the formation of the metal-rich component was completed well before $z\sim2$, and that such stars are present in the zCOSMOS-deep galaxies but already old enough so as not to contribute to the ultraviolet light. 
Such a scenario seems implausible to us for several reasons.  Pushing star-formation to earlier epochs will clearly tend to {\it increase} the $\alpha$-enhancement, not decrease it, exacerbating the problems identified in the previous paragraph. Not least, the stars would have had to form in a still shorter time at even higher sSFR.

Lastly, we should also consider a bias in the selection of our sample, namely that galaxies containing metal-rich star formation may be not represented in our blue-selected ($B_\mathrm{AB} \lesssim 25$) zCOSMOS-deep sample, which excludes objects with low FUV continuum.  It is naturally expected that galaxies either dominated by metal-rich star formation or having experienced starbursts are dust rich and thus their rest-frame UV emission may well disappear below the selection limit.  A highly-complete study, including UV faint sources, is desired to gain a more robust conclusion.

We return to emphasize the basic difficulties (required in both the second and third possibilities discussed above) of achieving low (solar) $\alpha$-enhancement at $z \sim 2$, at least using our fiducial SN~Ia DTD that provides the satisfactory match presented in Section \ref{sec:modeling}.  
We showed in Section \ref{sec:model_alpha-metallicity} that the SFHs that are obtained by integrating the main sequence simply do not yield gas abundances with low (i.e. solar) $\alpha$-enhancements at these redshifts. This is ultimately because of the high sSFR of the main sequence at $z \sim 2$.  Except in extremely contrived scenarios, low $\alpha$-enhancement stars must be formed at low sSFR. If a significant mass of stars is to be formed, then this star-formation must be maintained over an extended period of time, i.e. over timescales of order sSFR$^{-1}$. This is naturally achieved at late epochs in the universe, but it is hard to see how this can happen at much earlier times. 
Quite independent of our own observational results, this conundrum represents a basic puzzle about the origin of the low $\alpha$-enhancement metal-rich central population identified in the central kpc of our Galaxy by \citet{2020A&A...638A..76Q} if these stars are truly over 10~Gyr old rather than $\lesssim 4$~Gye old (i.e., formed after $z\sim0.5$) as indicated by our model (Figure \ref{fig:models_Z_vs_OFe}).

In this context, there is another possibility to be mentioned, namely that the SN~Ia DTD may shift to shorter delays with increasing metallicity, in particular in the supersolar regime.  For example, in the double-degenerate scenario for SN~Ia progenitors, the event results from the merging of two white dwarfs (WD) as they spiral-in due to gravitational wave radiation. Thus, the DTD is controlled by the distribution of the binary WD separations as they emerge from their last common envelope event.  As the delay time scales as the fourth power of such separation \citep[e.g.,][]{2005A&A...441.1055G}, a $\sim 30\%$ reduction in the distribution of WD separations would result in a shift by a factor of $\sim 5$ in the DTD, e.g., moving the median delay time from, say, $\sim 1~$Gyr to $\sim 200$~Myr.
The amount of orbital shrinkage during common envelope events is notoriously hard to predict, nevertheless we note that stellar sizes are larger at high metallicity, hence stars fill their Roche lobe at an earlier evolutionary phase compared to the case at lower metallicity.  An effect of metallicity on the WD orbital separations and on the resulting DTD is therefore to be expected, though hard to quantitatively predict, and the few attempts in this direction remain inconclusive \citep{2011PASJ...63L..31M,2012A&A...543A.137M,2013ApJ...770...88K}.
The old ages, high metallicity and low [$\alpha$/Fe] of bulge stars could then be reconciled if the DTD were to move to substantially shorter delays in the supersolar regime.

We should also mention that, by construction, our models with representative SFHs produce one continuous sequence of stars, rather than the two distinct components as observed in the $\alpha$/Fe--metallicity planes \citep{2020A&A...638A..76Q}.  This suggests that the construction of the MW was a more complex phenomenon.

It is clear to us that integrated light spectroscopy of high-redshift galaxies, such as presented in this paper, can highlight the problem, but is unlikely to enable us to understand the complexities revealed by individual star abundances across the body of the MW. It is possible that high-resolution imaging and integrated field spectroscopy with the forthcoming JWST instruments will finally give clues to the solution of this puzzle.

\section{Summary}
\label{sec:summary}

We have measured the stellar metallicities of galaxies for 1336 star-forming galaxies at $1.6\le z\le 3.0$ using high signal-to-noise stacked low-resolution ($R\sim200$) rest-frame FUV spectra from the zCOSMOS-deep survey.  The metallicities were estimated using fits of high-resolution model spectra constructed from stellar population synthesis models across a range of the stellar metallicity. These metallicity estimates enabled us to construct the relationship between stellar mass and stellar metallicity, i.e., the stellar MZR at $z \sim 2.2$.

The measured stellar metallicities, which mostly reflect the iron abundance, range between $-1.3 \lesssim \log Z_\ast/Z_\odot \lesssim -0.4$ across the stellar mass range $10^9 \lesssim M_\ast/M_\odot \lesssim 10^{11}$. Because they are based on the spectra of short-lived massive stars, we argue that this iron abundance should be representative of the gas phase in these galaxies.  Our measurements are consistent with the one previous work on similarly high-$z$ galaxies \citep{2019MNRAS.487.2038C}.

A clear positive correlation between stellar mass and stellar metallicity is established, in which the metallicity scales as $\log (Z_\ast / Z_\odot) = -0.81 + 0.32 \log (M_\ast/10^{10} M_\odot)$ (Equation \ref{eq:MZ_linear}).  
Comparing with iron metallicity data at $z \sim 0$ from the literature, we find that the $z\sim2.2$ stellar MZR is offset by $\sim0.8$~dex below the local relation.  

Adding published [O/H] measurements at both high and low redshifts, we implied [O/Fe] ratios (i.e., $\alpha$-enhancement) to be $\approx 0.47\pm 0.12$ at $z \sim 2$ for $M_\ast\sim 10^{10}~M_\odot$ galaxies. This is considerably enhanced against the local value of $\mathrm{[O/Fe]}\sim0$ and, in fact, approaches the $\mathrm{[O/Fe]}\sim0.6$ limit imposed by the yields of core-collapse supernovae.  This indicates that SNe Ia have not yet contributed very much to the iron supply at these epochs.

These results are then compared with the expectations of ``flow-through'' gas-regulator models, especially in the context of evolutionary tracks in the [O/Fe]--metallicity plane.   In constructing these, it is assumed that the SFHs follow those implied by the evolving main sequence of star-forming galaxies.  Adjusting the regulator parameters and iron yields within very reasonable ranges, it is found that the models can reproduce the evolution of [O/Fe] and [Fe/H] (or [O/H]) from $z\sim2$ to $z\sim0$.  The models predict that galaxies at $M_\ast \sim 10^9\textrm{--}10^{11}$ locally should lie on a relatively narrow locus in the [O/Fe]--[Fe/H] (or [O/H]) plane if they have been continuously forming stars and following the main sequence in the past.

An important insight obtained from this modeling is that the instantaneous (gas-phase) $\alpha$-enhancement is determined almost entirely by the instantaneous sSFR of the galaxy.  This is because the sSFR is a good proxy of the instantaneous number ratio of CCSNe to SNe~Ia which is what effectively determines the gas-phase $\alpha$-enhancement in flow-through models.  The variations in [O/Fe] among a galaxy population at a single epoch arise due to the mass-dependence of the main-sequence sSFR, while the variations in the evolutionary tracks in the [O/Fe]--metallicity planes arise due to the dependence of the regulator parameters (SFE and mass-loading $\eta$) on the stellar mass.  

We show that $z\sim2$ galaxies at a representative mass of $10^{10}~M_\odot$ have similar $\alpha$-enhancement and metallicity as the low metallicity thick disk stars in our own Galaxy.  These Galactic stars were probably formed around 10~Gyr ago, when the Milky Way presumably had a stellar mass of around this same value, $10^{10}~M_\odot$.  This observation therefore provides an unusually direct concordance between the results of Galactic Archaeology and observations of presumed Milky Way progenitors at high redshift.

There remains, however, an open question about the formation of the population of old metal-rich stars seen in the MW bulge with low (roughly solar) $\alpha$-enhancement.  Our rest-frame FUV data at $z \sim 2$ shows no evidence of such high metallicities and low-$\alpha$ enhancements. 
We discuss three possible explanations of this observational discrepancy: (i) that these central Galactic stars are not as old as so far estimated with a variety of methods, and formed well after $z\sim 2$. Besides being in contrast with current age estimates, this option would predict the existence of supersolar star-forming galaxies at lower redshifts, that have not been observed so far; (ii) that they may indeed be forming at $z \sim 2$ but doing so in highly obscured environments as one may expect in a supersolar metallicity regime; or (iii) that they may have already formed well before $z\sim 2$.  We discuss how both the second and third possibilities are problematic in terms of achieving low $\alpha$-enhancements by $z\sim 2$, and, as a possible solution, the possibility that the SN~Ia delay time distribution shifts to substantially shorter delays in the supersolar metallicity regime.

Spatially-resolved imaging and spectroscopy, which will be enabled by JWST, may be able to constrain the formation scenarios of the possible metal-rich bulges and the detailed process of chemical evolution through cosmic history.  Furthermore, the forthcoming multi-object spectrographs, i.e., Subaru/PFS and VLT/MOONS, will enable less biased studies with highly complete sample, including UV faint sources which were excluded in this study.

\begin{acknowledgments}
This research is based on observations undertaken at the European Southern Observatory (ESO) Very Large Telescope (VLT) under the Large Program 175.A-0839 and has been supported by the Swiss National Science Foundation (SNF) and JSPS KAKENHI Grant Number JP21K13956. This work made use of v2.2.1 of the Binary Population and Spectral Synthesis (BPASS) models as described in \citet{2017PASA...34...58E} and \citet{2018MNRAS.479...75S}.
YP acknowledges National Science Foundation of China (NSFC) Grant No. 12125301, 11773001 and 11991052.
\end{acknowledgments}


\bibliography{ads}
\bibliographystyle{aasjournal}



\appendix

\section{Offset between gas-phase and stellar Fe/H}
\label{sec:offset_Fe_H}
It is straightforward to calculate mass-weighted stellar metallicities, $Z_\mathrm{Fe,\ast}^M$, once $\mathrm{SFR}(t)$ and (gas-phase) $Z_\mathrm{Fe}(t)$ are specified:
\begin{eqnarray}
    Z_\mathrm{Fe,\ast}^M(t) = \frac{\int_0^t Z_\mathrm{Fe}(t^\prime) \mathrm{SFR}(t^\prime)dt^\prime}{\int_0^t \mathrm{SFR}(t^\prime)dt^\prime}.
    \label{eq:Zstar_Fe_MW}
\end{eqnarray}
To calculate luminosity-weighted stellar metallicities, we additionally need a library of the spectra for a single stellar population as a function of age.  We adopted the same BPASSv2.2.1 template spectra as in the main analysis and calculate the stellar metallicity weighted by the average luminosity around 1500~{\AA} (FUV) and 5500~{\AA} (optical):
\begin{eqnarray}
    Z_\mathrm{Fe,\ast}^L(t) = \frac{\int_0^t Z_\mathrm{Fe}(t^\prime) L^\mathrm{SSP}_\lambda (t-t^\prime) \mathrm{SFR}(t^\prime)dt^\prime}{\int_0^t L^\mathrm{SSP}_\lambda (t-t^\prime) \mathrm{SFR}(t^\prime)dt^\prime},
    \label{eq:Zstar_Fe_LW}
\end{eqnarray}
where $L^\mathrm{SSP}_\lambda(t)$ ($\lambda=1500$ or 5500~{\AA}) is the luminosity density at a particular wavelength of a single stellar population of age $t$.  Note that, in this equation, we ignore the effects of possible differential dust attenuation between the younger and longer-lived stellar components.

\begin{figure}[tbp]
\begin{center}
\includegraphics[width=3.5in]{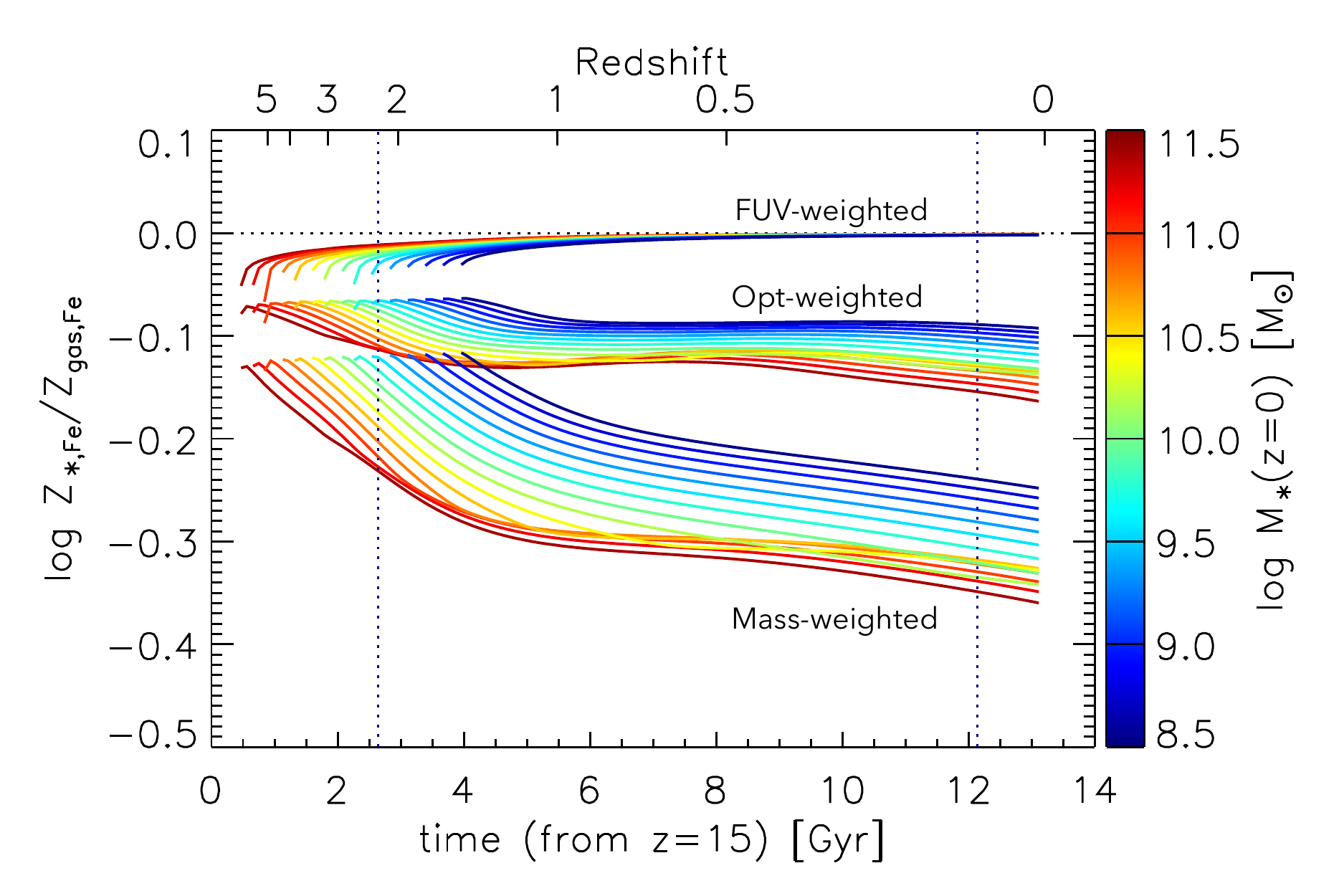}
\caption{
The stellar-to-gas iron metallicity ratio as a function of cosmic time for representative SFHs adopted in Section \ref{sec:modeling}.  The stellar $Z_{\ast,\mathrm{Fe}}$ values are inferred either weighting by stellar mass, FUV-, or optical luminosity, as labeled.  The evolutionary tracks for the individual SFHs are color-coded by the present-day stellar mass.  Two redshifts, $z=2.2$ of our high-$z$ sample and $z=0.08$ of the local sample, are marked by vertical dotted lines.  The horizontal dotted line marks $Z_\mathrm{\ast,Fe}=Z_\mathrm{gas,Fe}$.
\label{fig:hist_weighted_Fe_H}}
\end{center}
\end{figure}

We calculated Equations (\ref{eq:Zstar_Fe_MW}--\ref{eq:Zstar_Fe_LW}) for the same evolutionary tracks obtained in Section \ref{sec:modeling}.  Figure \ref{fig:hist_weighted_Fe_H} shows the ratios of either the stellar [Fe/H], weighted by either mass, $L_{1500}$ (FUV), or $L_{5500}$ (optical), to the instantaneous gas-phase [Fe/H] as a function of cosmic time.  The different evolutionary tracks correspond to the different SFHs and are color-coded by the present-day stellar mass.  The two redshifts $z=0.08$ and 2.2 are marked.

As expected, the $L_{1500}$-weighted [Fe/H] is similar to the gas-phase value within $\sim 0.02$~dex at $z=2.2$, and almost equivalent at $z=0$ when the metallicity change is slow.  In contrast, the mass-weighted and $L_{5500}$-weighted [Fe/H] values show substantial offsets and some scatters for different SFHs.  At $z=0.08$, the offsets in the $L_{5500}$-weighted values are in a range of $-(0.08\textrm{--}1.5)$~dex, being larger (in negative) for larger masses, i.e., lower sSFR.   We adopted $-0.1$~dex, the value for $M_\ast(z\sim 0)=10^{10}~M_\odot$, as a representative value for correcting the local stellar [Fe/H] from \citet{2017ApJ...847...18Z} so that they reflect better instantaneous values for comparison with our FUV-based [Fe/H] and the model values in Section \ref{sec:O/Fe} and later.

\end{document}